\documentclass[a4paper,11pt]{article}
\usepackage[a4paper,
  left=2.5cm, right=2.5cm,
  top= 3cm, bottom=4cm]{geometry}
\usepackage{amsmath}
\usepackage{oldgerm}
\usepackage{amssymb}
\usepackage{hyperref}
\usepackage[dvips]{graphicx}
\usepackage{epsfig}
\usepackage{color}
\usepackage{epic}
\usepackage[numbers, square, comma, sort&compress]{natbib}

\newcommand{\be}{\begin{equation}}
   \newcommand{\ee}{\end{equation}}

\newcommand{\rem}[1]{} 

\newcommand{\nn}{\nonumber}

\def\Z{\mathbb{Z}}

\def\P{\mathbb{P}}
\def\L{\mathcal{L}}

\def\Hirz[#1]{\mathbbm{F}_{#1}}
\def\o[#1]{\overline{#1}}

\def\rem{$\clubsuit$}
\def\ch{\mbox{ch}}
\def\cale{\mathcal{E}}
\def\call{\mathcal{L}}

\numberwithin{equation}{section}

\title{}

\begin{document}

\thispagestyle{empty}
\vspace*{.2cm}
\noindent
\begin{flushright}{TUW-11-19\\
LMU-ASC 33/11\\
ZMP-HH/11-13}
\end{flushright}

\vspace*{2.0cm}

\begin{center}
{\Large\bf $G$-flux in F-theory and algebraic cycles}
\\[1.5cm]
{\large A.~P.~Braun$^a$, A.~Collinucci$^b$ and R.~Valandro$^c$\\[.5cm]}
{\it$^a$ Institute for Theoretical Physics,
Vienna University of Technology, Austria }
\\[.5cm]
{\it$^b$ Arnold Sommerfeld Center for Theoretical Physics, LMU Munich, Germany}
\\[.5cm]
{\it$^c$ II. Institut fuer Theoretische Physik, Hamburg University, Germany}
\\[.5cm]
{\small\tt (\,abraun@hep.itp.tuwien.ac.at}{\small ,} {\small\tt
\,andres.collinucci@physik.uni-muenchen.de}{\small ,} {\small and} {\small\tt
\,roberto.valandro@desy.de)}
\\[2.0cm]
{\bf Abstract}\\
\end{center}

We construct explicit $G_4$ fluxes in F-theory compactifications. Our
method relies on identifying algebraic cycles in the Weierstrass equation of elliptic Calabi-Yau fourfolds. We 
show how to compute the $D3$-brane tadpole and the induced chirality indices directly in F-theory. Whenever a weak coupling limit is available, we compare and successfully match our findings 
to the corresponding results in type~IIB string theory. Finally, we present some 
generalizations of our results which hint at a unified description of the elliptic Calabi-Yau fourfold together with the 
four-form flux $G_4$ as a coherent sheaf. In this description the close link between $G_4$ fluxes and algebraic cycles
is manifest.

\newpage

\tableofcontents

\section{Introduction}

In the recent years there has been increasing interest in model building with 7-branes in type~IIB/F-theory 
\cite{Beasley:2008dc,Beasley:2008kw,Donagi:2008ca,Donagi:2008kj,Donagi:2009ra,Hayashi:2008ba,Hayashi:2009ge,Blumenhagen:2008zz,Blumenhagen:2009yv}.
The main advantage of such constructions is that it is in principle possible to both realize moduli stabilization \cite{Giddings:2001yu,Dasgupta:1999ss}
and construct particle physics models at the same time.

F-theory \cite{Vafa:1996xn,Sen:1996vd} is the most suitable setup to describe type~IIB backgrounds with 7-branes, 
as it combines both the internal compactification manifold and the axio-dilaton profile into one geometric entity.
For a review on F-theory,
see for example \cite{Denef:2008wq,Weigand:2010wm,Blumenhagen:2010at,Braun:2010ff}.

However, F-theory does not geometrize all the relevant data of a type~IIB model with 7-branes. Besides the construction of an 
elliptically fibered CY fourfold, an important ingredient is the presence of fluxes: Three-form fluxes are needed to stabilize 
part of the geometric moduli of $B_3$, while two-form gauge fluxes on the branes are necessary to generate four dimensional chiral 
matter at the 7-brane intersections. The latter are also important, as they control the presence of chiral zero-modes localized 
wherever a Euclidean D3-instanton intersects a matter brane. Such modes are discussed in 
\cite{Ganor:1996pe,Blumenhagen:2006xt,Ibanez:2006da,Florea:2006si,Argurio:2007qk,Argurio:2007vqa,Blumenhagen:2007sm,Blumenhagen:2010ja,Cvetic:2011gp,Bianchi:2011qh}.
In F-theory, both kinds of fluxes are encoded in a four-form flux $G_4$ on $X_4$ (see also \cite{Denef:2008wq,Weigand:2010wm}).
Both the geometry ($X_4$) and the flux ($G_4$) can be given physical interpretations in M-theory. 
In the limit of vanishing elliptic fiber, M-theory compactified on $X_4$ is dual
to type~IIB compactified on $B_3$. 
This duality can also be used to get information on the low energy effective theory of F-theory 
compactifications with fluxes: One considers the reduction of the M-theory action \cite{Becker:1996gj,Haack:2001jz} and 
then takes the F-theory limit \cite{Dasgupta:1999ss,Grimm:2010ks,Valandro:2008zg,Braun:2008pz}. 

In the last years, there has been much effort to construct GUT models using F-theory setup. The GUT group is realized by an
$ADE$-singularity along a surface $S$ in the base space $B_3$. The matter fields reside on curves of $S$ over which the 
singularity enhances \cite{Katz:1996xe}. 
Since the gauge dynamics is localized on the surface $S$, most of the properties of the gauge theory are already encoded in the
neighborhood of $S$ in the fourfold. Such a local approach has been intensively
studied in the last few years. It allowed to achieve many phenomenological requirements, see 
\cite{Beasley:2008dc,Beasley:2008kw,Donagi:2008ca,Donagi:2008kj,Donagi:2009ra,Hayashi:2008ba,Hayashi:2009ge, 
Heckman:2008es,Heckman:2008qa,Heckman:2009mn,Marsano:2009gv,Cecotti:2009zf,Hayashi:2009bt,Tatar:2009jk,
Hayashi:2010zp,Dudas:2009hu,Dudas:2010zb,Conlon:2009qq,Font:2008id,Font:2009gq,Aparicio:2011jx,Chen:2010tp,Dolan:2011iu, Oikonomou:2011ba}
for an (incomplete) list of references.

To have 4D chiral fields living on matter curves, one has to introduce a two-form flux $F_2$ for the gauge theory living on $S$.
In the fourfold, this kind of flux must of course come from a flux $G_4$, requiring an understanding of the whole fourfold $X_4$
in general. In spite of the success of the local models, it is not always 
clear whether a local model can be embedded into a globally defined one \cite{Hayashi:2009bt,Cordova:2009fg,Ludeling:2011en}. In 
\cite{Blumenhagen:2009yv,Marsano:2009ym,Marsano:2009wr,Grimm:2009yu, Chen:2010ts,Chung:2010bn,Chen:2010tg,Cvetic:2010rq,Knapp:2011wk,Knapp:2011ip} 
such embeddings have been constructed, leading to interesting global realizations of GUT models in F-theory.
Note that aspects like the tadpole cancellation conditions, which are relevant for consistent models, can only be addressed in a
global model.

The geometry of the fourfold used to realize such global models is fairly well understood. However, the four-form flux needed to have
4D chiral matter (and for moduli stabilization) is much less under control. Following \cite{Friedman:1997ih,Curio:1998bva},
the powerful technique of the spectral cover was introduced in \cite{Donagi:2009ra,Hayashi:2009ge} to describe the gauge flux in F-theory,
and to compute the number of chiral modes generated by this flux on the matter curves. This technique is mutated from the dual Heterotic String Theory.
Basically, this four-form flux is defined using a line bundle on the spectral cover of the GUT surface $S$. In spite of several successful results, it is still not completely clear whether this approach captures all the global aspects of the four-form flux. 
In fact, this construction is local in the sense that the flux is defined near the surface $S$. Only when $X_4$ is a K3 fibration, 
so that a Heterotic dual exists, the flux is guaranteed to be extendable to the whole $X_4$. 
In building global models it has been essentially assumed that this can be done also when $X_4$ does not admit an Heterotic dual. 
Some tests of this assumption have been successful, e.g. a way to compute the $D3$-tadpole has been tested in situations without Heterotic dual in 
\cite{Blumenhagen:2009yv}. A proposal to generalize the spectral cover construction of the flux to backgrounds that do not have an Heterotic dual
has been given in \cite{Marsano:2010ix} and refined in \cite{Marsano:2011nn}. 
Recently, a global description of $G_4$ fluxes corresponding to the diagonal $U(1)$ within a $U(N)$
GUT group has been given in \cite{Grimm:2011tb}.

\vskip 3mm

In this paper we present a new way to construct $G_4$ flux in F-theory compactifications related to gauge flux on $7$-branes. 
Our fluxes are globally defined from the start, and do not rely in any way upon the extension of local fluxes. Our technique 
allows us to explicitly compute the induced D3-tadpole, and the induced chirality along the intersection of the $I_1$ locus 
with some $SU(N)$ enhancement.

We present a geometric characterization of the four-form flux in terms of its Poincar\'e dual four-cycle. This has several
advantages: The four-cycle related to the flux is defined in an algebraic way, so that the flux is automatically of type $(2,2)$. Of course, the introduction 
of such a flux will fix some complex structure moduli; we will be explicit on such a stabilization. When we require that the flux generates chiral modes, the restriction in the complex structure
moduli space is the same as the one found in \cite{Grimm:2010ez} to have massless $U(1)$s. In addition, the algebraic origin of the flux makes it easy to 
determine its quantization properties and compute the $D3$-tadpole and the number of chiral modes. 

The tadpole $-\frac12\int_{X_4}G_4\wedge G_4$ is directly obtained by computing the self-intersection of the corresponding four-cycle.
In case the weak coupling limit is smooth, we are able to map the $G_4$ flux to a two-form flux on the brane and to 
verify that they induce the same $D3$-charge.

The number of 4D chiral modes on a matter curve is computed by integrating the four-form flux on an appropriate four-cycle. This four-cycle is related to 
the matter curve in the following way: The matter curve is the locus in $B_3$ over which there is an enhancement of the singularity. This implies that 
a new two-cycle, which is fibered over the matter curve, appears in the resolution. This fibration gives a new four-cycle called the matter surface. This was 
already suggested by the duality with Heterotic String Theory in \cite{Donagi:2008ca,Hayashi:2008ba} and was related to the spectral cover computations \cite{Marsano:2010ix,Marsano:2011nn}. 
In this paper we show how to explicitly identify matter surfaces and compute integrals of the fluxes. In situations where 
the weak coupling limit is smooth, we compare this number with the one obtained by the index theorem in perturbative type~IIB.

We will present several examples. In particular, in our final example, we describe a situation with a non-abelian singularity on a surface $S$. 
We study four-form fluxes related both to type~IIB fluxes along the Cartan of the non-abelian gauge group 
and to fluxes that do not live on the brane $S$. The last ones are important because they do not break the GUT group, 
and can give chirality both to charged matter and to GUT singlets.

\section{Main idea and summary} \label{sec:mainideaandsummary}

As the bulk of this paper is somewhat technical, we summarize our main ideas and results here.
To motivate the subsequent discussion, let us start by compiling the crucial properties of cycles
that can carry supersymmetric $G_4$ fluxes. In a supersymmetric minimum of a compactification of 
F-theory on an elliptic Calabi-Yau fourfold, $G_4$ flux has to be of type $(2,2)$, 
primitive \cite{Becker:1996gj} and have one leg in the fiber \cite{Dasgupta:1999ss}. 
The last condition has to be satisfied in order for the flux not to break 
Lorentz symmetry in the four-dimensional effective theory. We can rephrase this condition by
demanding that the four-form $G_4$ should integrate to zero on any divisor of the base or
the elliptic fibration over a curve in the base. Hence the flux cannot be simply the Poincar\'e dual of a complete
intersection of the Weierstrass equation with two divisors of the ambient space.

The condition that the flux should be of type $(2,2)$ can be easily satisfied if we describe 
the four-cycle which is Poincar\'e dual to $G_4$ by algebraic equations. We are hence interested in situations
in which fourfolds described by a Weierstrass equation gain extra algebraic cycles.
To investigate under which conditions we can find such extra cycles we rewrite the Weierstrass model
in a specific form, as is discussed below.

We start with the Tate form of the elliptic fibration:
\begin{equation}\label{weiertateintro}
Y^2+a_1XYZ+a_3YZ^3=X^3+a_2X^2Z^2+a_4XZ^4+a_6Z^6 \, .
\end{equation}
One can shift $X$ and $Y$ to bring this into the Weierstrass form 
\begin{equation}\label{weier}
 y^2=x^3+xz^4 f+z^6 g \, .
\end{equation}
The shift of coordinates taking us from the
Weierstrass to the Tate form yields $f$ and $g$ as functions of the $a_i$, 
see appendix \ref{appweier}. We will use this parametrization in the following, as it allows
us to use the simpler Weierstrass form while still having all the information contained
in \eqref{weiertateintro}. This parametrization has the further advantage of being
equivalent to the parametrization used in Sen's weak coupling limit \cite{Sen:1997kw, Sen:1997gv}.

Our construction of fluxes starts from using the parametrization of $f$ and $g$ in terms of
the sections $a_i$ appearing in the Tate form and rewriting the Weierstrass model as
\begin{equation}\label{weierform}
Y_-Y_+-z^6a_6= X Q \, .
\end{equation}
Here we have defined the quantities
\begin{align}
Y_\pm&=y\pm\tfrac{1}{2}z^3a_3 \nn \\
X&=x-\tfrac{1}{12}z^2b_2 \nn\\
Q&=(x-\tfrac{1}{12}z^2b_2)(x+\tfrac{1}{6}z^2b_2)+ \tfrac{1}{2}z^4b_4 \nn\\
&=X(X+\tfrac{1}{4}z^2b_2)+\tfrac{1}{2}z^4b_4  \label{defYXQ} \, .
\end{align}
We have collected the details of this reformulation in appendix \ref{appweier}.

For our cases of interest, \eqref{weierform} defines an elliptic Calabi-Yau fourfold as a hypersurface in a 5-dimensional
ambient space. When $a_6$ factorizes in a non-trivial way, $a_6\equiv\rho\tau$, we find that the fourfold gains extra algebraic
$(2,2)$ cycles. These cannot be written as complete intersections of the Weierstrass equation with divisors of the ambient
space, but take for instance the form
\begin{equation}\label{sigmasect2}
 \sigma_\rho:\quad  \{ Y_-=0\} \quad\cap\quad \{X=0\} \quad\cap\quad \{\rho=0\}  \, 
\end{equation}
in the ambient space. We may then construct a four-cycle which has one leg in the fiber, i.e. is orthogonal to
horizontal and vertical divisors, by defining
\begin{equation}\label{gsect2}
\gamma_\rho \equiv \sigma_\rho - [\rho] \cdot F\,.
\end{equation}
Here, $F$ is the hyperplane section of the $\P^2_{123}$ the elliptic fiber is embedded in. It is chosen such that $x$ and $y$ 
are sections of $F^2$ and $F^3$, respectively. Constructions of this type can also be performed in many situations in which 
the polynomials $a_i$ already factorize such that we have non-abelian gauge groups. Due to the form \eqref{gsect2}, the flux we have 
constructed cannot be represented as the wedge product of two two-forms inside $X_4$.

We claim that putting a flux $G_4=\gamma_\rho$ corresponds to having a supersymmetric flux $F_2$ in the gauge theory 
on the $7$-branes. Note that $\gamma_\rho$ is defined using a divisor class $[\rho]$ of the base, which 
intersects the $7$-branes in curves. The $7$-brane divisors have the form
\be
\eta^2-\xi^2\,(\psi^2-\rho\,\tau)=0
\ee
where $\eta$ is a polynomial, and $\xi \rightarrow -\xi$ and the orientifold involution. Then, these curves are given by $\rho=0 \cap \eta=\pm \xi\,\psi$.
They are Poincar\'e dual to two-forms of 
type $(1,1)$, so that it seems natural to identify 
\begin{equation}\label{FandGsect2}
 G_4= \sigma_\rho - [\rho] \cdot F\quad \longleftrightarrow \quad F_2 = PD(\{\rho=0 \cap \eta=\pm \xi\,\psi \}) \, .
\end{equation}

Using our explicit realization of $G_4$, we can compute the $D3$ tadpole by integrating
\begin{equation}
 -\tfrac12\int_{CY_4}G_4\wedge G_4 \, .
\end{equation}
Using our proposed relation to the gauge flux on the $D7$-branes, we compute its tadpole also
from the type~IIB perspective. We compare the results whenever the weak coupling limit is available,
finding complete agreement.

A gauge flux $F_2$ can also lead to chirality for charged matter. In F-theory, charged matter can 
reside at the intersections of branes. Over such matter curves, there are extra vanishing spheres which encode the
representations of the matter multiplets. Via small resolutions of these singularities to make those spheres finite, 
one naturally produces a four-cycle $\hat{C}$ which is
a collection of spheres fibered over the matter curve. 
The conservation of induced D3-charge implies that, upon resolving the singularity, 
our $G_4$ flux in \eqref{FandGsect2} will itself undergo a transition to a new flux $\tilde G_4$.
It seems natural \cite{Donagi:2008ca,Hayashi:2008ba} that the chirality induced by this new flux 
can be computed by integrating
\begin{equation}
 \int_{\hat{C}} \tilde G_4 \, .
\end{equation}
Again, we use the identification \eqref{FandGsect2} and find agreement between the results obtained in F-theory in the 
weak coupling limit and type~IIB for all configurations we consider.

In models with non-abelian gauge groups, one can get four-cycles of type $(2,2)$ by appropriately resolving
the singularities. These four-cycles are obtained by fibring the exceptional curves of the resolution over a divisor of the
non-abelian brane stack. Fluxes constructed in this way correspond to $F_2$ in the Cartan of the non-abelian gauge group.
Even though they can generate a chiral spectrum, they also necessarily break the gauge group to the commutant subgroup.

Our construction is fundamentally different in this regard: The flux \eqref{gsect2} induces chirality but does not break 
the non-abelian gauge group. Moreover, because we can build a global flux over the $I_1$ locus of the discriminant, it can 
generate chiral modes for singlets of the non-abelian (GUT) group.

On the type~IIB side, we make extensive use of the technique of $D7$-brane `deconstruction', introduced in \cite{Collinucci:2008pf}.
Besides allowing us to treat situations with a singular $D7$-brane locus in an elegant way, the main
conceptual advantage of this description is the unification of the $D7$-branes and their gauge bundles into a
single object. From a mathematical point of view, this means that we are describing the $D7$-brane as a
coherent sheaf $\cale$ defined by the short exact sequence
\begin{equation}
0 \rightarrow E \stackrel{T}\longrightarrow F \rightarrow \cale \rightarrow 0  \, .
\end{equation}
Here, $E$ ($F$) is a vector bundle on a stack of $D9$ ($\overline{D9}$)-branes and $T$ is the tachyon 
that makes this unstable configuration collapse into a $D7$-brane. Consequently, the locus of the $D7$-brane
is given as the determinant of the tachyon matrix. The flux, which is determined by the bundles $E$ and $F$,
has a natural algebraic description in this setting.

This work indicates that a similar structure exists in F-theory compactifications on elliptic
Calabi-Yau fourfolds. We can write down an exact sequence which gives the Calabi-Yau fourfold
together with the $G_4$ flux as a coherent sheaf. The role of the determinant of the tachyon matrix
is played by the Weierstrass equation. It turns out that the latter can be written as the Pfaffian
of an anti-symmetric matrix. Hence its rank goes down by multiples of two (instead of one in the IIB case) so that we end up 
with a vector bundle on the elliptic Calabi-Yau fourfold. The desired $G_4$ flux can then be
constructed from the second Chern class of this vector bundle.

This paper is organized as follows. In section \ref{sectk3} we explain our ideas in the 
case of elliptic K3 surfaces, which already captures some of the salient ingredients of
our construction. In this case, we aim for non-singular Weierstrass models for which further
integral cycles of type $(1,1)$ are present.

In section \ref{sect4folds} we come to elliptic fourfolds, which are our main object of interest. 
After explaining the construction of fluxes in this case, we compute the $D3$ tadpole and
the number of chiral modes in various models. In particular, we consider the 
following configurations: We start with a single 7-brane, so that the elliptic fourfold is smooth.
We then describe a fourfold for which $a_6\equiv 0$. In this case, which has been dubbed $U(1)$ restricted
Tate model in \cite{Grimm:2010ez}, there is an intersection of the 7-brane with itself. This leads to
matter localized at the intersection which becomes chiral for non-zero fluxes. After this, we turn
to models with non-abelian gauge symmetries. As toy models for F-theory GUTs, we consider 
configurations with `non-split' $SU(2)$ ($Sp(1)$) and `split' $SU(2)$ singularities. In these cases, flux of the type \eqref{gsect2} 
induces chirality in the case of $U(1)$ restricted Tate models.

In section \ref{sectIIB} we study the weak coupling limits of the configurations considered in
section \ref{sect4folds}. We compute the $D3$ tadpoles and the chiral indices from the type~IIB
perspective using the methods developed in \cite{Collinucci:2008pf} and find complete agreement
with the results obtained in F-theory.

Finally, in section \ref{sectpfaff}, we comment on a description of elliptic fourfolds together with $G_4$ fluxes  in terms of coherent sheaves, and find an interesting link to the techniques of matrix factorizations. 
To show the power of this description, we reproduce the D3-tadpole computation in a very quick way, via a simple application of the 
Grothendieck-Riemann-Roch formula for the push-forward of a rank two vector bundle.

\section{Algebraic cycles in the K3 surface}\label{sectk3}

In this section we develop our ideas in the simplified setting of K3. For compactifications of F-theory on K3$\,\times\,$K3, 
one of the two K3 surfaces has to be elliptic. In this model, all branes are points in the elliptic K3 and wrap the
other K3 completely.We are interested in a $G_4$ flux of the form
\begin{equation}
 G_4=F_2\wedge \gamma \, ,
\end{equation}
where $F_2$ is a two-form flux on a brane.

As supersymmetry demands that $G_4$ is of type $(2,2)$ and $F_2$ is of type $(1,1)$, $\gamma$ must be an element of the 
Picard group of the elliptic K3:
\begin{equation}
\operatorname{Pic}(X)\equiv H^{1,1}(X)\cap H^{2}(X,\mathbb{Z}) \ .
\end{equation}
It is well-known that the rank of the Picard group, which is called the Picard number, is
zero for generic K3 surfaces. Even though there are $20$ independent harmonic $(1,1)$ forms
for any K3, none of them is in the \emph{integral} cohomology generically. This is not 
the case if the K3 surface in question is embedded in a projective space or toric variety:
intersections with divisors of the ambient space will descend to non-trivial elements of the 
Picard group. For a given hypersurface or complete intersection, the rank of the Picard group 
can be further enhanced by appropriately choosing a non-generic form for its defining equations.
This will be the crucial ingredient in our construction.

For an elliptic K3 surface described by a Weierstrass model, 
the two independent toric divisors of the ambient space give rise to a two-dimensional
Picard group. Its generators can be chosen to be the base and the fiber of the elliptic fibration. 
Hence the dual two-forms will either have two or no legs along the fiber directions. Hence we break 
Lorentz symmetry after taking the F-theory limit if we use one of those forms for $\gamma$. Thus, we are 
interested in finding K3 surfaces which have further integral $(1,1)$ cycles. We now discuss those enhancements 
of the Picard group as well as their physical interpretation.

Generically, the elliptic fiber degenerates over $24$ points in the base $\P^1$ of the elliptically 
fibered K3. If some of these points come together, this worsens the singularity 
of the elliptic fiber and leads to ADE singularities of the K3 surface. From the physical point of view, these 
singularities signal the appearance of non-abelian gauge symmetries due to coincident branes. The vanishing cycles 
of these singularities precisely correspond to the Cartan generators of the gauge groups. One can then
enhance the Picard group by blowing up these singularities. Physically, this means moving onto the Coulomb branch
of the gauge theory. The exceptional cycles are obviously integral and of type $(1,1)$. Furthermore,
their intersection numbers with the base and the fiber of the elliptic fibration are vanishing. Hence these
cycles can be used to construct $G_4$ flux corresponding to $F_2$ along the elements of the Cartan 
subalgebra. 

The process of blowing up does not physically separate the branes in F-theory: Taking the F-theory limit (i.e. vanishing fiber size), the exceptional
curves are blown down again \cite{Braun:2009wh}. From the perspective of type~IIB, it is clear that we can have a 
supersymmetric two-form flux $F_2$ also along the $U(1)$s of separate $D7$-branes, without having any non-abelian
gauge symmetries, however. Hence this situation is fundamentally different from the one considered above. 

We now discuss how \eqref{weierform} can gain extra algebraic cycles without gauge enhancement
and blow-ups. The crucial observation is that the Weierstrass model is the sum of two factors,
\begin{equation}\label{weieralgk3}
Y_-Y_+= X Q \, .
\end{equation}
if $a_6\equiv 0$, i.e. when $b_6$ becomes a square. This means we restrict the complex structure moduli
of K3 in a certain way. One can check that the locus where $a_6\equiv 0$ is at codimension one in moduli
space.

The curves $\sigma_\pm$ defined by taking the two equations 
\begin{align}\label{exk3cyc}
Y_\pm=0 \quad\text{and} \quad   X=0 \, ,
\end{align}
in the ambient space are automatically inside K3 for the configuration we consider. As we can 
explicitly represent these curves by algebraic equations, it follows that the corresponding cycles 
are of type $(1,1)$ and integral. Note that these curves cannot be written as complete intersections
of some divisor with the K3 surface, so that they are not equivalent to some
linear combinations of the base and the fiber of the elliptic fibration. 
The K3 hypersurface generically remains smooth for $a_6\equiv 0$ because a 
singularity only occurs if all four factors in the ambient threefold vanish simultaneously.

If we move away from the locus in moduli space where $a_6\equiv 0$, the integral cycle we have constructed
ceases to be of type $(1,1)$. Consequently, we loose the description \eqref{exk3cyc}, as these curves are 
not inside K3 for $a_6\neq 0$.

Let us now study some of the properties of these curves. First note that they do not intersect the standard section which sits
at $z=0$, $y^2=x^3$: Setting $z=0$ forces $y=x=0$ which is part of the exceptional set of the ambient toric variety. Second,
\eqref{exk3cyc} determines a single point in the fiber for every point on the base, so that it intersects the fiber cycle exactly once.
Hence the cycles $\sigma_{\pm}$ have all the properties one expects from a section of the elliptic fibration. As they are holomorphic sections 
of the elliptic fibrations, $\sigma_{+}$ and $\sigma_{-}$ both have the same topology as the base, i.e. each is a $\P^1$. Furthermore, as $\sigma_{\pm}$ 
are curves in K3, their self-intersection numbers must be equal to $-2$. 

Elliptic fibrations of the form \eqref{weieralgk3}, which allow for a second section, have been studied
in \cite{Aldazabal:1996du,Klemm:1996hh,Klemm:1996ts,Candelas:1997pv,Bershadsky:1998vn,Berglund:1998va}.
In particular, they are related to fibrations of $E_7$ type, for which the elliptic fibre is
embedded in $\P^2_{1,1,2}$ instead of $\P^2_{2,3,1}$. This has been discussed in the context of
elliptic fourfolds in \cite{Grimm:2010ez} recently.

We can now easily construct integral cycles of type $(1,1)$ which are orthogonal to base $\beta$ and fiber $\phi$ by defining
\begin{equation}\label{defs}
 \gamma_{\pm}=\sigma_{\pm}-\beta-2\phi \, ,
\end{equation}
so that\footnote{Remember that base and fiber have the following intersection numbers: $\beta^2=-2$, $\phi^2=0$, $\beta\cdot \phi=1$.}
\begin{align}
&\gamma_{\pm}\cdot \phi=0 \, ,\quad \gamma_{\pm}\cdot \beta=0 \, .
\end{align}

Using \eqref{defs}, one can also compute that $\gamma_\pm\cdot \gamma_\pm=-4$. As there cannot 
be any connected algebraic curves inside K3 which have a self-intersection number smaller than $-2$, the last 
equation shows that the cycles $\gamma_\pm$ can only be represented by the disjoint union of at least two curves. 

The observation of the last paragraph is connected with a physical interpretation of the constraint $a_6\equiv 0$. By going to the weak coupling limit,
such configurations are realized when two pairs of $D7$-branes align such that their displacements in $\P^1$ are pairwise equal.
As these displacements are measured by periods (see e.g. \cite{Lust:2005bd,bht08}) of the holomorphic two-form of K3, we can find two pairs of 
integral cycles $\gamma_1$, $\gamma_1'$ and $\gamma_2$,  $\gamma_2'$ for which
\begin{equation}
 \int_{\gamma_1}\Omega^{2,0}= \int_{\gamma_1'}\Omega^{2,0} \, ,\hspace{1cm}   \int_{\gamma_2}\Omega^{2,0}= \int_{\gamma_2'}\Omega^{2,0}.
\end{equation}
The cycles $\gamma_1-\gamma_1'$ and $\gamma_2-\gamma_2'$ are hence orthogonal to $\Omega$, so that they are integral cycles 
of type $(1,1)$. We can identify these cycles with the curves $\gamma_\pm$ we have found from the Weierstrass model.
Indeed one can check that also their intersection numbers agree. Furthermore, we only have to adjust the position of
a single $D7$-brane to get to the desired configuration. Hence the extra integral cycles of type $(1,1)$ also appear
at codimension one in complex structure moduli space from this point of view.

\section{Algebraic cycles and fluxes for fourfolds}\label{sect4folds}

Before describing the analogue of the algebraic cycles we have constructed for K3 in the last section, let us make some 
definitions to fix our notation. Let $B_3$ be the base of $X_4$, $K$ its canonical and $\bar K$ its anti-canonical bundle. 
The first Chern class of the base is $c_1(B_3)=\bar{K}$. 
We are using the same symbol for a line bundle, its first Chern class, and its associated divisor.
We consider the Weierstrass model \eqref{weierform} as a hypersurface in an ambient space $X_5$ which 
is the total space of the weighted projective bundle:
\begin{equation}
X_5 \equiv \mathbb{P}_{2,3,1}\left( \mathcal{O}_{B_3} \oplus  \mathcal{O}_{B_3} \oplus K \right)\,.
\end{equation}
In other words, we define the three coordinates $(x,y,z)$ to be sections of $\mathcal{O}_{B_3},  \mathcal{O}_{B_3}, K$, respectively, and then 
quotient the space by a $\mathbb{C}^*$ action with the weights $(2,3,1)$. Let $F$ denote the divisor class corresponding to the projective 
action in the fiber directions, so that $x, y, z$ are sections of $F^{\otimes 2},\, F^{\otimes 3},\, F\otimes K$, respectively. 
This is summarized in table \ref{tab:ambientx5}

\begin{table}[h] 
\begin{centering} 
\begin{tabular}{|c|c|c|}  
\hline 
 $x$ & $y$ & $z$\tabularnewline
\hline
\hline 
0 &0 & $K$ \tabularnewline
\hline 
$2$ &$3$ &$1$  \tabularnewline

\hline
\end{tabular}
\par
\end{centering}
\caption{The rows indicate the projective weights of the coordinates under the toric $\mathbb{C}^*$ 
action on the fiber coordinates of the $X_5$ space, and their transformation properties w.r.t. transitions along the base. $K$ denotes the canonical bundle of $B_3$.}
\label{tab:ambientx5}
\end{table} 
The Stanley-Reisner ideal of $X_5$ is simply given by the element $x\,y\,z$. From it, we deduce the relation $F^3=F^2\,\bar K$. 

Furthermore, we have that
\begin{equation}
6\,F^2 = PD(B_3 \subset X_5)\,.
\end{equation}
This corresponds to the fact that fixing the values of $x$ and $y$ gives us a copy of $B_3$ in $X_5$.

As before, we write the Weierstrass equation in the form
\begin{equation}
Y_-Y_+-z^6a_6-XQ=0 \, .
\end{equation}

\subsection{Construction of the flux}

We now generalize the construction of the last section to the case of elliptically fibered Calabi-Yau fourfolds. 
In this case, the form \eqref{weieralgk3} for the Weierstrass model would lead to a conifold singularity fibered over a curve in the base. 
We can, however, extend the structure used in the last section by demanding $a_6$ to factorize in a non-trivial way: $a_6\equiv\rho\tau$. 
In this case, the Weierstrass equation for the elliptic fourfold $X_4$ can be written as
\begin{equation}\label{X4}
Y_-Y_+-z^6\rho\tau -XQ = 0 \, .
\end{equation}
As before, this equation generically describes a smooth fourfold: In order for the equation and its
gradients to vanish one would have to solve $6$ equations in the ambient fivefold, which generically is not possible for dimensional reasons.

We can now find an algebraic four-cycle which is automatically inside $X_4$ by imposing three equations in the ambient fivefold. 
Let us consider the following\footnote{We can also pick the inequivalent branch with $Y_+=0$. The sum of the two branches $Y_-$ and $Y_+$ is
a complete intersection of the Weierstrass model with $X=0$ and $\rho=0$.}:
\begin{align}\label{sigmadef}
\sigma_\rho:\quad& \{Y_-=0\} \quad\cap\quad \{X=0\} \quad\cap\quad \{\rho=0\} \quad \subset X_5\,.
\end{align}

The first two equations eliminate the coordinates $y$ and $x$. In addition, they forbid $z$ from vanishing, 
since $x y z$ is in the SR-ideal. Hence, the first two equations define a threefold that is isomorphic to 
$B_3$ (but not contained in $X_4$). Therefore, $\sigma_\rho$ is a `horizontal' surface that is isomorphic to the 
hypersurface $\rho=0 \subset B_3$. Note that $\sigma_\rho$ does not intersect the `old' section of the fibration given by $z=0$,
and hence none of the `horizontal' surfaces. As $\sigma_\rho$ is a section of the elliptic fibration over $\rho=0 \subset B_3$,
it intersects each fiber and hence also four-cycles given by the intersection of two equations in $B_3$.

We now construct the fourfold analogue of \eqref{defs}. Our goal is to create a cycle $\gamma$ which satisfies
\begin{equation}\label{condgamma}
\gamma \cdot D_i \cdot D_j = 0\,, \quad \forall \, i, j \quad \text{and} \quad \, \gamma \cdot D_i \cdot (F+K) = 0\,, 
\end{equation}
for any divisor $D_i$ of the base. Note that $F+K$ is the class corresponding to the section at $z=0$ of
the Weierstrass model. As $K$ is also a divisor of the base, these equations imply that $\gamma\cdot D_i\cdot F=0$.

As $\sigma$ does not meet the section at $z=0$, we find that $\sigma_\rho \cdot D_i \cdot (F+K) = 0$. 
However, using the fact that $\sigma$ meets every fiber over $B_3\cap [\rho]$ in a single point, one can deduce that
\begin{equation}\label{sigmaintDD}
\int_{X_4} \sigma_\rho \cdot D_i \cdot D_j = \int_{B_3} [\rho] \cdot D_i \cdot D_j\,,
\end{equation}
where $[\rho]$ is the Poincar\'e dual to $\rho=0$. In order to eliminate this intersection, one immediate remedy is to define the cycle
$\sigma_\rho - [\rho]\cdot F$, since
\begin{eqnarray}
\int_{X_4} F \cdot [\rho]\cdot D_i \cdot D_j\,, &=& \int_{X_4} (F+K) \cdot [\rho]\cdot D_i \cdot D_j\,,\\
&=&\int_{B_3} [\rho]\cdot D_i \cdot D_j\,.
\end{eqnarray}
The first equality follows from the fact that $K,\rho, D_i,D_j$ are all elliptic fibrations over divisors in the base,
so that the intersection between all four of them vanishes. 

Since the relation $F \cdot (F+K)=0$ holds on $X_4$, this new term will not generate any unwanted intersections with `horizontal' 
four-cycles. Therefore, our coveted\footnote{Performing the same construction with a cycle that is given by
a divisor $D$ in the base intersected with the section $z=0$ yields the four-cycle $(F+K)\cdot D-D\cdot F=K\cdot D$.
This four-cycle clearly is again an elliptic fibration over a curve in the base, so that it has two legs along the fiber.
Such a state of affairs can be expected from the fact that \eqref{sigmaintDD} does not hold in this case.} 
four-cycle is given by
\begin{equation}\label{4formFlux}
\gamma_\rho \equiv \sigma_\rho - [\rho] \cdot F\,.
\end{equation}

If we evaluate this expressions in the ambient space $X_5$, we find that
\begin{align}
 \gamma_\rho&=  3F\cdot 2F\cdot[\rho]-[\rho]\cdot F \cdot PD(X_4\subset X_5) \nn\\
&=6[\rho]\cdot F\cdot F-[\rho]\cdot F \cdot 6F=0 \, .
\end{align}
Hence $\gamma_\rho$ is a trivial cycle in $X_5$. This does, however, not imply that it is also trivial
inside $X_4$. The cycle $\sigma_\rho$ cannot be written as the intersection of two divisors inside $X_4$,
whereas the part we substract has this form. Hence the two terms can never cancel inside $X_4$. 
In particular, $\gamma_\rho$ cannot be written as the wedge product of two
two-forms of $X_4$. As explained in appendix \ref{appendixB}, triviality in the ambient space $X_5$ ensures 
that our flux is orthogonal to all divisors in $X_4$ which descend from $X_5$.

Note that choosing $\tau$ to be a constant, which corresponds to going back to choosing a completely
generic $a_6$, means that $\sigma_\rho$ is an intersection of divisors in $X_4$: 
\begin{equation}
\sigma_\rho=\{Y_-=0\}\cap\{X=0\}\cap\{W=0\}=3F\cdot 2F \quad \subset X_4 \, .
\end{equation}
Furthermore, $[\rho]=6\bar{K}$ in this case. Hence we find $\gamma_\rho=6F\cdot F-6\bar{K}\cdot F$.
As we have seen before, $F^2-\bar{K}F$ vanishes, as it corresponds to an element of the Stanley-Reisner ideal of $X_4$, so that $\gamma_\rho$ 
is trivial in homology.

Following the same steps as for $\gamma_\rho$, we can define another flux 
\begin{equation}
\gamma_\tau \equiv \sigma_\tau - [\tau] \cdot F\,.
\end{equation}
As $a_6=\rho\tau=0$ is a complete intersection of the Weierstrass equation 
with $Y_+=0$, $X=0$, we find that
\begin{equation}
G_{\rho}+G_{\tau}=0 \, ,
\end{equation}
so that the four-cycle $\gamma_\tau$ related to $\tau$ is equal to minus the four-cycle $\gamma_\rho$.

The main difference between divisors in K3 and in Calabi-Yau threefolds is that, in the former, it depends on the complex structure 
whether an integral two-cycle is of type $(1,1)$. A similar situation 
arises in $H^4(X_4,\mathbb{Z})$. For Calabi-Yau fourfolds, the space $H^{2,2}$ splits into the primary 
horizontal and the primary vertical subspace, $H^{2,2}=H_H^{2,2}\oplus H_V^{2,2}$ \cite{Greene:1993vm}. Similar to 
the case of Calabi-Yau threefolds, derivatives 
of $\Omega^{4,0}$ span the whole of $H^{3,1}$. Second derivatives of $\Omega^{4,0}$, however, fail to reach all of 
$H^{2,2}$ \cite{Strominger:1990pd}, but only map to the primary horizontal subspace $H_H^{2,2}\subset H^{2,2}$. 
Furthermore, the primary vertical subspace $H_V$ is spanned by products of elements of $H^{1,1}$ \cite{Greene:1993vm}.

The cycle we have constructed, \eqref{sigmadef}, cannot be written as the wedge of two two-forms, so that it must belong to $H_H$. This also clarifies why we have to fix some of the complex structure moduli: the lattice
\begin{equation}
H_H^{2,2}(X_4)\cap H^4(X_4,\mathbb{Z}) \, ,
\end{equation}
which plays a similar role as the Picard lattice of K3, does not contain any elements generically~\footnote{In contrast, 
$H_V^{2,2}(X_4)$ does not depend on the complex structure moduli and one can find a basis of integral cycles.}. 
When we fix some of the complex structure moduli, some integral four-forms can become of type $(2,2)$.
This is precisely what happens for the integral cycle \eqref{4formFlux}, which becomes of type $(2,2)$ when $a_6$
factorizes as $a_6=\rho\tau$.

\subsubsection{The D3-brane tadpole}

Turning on a flux $G_4  = \gamma_\rho$ will induce some $D3$-charge equal to $-\tfrac{1}{2}\,\int G_4 \wedge G_4$. 
Using our explicit description of $G_4$, we will now show how to compute the tadpole in terms of intersections
of divisors in the base. In the weak coupling limit, these expressions are reproduced
in the corresponding type~IIB computation. For the sake of brevity, we shall drop the index $\rho$ from the cycles 
$\gamma$ and $\sigma$ in this section.

Let us therefore proceed to compute the self-intersection number of $\gamma$. As $\gamma\cdot D_i\cdot F=0$,
we have
\begin{equation}
\int_{X_4} \gamma^2 =\int_{X_4} \gamma \cdot \sigma = \int_{X_4} \sigma^2-\sigma \cdot F \cdot [\rho] \, .
\end{equation}
We may now use that $\sigma$ defines a section of the elliptic fibration over the locus $\rho=0$
to pull the second term down to the base. Hence we find
\begin{equation}
\int_{X_4} \gamma^2= \int_{X_4} \sigma^2+\int_{B_3} K \cdot [\rho]^2 \, . 
\end{equation}

Since $\sigma$ is not a complete intersection of two divisors within $X_4$, we must use rather indirect techniques in order to calculate $\sigma^2$. We first note 
that $\sigma$ can be thought of as the zero locus of a section of some rank-two vector bundle $E$ over $X_4$ which restricts to 
the normal bundle of $\sigma$ as follows:
\begin{equation}
E\lvert_{\sigma} = N_{\sigma \subset X_4}\,.
\end{equation}
Then, we can express the self-intersection of $\sigma$ in terms of the second Chern class of the normal bundle as follows:
\begin{equation}
\sigma \cdot \sigma  = \int_{\sigma} \sigma = \int_{\sigma} c_2(N_{\sigma})\,.
\end{equation}
Since the surface is not given by two equations in $X_4$, some work is required to compute this Chern class. The shortest route is to use the 
following exact sequence of normal bundles:
\begin{equation}
0 \rightarrow N_{\sigma \subset X_4} \rightarrow N_{\sigma \subset X_5} \rightarrow N_{X_4 \subset X_5} \rightarrow 0\,.
\end{equation}
We know that $\sigma$ is given by three equations in $X_5$, as defined in \eqref{sigmadef}. Hence, we can express the middle bundle as follows:
\begin{equation}
N_{\sigma \subset X_5} = F^{\otimes 2} \oplus F^{\otimes 3} \oplus [\rho]\,.
\end{equation}
Similarly, we can express the right-most bundle as
\begin{equation}
N_{X_4 \subset X_5} = F^{\otimes 6}\,.
\end{equation}
The exact sequence tells us that
\begin{eqnarray}
c(N_{\sigma \subset X_4}) &=& c(N_{\sigma \subset X_5})/c(N_{X_4 \subset X_5})\nonumber \\
&=& 1+\left([\rho]-F\right) + F\cdot (12\,F-[\rho])\,.
\end{eqnarray}
hence,
\begin{equation}
c_2(N_{\sigma \subset X_4}) = F\cdot (12\,F-[\rho])\,.
\end{equation}
Integrating this over $\sigma$ gives us
\begin{equation}
\int_{X_4} \sigma \cdot \sigma = \int_{X_5} 6\,F^2\cdot[\rho]\cdot F\cdot (12\,F-[\rho]) = \int_{B_3}  \bar{K}\cdot [\rho] \cdot (12\,\bar{K}-[\rho])\,.
\end{equation}
Therefore, the self-intersection of $\gamma$ is 
\begin{equation}
\gamma \cdot \gamma  = 2\,\int_{B_3} \bar{K} \cdot [\rho] \cdot (6\,\bar{K}-[\rho])\:.
\end{equation}
\vskip 3mm

If we turn on a flux $G_4$ that is Poincar\'e dual to $\gamma$, this will hence induce a $D3$-charge given by:
\begin{equation} \label{g4tadpole}
Q^F_{D3} = -\tfrac{1}{2}\,\int_{X_4} G_4 \wedge G_4 = -\int_{B_3} c_1(B_3) \cdot [\rho] \cdot (6\,c_1(B_3)-[\rho])=-\int_{B_3} c_1(B_3) \cdot [\rho] \cdot[\tau] \,,
\end{equation}
where we used $\bar{K}=c_1(B_3)$. We will confirm this result from the corresponding type~IIB computation in the weak coupling limit in section \ref{sectIIBwhitney}.

\subsection{Recombination and flux change}\label{sectFsplit}

When a smooth brane supporting two-form flux splits into two (or more) intersecting pieces, the flux quanta change \cite{Collinucci:2008pf}. 
This is to be expected from the perspective of F-theory: As the Euler numbers
of the (resolved) corresponding fourfolds change, the geometric part of the $D3$-brane tadpole must also change in this process.
This change is compensated by a change in the four-form flux $G_4$, which is connected to the flux on the brane
world-volume.

Let us discuss this in the present context. We start with a smooth fourfold \eqref{X4}, so that there is one smooth
recombined 7-brane, and a four-form flux $G_{\rho}$. If we set $a_6\equiv 0$ in \eqref{X4}, the CY fourfold becomes:
\begin{equation}\label{X4bib}
Y_-Y_+ = X Q \, .
\end{equation}
This fourfold is clearly singular over the curve given by $Y_+=Y_-=X=Q=0$. As the singularity occurs at codimension
two in the base, it must be due to intersections between branes. The discriminant locus, however,
is given by one connect surface. The intersection of this brane with itself leads to matter charged under its $U(1)$.
The connection of configurations with $a_6\equiv 0$ and an extra $U(1)$ was discussed in \cite{Grimm:2010ez}, 
(see also \cite{Aldazabal:1996du,Klemm:1996hh,Klemm:1996ts,Candelas:1997pv,Bershadsky:1998vn,Berglund:1998va}) where the
corresponding elliptic fibration was called a `$U(1)$ restricted Tate model'. We will also use this terminology in the following. 
In the weak coupling limit this configuration corresponds to splitting the otherwise connected $D7$-brane into a
brane and its orientifold image in the $CY_3$ double cover. 

One can compute the geometric tadpole of $X_4$ by appropriately resolving its singularities, which we will do in section \ref{sectsmallres}. 
The result differs from the geometric tadpole of the smooth fourfold we have started with. 
In the present situation, the difference is given by (see also \cite{Grimm:2010ez}):
\begin{equation}
 9\int_{B_3}\bar{K}^3 \:.
\end{equation}
As the tadpole has to be canceled both before and after the transition, the flux
must also change such that the total tadpole remains invariant. If we denote the flux in the 
situation with intersecting branes by $\tilde{G}_{\rho}$, we have that
\begin{equation}
 \frac{\chi(X_4)}{24}-\frac{1}{2}\int_{X_4}G_\rho\wedge G_\rho=\frac{\chi(\tilde{X}_4)}{24}-\frac{1}{2}\int_{\tilde{X}_4}\tilde{G}_\rho\wedge\tilde{G}_\rho \, .
\end{equation}
As 
\begin{equation}
\frac{\chi(X_4)}{24}-\frac{\chi(\tilde{X}_4)}{24}=9\int_{B_3}\bar{K}^3 \, ,
\end{equation}
it must be that
\begin{equation}
-\frac{1}{2}\int_{\tilde{X}_4}\tilde{G}_\rho\wedge\tilde{G}_\rho+\int_{B_3}\bar{K}\cdot[\rho]\cdot(6\bar{K}-[\rho])=+9\int_{B_3}\bar{K}^3 \, .
\end{equation}
Hence the tadpole contribution of the remaining flux on the resolved fourfold $\tilde{X}_4$ 
can be written as
\begin{equation}
\tilde{Q}_\rho=\int_{B_3}\bar{K}\cdot([\rho]-3\bar{K})\cdot ([\rho]-3\bar{K}) \, .\label{Qalpha}
\end{equation}

\subsubsection{A small resolution}\label{sectsmallres}

Let us now explicitly resolve the singularity discussed in the previous segment. The singularity of the hypersurface \eqref{X4bib} has 
the structure of a conifold singularity fibered over the curve 
\begin{equation} \label{Csmallres}
C: \{Y_- = 0\} \quad \cap \quad \{Y_+ = 0\} \quad \cap \quad \{X = 0\} \quad \cap \quad \{Q=0\} \quad \subset X_5\,.
\end{equation}
Going back to the definitions of these quantities in \eqref{defYXQ}, one sees that this curve can be
equivalently described by
\begin{equation}
C: \{y = 0\} \quad \cap \quad \{a_3 = 0\} \quad \cap \quad \{X= 0\} \quad \cap \quad \{b_4=0\} \quad \subset X_5\,.
\end{equation}

In order to define the flux and perform a computation of the induced $D3$-tadpole, we need
to appropriately resolve this singularity. We could do this via a blow-up, however, it is more natural to perform a small resolution by tagging a 
$\mathbb{P}^1$ with homogeneous coordinates $[\lambda_1: \lambda_2]$ to the ambient fivefold, resulting in an ambient sixfold $X_6$ described in table \ref{tab:ambientx6}, and imposing 
\begin{equation} \label{twosmallres}
\tilde X_4: \quad Y_-\,\lambda_2 =Q\,\lambda_1 \quad \cap \quad Y_+\,\lambda_1 = X\,\lambda_2\, \quad \subset X_6\,.
\end{equation}

In \cite{Esole:2011sm}, the necessity for small resolutions in F-theory was stressed. Hence our resolved fourfold $\tilde{X}_4$ is given by two equations in an ambient sixfold $X_6$. The divisor classes
of the different homogeneous coordinate used in the construction of $X_6$ are
\begin{align}
 [Y_\pm]=3F \qquad [X]=2F \qquad [z]=F+K \nn \\
[\lambda_1]=H \qquad [\lambda_2]=F+H  \, .
\end{align}
We have collected this information in table \ref{tab:ambientx6}. 
\begin{table}[h] 
\begin{centering} 
\begin{tabular}{|c|c|c|c|c|}  
\hline 
 $x$ & $y$ & $z$ & $\lambda_1$ & $\lambda_2$ \tabularnewline
\hline
\hline 
0 &0 & $K$ &0 &0 \tabularnewline
\hline 
$2$ &$3$ &$1$ & $1$ & $0$ \tabularnewline
\hline 
$0$ &$0$ &$0$ & $1$ & $1$ \tabularnewline
\hline
\end{tabular}
\par
\end{centering}
\caption{The rows indicate the projective weights of the coordinates under the two toric $\mathbb{C}^*$ actions for the $X_6$ space, and the transitions functions w.r.t. patches on the base. 
$K$ denotes the canonical bundle of $B_3$.}
\label{tab:ambientx6}
\end{table} 

The exceptional sets and the SR-ideal are given by 
\begin{align}
\{y=x=z=0\}\leftrightarrow F^2\cdot(F+K)=0 \nn\\
\{\lambda_1=\lambda_2=0\}\leftrightarrow H\cdot(F+H)=0 \, .
\end{align}
The fourfold $\tilde{X}_4$ inside $X_6$ is given as $(4F+H)(3F+H)=6F(2F+H)$. On $\tilde{X}_4$, the first 
relation becomes
\begin{equation}
F(F+K)=0 \, .
\end{equation}

The standard section of the Weierstrass model at $z=0$ is still present and sits at $y=X=\lambda_1=\lambda_2=1$.
Furthermore, there is now a second section at $\lambda_1=X=Y_-=0$. Such models should be related to
elliptic fibrations of $E_7$ type, which generically have a second 
section \cite{Aldazabal:1996du,Klemm:1996hh,Klemm:1996ts,Candelas:1997pv,Bershadsky:1998vn,Berglund:1998va}.
In \cite{Grimm:2010ez}, the appearance of a fibration of $E_7$ type was shown by blowing up the singular fourfold. 
It would be interesting to establish a similar result also for the small resolution considered here.

\subsubsection{The flux after the transition}

We now want to see how to construct a four-form flux with one leg in the fiber in the case of $\tilde{X}_4$. We can consider 
cycles given by
\begin{equation}\label{sigmaalpha}
\sigma_{\alpha}=\{\lambda_1=0\}\cap\{Y_- = 0\}\cap\{X=0\}\cap\{\alpha=0\}  \quad \subset X_6\,.
\end{equation}
This is the intersection of the new section of the elliptic fibration with the locus $\alpha=0$. 

First, we demand that the corresponding flux have a vanishing intersection with any two divisors of the base. 
As before we have
\begin{equation}\label{sdd}
 \int_{\tilde{X}_4} D_i\cdot D_j \cdot \sigma_{\alpha}=\int_{B_3}D_i\cdot D_j \cdot[\alpha] \, .
\end{equation} 

Furthermore, we want the flux to be orthogonal to the $z=0$ section. We still have
\begin{equation}\label{sdf}
 \sigma_{\alpha} \cdot (F+K) = 0 \, ,
\end{equation}
as $x=y=z=0$ is part of the exceptional set.

As before, we can cancel the term arising in \eqref{sdd} by substracting $[\alpha]$ times $F$:
\begin{equation}\label{tildeG}
 \gamma_{\alpha}=\sigma_\alpha - F\cdot[\alpha] \, ,
\end{equation}
because
\begin{equation}
 \int_{\tilde{X}_4} F\cdot[\alpha]\cdot D_i\cdot D_j=(F+K)\cdot[\alpha]\cdot D_i\cdot D_j=\int_{B_3}[\alpha]\cdot D_i\cdot D_j \, ,
\end{equation}
where we have used the section of the elliptic fibration to restrict the integral to the base.

Furthermore,
\begin{equation}
F\cdot[\alpha]\cdot(F+K)=0 \, 
\end{equation}
holds inside $\tilde{X}_4$, so that $\gamma_{\alpha}$ fulfills both requirements \eqref{sdd}, \eqref{sdf}.
Note that even though $\gamma_\rho$ was trivial in the ambient space $X_5$, 
$\gamma_{\alpha}$ is non-trivial in the ambient space $X_6$.

Let us now compute the tadpole of a flux $G_\alpha=\gamma_\alpha$ on $\tilde{X}_4$.
First note that 
\begin{equation}
\sigma_\alpha\cdot(F+K)=0 \, .
\end{equation}
This equation is obtained as before: $F+K$ forces $z=0$, whereas $\sigma_\alpha$ is along $z=1$.
As we also have $G_\alpha\cdot D_i\cdot D_j=0$, it again follows that $G_\alpha\cdot F=0$. Proceeding as before, we find
\begin{align}
Q_\alpha&=-\frac{1}{2}\int_{\tilde{X}_4}G_\alpha\wedge G_\alpha=-\frac{1}{2}\int_{\tilde{X}_4}G_\alpha\wedge \sigma_\alpha\nn \\
&=-\frac{1}{2}\int_{\tilde{X}_4}\left(\sigma_\alpha\wedge \sigma_\alpha - \sigma_\alpha\cdot F\cdot [\alpha]\right)\nn\\
&=-\frac{1}{2}\int_{\tilde{X}_4}\sigma_\alpha \wedge \sigma_\alpha +\frac{1}{2}\int_{B_3} \bar{K}\cdot [\alpha]^2 \, .
\end{align}

To compute the self-intersection of $\sigma_\alpha$ inside $\tilde{X}_4$, we note (as before) that 
\begin{equation}
 \int_{\tilde{X}_4}\sigma_\alpha\wedge\sigma_\alpha=\int_{\sigma_\alpha}c_2(N_{\sigma_\alpha\subset \tilde{X}_4}) \, .
\end{equation}
We can compute the second Chern class of the normal bundle of $\sigma_\alpha$ by using $N_{\tilde{X}_4\subset X_6} =(4F+H)\oplus(3F+H)$,
$N_{\sigma_\alpha\subset X_6}=2F\oplus 3F\oplus [\alpha]\oplus H$ and
\begin{align}
 c(N_{\sigma_\alpha\subset \tilde{X}_4})&= c(N_{\sigma_\alpha\subset X_6})/c(N_{\tilde{X}_4\subset X_6} )
 \end{align}
which yields 
\be
 c(N_{\sigma_\alpha\subset \tilde{X}_4})_2=8\,F^2 + 9 \,F \cdot H + H^2 - 2\,F\cdot \alpha - H \cdot \alpha
\ee
Hence
\begin{align}
 \int_{\sigma_\alpha}c_2(N_{\sigma_\alpha\subset \tilde{X}_4})&=\int_{X_6}6F^2\cdot H\cdot[\alpha]\cdot c_2(N_{\sigma_\alpha\subset \tilde{X}_4})\nn\\
&=-\int_{X_6}6F^2\cdot H\cdot[\alpha]\cdot \bar{K} \cdot [\alpha]\nn\\
&=-\int_{B_3}[\alpha]^2 \bar{K} \, ,
\end{align}
where we have repeatedly used Poincar\'e duality and the relations coming from the SR-ideal.

Putting everything together, and using $\bar{K}=c_1(B_3)$,
\begin{equation}\label{qFsplit}
 Q_\alpha=-\frac{1}{2}\int_{\tilde{X}_4}\sigma_\alpha\wedge \sigma_\alpha +\frac{1}{2}\int_{B_3} \bar{K}\cdot [\alpha]^2=\int_{B_3} c_1(B_3)\cdot [\alpha]^2 \, .
\end{equation}

Looking back at \eqref{Qalpha}, we conclude that after the transition to intersection branes, \eqref{X4bib}, a flux $G_{\rho}$
on $X_4$ is replaced by a flux $G_\alpha$ on $\tilde{X}_4$ as in \eqref{tildeG}, with $[\alpha]=[\rho]-3\bar{K}$. We will confirm this result 
from the perspective of type~IIB string theory in section \ref{sectIIBsplit}.

\subsection{Chirality}

As two-form fluxes on branes can induce chirality, and the flux $G_4$ that we have constructed is related
to such fluxes, we expect it to also induce chirality. In F-theory,
matter resides at the intersections of branes, the so-called matter curves. Over these curves, the
ADE singularity sitting over a 7-brane is enhanced. This enhancement is related to extra vanishing $\P^1$s over the matter curve. They extend the Dynkin diagram of the ADE gauge group on the 7-brane,
which allows to determine the representation of matter that sits on the matter curve.
 
If we fiber the vanishing $\P^1$s responsible for charged matter over their matter curves $C_i$, we 
obtain four-cycles $\hat{C}_i$ of $X_4$. It hence seems natural that the chirality can be obtained 
as the integral of $G_4$ over these curves \cite{Donagi:2008ca,Hayashi:2008ba}.
In this section, we will test this idea by explicitly computing the integral of our proposed fluxes,
\begin{equation}
 I=\int_{\hat{C}_i}G_4 \, ,
\end{equation}
after appropriately resolving the matter curves.

\subsubsection{A simple example}\label{chirsplit}

The simplest situation in which a matter curve arises is the configuration considered in section \ref{sectFsplit}.
In the base, the matter curve is simply given by $a_3=b_4=0$. The corresponding four-cycle $\hat{C}$ is the fibration of the
$\P^1$ introduced in \eqref{twosmallres} over this curve. It is given by $a_3=b_4=X=y=0$ in the ambient sixfold $X_6$ introduced
in section \ref{sectsmallres}. 

Let us hence consider a flux $G_4=\gamma_\alpha$ as before and compute the intersection number 
\begin{equation}
 \int_{\hat{C}}\gamma_\alpha =  \int_{\hat{C}}\left(\sigma_\alpha - F\cdot[\alpha]\right) \, .
\end{equation}
We first consider the second term. Since $xyz$ is in the SR-ideal of $X_5$, the intersection of $\hat C$ with $z=0$ is empty. Hence, $\hat C \cdot (F+K) = 0$,
so that the second term is equal to $\int_{X_4}\hat C \cdot [\alpha] \cdot K$. Since this involves intersecting four polynomials that depend on $B_3$ only, 
this intersection must be empty. As both the cycle $\sigma_{\alpha}$, \eqref{sigmaalpha}, and the curve $\hat{C}$, lie within the section over $Y_-=X=0$, we find 
\begin{equation} \label{ftheoryindex}
\int_{\hat C} G_\alpha = \sigma_{\alpha} \cdot \hat C = 12\,\int_{B_3} c_1(B_3)^2 \cdot [\alpha]\,.
\end{equation}
In the presence of such $G_4$ flux, we claim that this is the chirality index that counts the massless M2-brane states wrapped on the $\mathbb{P}^1$ hovering over $C$. 
We will validate this result in the weak coupling limit by performing the corresponding computation in the IIB picture in section \ref{sectIIBsplit}.

\subsection{Non-abelian singularities, fluxes and chirality}

\subsubsection{$Sp(1)$ singularity}\label{SecWitSp1}

As a further example, we consider the Weierstrass equation in the case of an $Sp(1)$ singularity along the 
locus $P=0$ in the base\footnote{
Though $Sp(1)\cong SU(2)$, we will distinguish between an $Sp(1)$ and an $SU(2)$ singularity, meaning with the first a non-split $SU(2)$ and with the second a split one.
}. Hence, the discriminant factorizes into an $Sp(1)$-brane locus, and a remaining $I_1$-locus. This means that the polynomials $a_3,a_4,a_6$, and correspondingly $b_4$ and $b_6$, factorize in the following way:
\begin{equation}
 a_3 = a_{3,1}\cdot P \qquad a_4 = a_{4,1}\cdot P \qquad a_6 = a_{6,2}\cdot P^2 \qquad b_4 = b_{4,1}\cdot P \qquad b_6 = b_{6,2}\cdot P^2 \, .
\end{equation}
We will put the further constraint $a_{6,2}\equiv\rho\,\tau$ on the complex structure in order to define a four-form flux as in the previous sections.

To resolve the singularity over $P=0$, we follow \cite{Collinucci:2010gz} and introduce a new coordinate $s$ and a new equation $s = P$. 
Then the singular fourfold $X_4$ is given by the two equations:
\begin{equation}
 \left\{\begin{array}{l}
     Y_+\,Y_- - X \,Q - s^2\rho\tau \,z^6 = 0  \\ s = P \\
   \end{array}\right.
\end{equation}
with 
\begin{align}
Y_\pm&=y\pm\frac{1}{2}z^3a_{3,1}s \nn \\
X&=x-\frac{1}{12}z^2b_2 \nn\\
Q&=X(X+\frac{1}{4}z^2b_2)+\frac{1}{2}z^4b_{4,1}s \label{XYpYmQ} \, .
\end{align}

The resolved fourfold $\hat{X}_4$ is then given by the two equation
\begin{equation}\label{eqx6}
 \left\{\begin{array}{l}
      Y_+\,Y_- - X \,Q_v - s^2\rho\tau \,z^6 = 0  \\ s\,v = P \\
   \end{array}\right.
\end{equation}
in an ambient sixfold $X_6$. Note that homogeneity of the equations enforces that the 
definition of $Q$ is now changed to $Q_v=X(v\,X + \tfrac14 z^2 b_2) +  \tfrac12 b_{4,1}s\,z^4$. 
The new exceptional divisor is $E:$ $v=0$.
The coordinates $x,y,z,s,v$ belong to $F^2\otimes [E]^{-1}$, $F^3\otimes [E]^{-1}$, $F\otimes K$, $[P]\otimes [E]^{-1}$, $[E]$,
which is summarized in the table below. 

\begin{table}[h] 
\begin{centering} 
\begin{tabular}{|c|c|c|c|c|}  
\hline 
 $x$ & $y$ & $z$ & $s$ & $v$ \tabularnewline
\hline
\hline 
0 &0 & $K$ &$[P]$ &0 \tabularnewline
\hline 
$2$ &$3$ &$1$ & $0$ & $0$ \tabularnewline
\hline 
$-1$ &$-1$ &$0$ & $-1$ & $1$ \tabularnewline
\hline
\end{tabular}
\par
\end{centering}
\label{tab:ambientx6.2}
\end{table} 

The SR-ideal of the resolved ambient six-fold is 
\begin{equation}
\{ xyz,Xys,vz\} \, .
\end{equation}
From the SR-ideal we see that the following relations hold:
\begin{equation}
  F^2(F+\bar{K}) =0 \, ,\qquad (E-[P])(3F-E)(2F-E)\, , \qquad E\,(F+K) = 0 \, .
\end{equation}
In particular. we note that on $B_3$ we have $F=\bar{K}$ and $E=[P]$.

Let us now see what are the possible holomorphic four-cycles that we can have on the resolved fourfold.
First, we have a cycle like in the smooth case:
\begin{equation}
 \gamma_{\rho} = \sigma_\rho - [\rho] \left(F-\frac{E}{2}\right) [\hat{X}_4] 
\end{equation}
where
\begin{equation}
  \sigma_\rho=\{ \rho=0 \,, Y_-=0\,, X=0\,, vs=P\}|_{X_6} \subset \hat{X}_4 \, .
\end{equation}
We have chosen the substraction such that the flux is trivial in the ambient space, while it is non-trivial 
on the fourfold. This choice is motivated by the observation that this naturally happens for
the corresponding configuration without an $Sp(1)$ stack.
One can check that $\gamma_\rho$ is zero when $\rho$ has maximal degree or is a constant.

Let us compute the tadpole of a flux $G_4^{(I_1)}=\gamma_\rho$:
\begin{equation}\label{TadWSpG41}
 -\tfrac12 \int_{\hat{X}_4} G_4^{(I_1)}\wedge G_4^{(I_1)} = -\tfrac12 \int_{\hat{X}_4} 
\sigma_\rho\cdot \sigma_\rho -\tfrac12 \int_{\hat{X}_4} \sigma_\rho \cdot [\rho] \left(F-\frac{E}{2}\right) [\hat{X}_4] \, .
\end{equation}
We compute the second piece first:
\begin{equation}
-\tfrac12 \int_{\hat{X}_4} \sigma_\rho \cdot [\rho] \left(F-\frac{E}{2}\right) [\hat{X}_4] = -\frac12 \int_{B_3} [\rho]^2\left(\bar{K}-\frac{[P]}{2}\right)\:.
\end{equation}
The first piece in \eqref{TadWSpG41} is computed with the trick of normal bundles, i.e.
\begin{equation}
 -\tfrac12 \int_{\hat{X}_4}  \sigma_\rho\cdot  \sigma_\rho = -\tfrac12 \int_{ \sigma_\rho } c_2(N_{ \sigma_\rho\subset\hat{X}_4})\:,
\end{equation}
where the normal bundle is determined by the following exact sequence
\begin{equation}
 0\rightarrow N_{ \sigma_\rho\subset\hat{X}_4}\rightarrow N_{ \sigma_\rho\subset X_6}\rightarrow N_{\hat{X}_4\subset X_6} \rightarrow 0 \:.
\end{equation}
We find that:
\begin{eqnarray}
{\rm c}(N_{\sigma_\rho \subset \hat{X}_4}) & = &\frac{ {\rm c}(N_{\sigma_\rho \subset X_6}) }{ {\rm c}(N_{\hat{X}_4 \subset X_6})} \\ 
     &=& 1+\left\{[\rho]-F\right\} + \left\{ 12 F^2 - 7 E F + E^2 - [\rho]F  \right\}       \, , \nonumber
\end{eqnarray}
consequently
\begin{equation}
 -\tfrac12 \int_{\hat{X}_4}  \sigma_\rho\cdot \sigma_\rho = -\tfrac12 \int_{B_3} [\rho]\left( 12\bar{K}^2-7[P]\bar{K}+[P]^2-[\rho]\bar{K} \right) \, .
\end{equation}
Putting everything together:
\begin{eqnarray}\label{TadWSpG41bis}
 -\tfrac12 \int_{\hat{X}_4} G_4^{(I_1)}\wedge G_4^{(I_1)} &=& \int_{B_3} \left( [\rho]\bar{K}  - \tfrac{1}{4}[\rho][P] - \tfrac{1}{2}[P]^2
		+ \tfrac72 [P]\bar{K} - 6 \bar{K}^2 \right) \nonumber\\
      &=& -\int_{B_3} \left(c_1(B_3)-\tfrac14[P]\right)\cdot [\rho] \cdot\left(6c_1(B_3)-2[P]-[\rho] \right) \, ,
\end{eqnarray}
where we have substituted $\bar{K}=c_1(B_3)$. Note that for $[P]=0$ we get the result of the previous section.

\

As we have resolved the $Sp(1)$ singularity over $P=0$, we should be able to construct a four-form flux
which corresponds to a flux in the Cartan. This flux has the form:
\begin{equation}
 G_4^{(\rm Sp)} = \tfrac12 E \wedge [q]
\end{equation}
where $[q]$ is the divisor $q=0$ for some polynomial $q$ of the base. 
This flux is integral if $[q]$ is even and half-integral otherwise. 
The cancellation of the Witten anomaly,
\begin{equation}
 G_4 + \frac{c_2(\hat{X}_4)}{2} \,\, \in \,\, H^4(\hat{X}_4,\mathbb{Z})\:,
\end{equation}
can enforce the introduction of this flux, i.e. it determines if $q$ is odd 
or even. In \cite{Collinucci:2010gz}, it was shown that any smooth elliptically fibered CY fourfold with a Weierstrass representation has an even $c_2$. It was also argued there that this implies that any odd contribution to $c_2$ must come from holomorphic submanifolds that do not survive the blow down map of the resolution of a singularity.

The flux $G_4^{(\rm Sp)}$ is Poincar\'e invariant: Its orthogonality to the section $z=0$ is implied by the 
SR-ideal, i.e. $E(F+K)=0$. The intersection of $G_4^{(\rm Sp)}$ with $D_i \cdot D_j$ is zero 
for the following reason: The divisor $E$ is a $\mathbb{P}^1$ fibration over a four-cycle on the base; 
the wedge product $[q]\wedge D_i\wedge D_j$ is Poincar\'e dual to the fiber on some 
point on the base that in general will not belong to the base of $E$. 
Hence the generic intersection is empty.

\

Let us compute the $D3$-charge of this flux:
\begin{eqnarray}\label{QG42F}
-\tfrac12\int_{\hat{X}_4} G_4^{(\rm Sp)}\wedge G_4^{(\rm Sp)}&=& -\tfrac12 \int_{\hat{X}_4} \tfrac14 E^2 [q]^2 \,\,=\,\, \int_{B_3} [P]\left(\frac{[q]}{2}\right)^2\:.
\end{eqnarray}

The flux $G_4^{(\rm Sp)}$ is orthogonal to $G_4^{(I_1)}$, i.e. $\int_{\hat{X}_4} G_4^{(\rm Sp)}\wedge G_4^{(I_1)} = 0$. Hence the total tadpole is 
the sum of the two.

\

The resolution we made creates a new four-cycle which is a fibration over the curve where the $Sp(1)$-brane intersects
the remaining component of the discriminant locus. This four-cycle is given by two equations in $\hat{X}_4$ 
(see appendix \ref{appendixC} for details):
\begin{equation}
 \hat{C}: \qquad \{ v=0 \}\, \cap\, \{b_{4,1}^2-b_2(a_{3,1}^2+4\rho\tau)=0  \} \, .
\end{equation}
The second equation is a section of $\bar{K}^8 [P]^{-2}$.

We claim that the number of 4D chiral states coming from the intersection of the $Sp(1)$ surface with 
the remaining brane is the integral of the four-form flux over the four-cycle $\hat{C}$. Indeed, this 
four-cycle arises, after resolution, at the locus where the singularity enhances.

If we integrate the flux $G_4^{(I_1)}$ on $\hat{C}$, we will get zero, as $\hat{C}$ coincides with its 
pushforward to the ambient space $X_6$. Hence the integral of $G_4=G_4^{(I_1)} + G_4^{(\rm Sp)}$ on $\hat{C}$ is equal to the integral~of~$G_4^{(\rm Sp)}$:
\begin{eqnarray}\label{chirfsp(1)}
 \int_{\hat{C}}G_4 = \int_{\hat{C}}G_4^{(\rm Sp)} &=& \tfrac12\int_{\hat{X}_4} E^2 [q] \, [b_{4,1}^2+b_2(4\rho\tau - a_{3,1}^2)] \nonumber \\  
      &=& -\int_{B_3} [P] \cdot [q] \cdot ( 8c_1(B_3)-2[P]) \, ,
\end{eqnarray}
where we have used that $[\hat{X}_4]\cdot E^2 = -2[P]\cdot[B_3]$.

\

Let us finally propose a formula for computing the bulk chiral matter, which is the matter living on the $Sp(1)$-brane
at ${P=0}$:
\begin{equation}
  \int_{E} G_4\wedge \left( c_1(E) - 2\bar{K} \right)
\end{equation}
in analogy with the type~IIB formula 
\begin{eqnarray}
 \tfrac12\left(\langle \Gamma',\Gamma\rangle +\tfrac14 \langle \Gamma_{\rm O7},\Gamma\rangle\right)
	      &=&  - \int_{[D7]} F\cdot \left(c_1([D7])+\bar{K}\right)\:. \nonumber
\end{eqnarray}

$E$ is the exceptional divisor that is a $\P^1$ fibration over the surface wrapped by the $Sp(1)$ brane. In our case:
\begin{equation}
\label{bulkchi}   \int_{E} G_4\wedge \left( c_1(E) - 2\bar{K} \right) = - \int_{\hat{X}_4} G_4^{(\rm Sp)} (E^2+2\bar{K}) = \int_{B_3} [q][P]([P]-\bar{K})\:.
\end{equation}

In the weak coupling limit, the results of this section are confirmed from the type~IIB perspective 
in section \ref{sectsp1iib}.

\subsubsection{$Sp(1)$ in a $U(1)$ restricted model}\label{sectsp1splitf}

Let us consider the same situation as in the last section and make a deformation such that $\rho\equiv 0\equiv \tau$.
This fourfold has an $Sp(1)$ singularity along the surface $S$: $X=y=P=0$ and an additional $SU(2)$ singularity along the curve $C$: $X=y=a_{3,1}=b_{4,1}=0$. \footnote{If we try to make a small resolution to cure the singularity 
on the curve $C$ (in the same way we did in section \ref{sectsmallres}, we also resolve the singularity along the 
surface $S$. 
One can look for remaining singularities on this resolved fourfold. 
Indeed, there is a singularity along the curve $X=y=P=b_{4,1} \lambda_1 - a_{3,1}\lambda_2=\lambda_2^2-b_2\lambda_1^2=0$.}

Let us resolve the fourfold along the $Sp(1)$ singularity like in Section \ref{SecWitSp1}. It is then described by the equations:
\begin{equation}
 \left\{\begin{array}{l}
      Y_+\,Y_- - X \,Q_v  = 0  \\ s\,v = P \\
   \end{array}\right.
\end{equation}
with $X,Y_\pm$ like in \eqref{XYpYmQ} and $Q_v=X(v\,X + \tfrac14 b_2 z^2) + \tfrac12 b_{4,1} z^4s$.
The new exceptional divisor is $E:$ $v=0$.
The coordinates $x,y,z,s,v$ belong to $F^2\otimes [E]^{-1}$, $F^3\otimes [E]^{-1}$, $F\otimes K$, 
$[P]\otimes [E]^{-1}$, $[E]$. 
\

One can see that the resolved fourfold still has a singularity along the curve $X=y=a_{3,1}=b_{4,1}=0$, which we cure with a small resolution. 
The resulting fourfold $\tilde{X}_4$ is given by
\begin{equation}\label{2resolvedX4}
\tilde{X}_4:\,\,\left\{\begin{array}{l}
    Q_v \lambda_1 = Y_- \lambda_2 \\ X \lambda_2 = Y_+ \lambda_1 \\ v\,s = P
   \end{array}\right. \,\,\, \subset X_7
\end{equation}
where $\lambda_1,\lambda_2$ are coordinates on a $\mathbb{P}^1$. Their divisor classes are $[\lambda_1]=H$ and $[\lambda_2]=H+F$. 
These satisfy $H(H+K)=0$.

We have summarized the divisor classes of the toric coordinates in the table below.
\begin{table}[h] 
\begin{centering} 
\begin{tabular}{|c|c|c|c|c|c|c|c|}  
\hline 
 $x$ & $y$ & $z$ & $s$ & $v$ & $\lambda_1$ & $\lambda_2$ \tabularnewline
\hline
\hline 
0 &0 & $K$ & $[P]$ &$0$  &0 &0 \tabularnewline
\hline 
$2$ &$3$ &$1$  & $0$ & $0$ & $0$ & $1$ \tabularnewline
\hline 
$-1$ &$-1$ &$0$  & $-1$ & $1$ & $0$ & $0$ \tabularnewline
\hline
$0$ &$0$ &$0$  & $0$ & $0$ & $1$ & $1$ \tabularnewline
\hline
\end{tabular}
\par
\end{centering}
\label{tab:ambientx7}
\end{table}

\

Let us see what are the possible holomorphic four-form fluxes that we can have on this resolved fourfold.
First, we can have an algebraic four-cycle as before
\begin{equation}\label{g41split}
 \gamma_\alpha = \sigma_\alpha - [\alpha](F-\tfrac{1}{2}E)[\tilde{X}_4]
\end{equation}
where
\begin{equation}
 \sigma_\alpha = \{\alpha=0 \,,\, Y_- =0 \,,\, X=0 \,,\, \lambda_1=0 \,,\, vs=P  \} \, .
\end{equation}
The four-form substraction is chosen such that the corresponding flux $G_4^{(I_1)}$ is Poincar\'e invariant 
and orthogonal to all fluxes of the type $G_4^{(\rm Sp)}$ (see below).

The form $\gamma_\alpha$ is not trivial in $X_7$, although it is Poincar\'e invariant. Its homology 
class in $X_7$ is:
\begin{equation}
 \gamma_\alpha = [\alpha]\left(H-(F-\tfrac{1}{2}E)\right)[\tilde{X}_4] \:.
\end{equation}
Moreover, when we compute the tadpole of this flux (using again the normal bundle trick), we find that
\begin{equation}\label{G1tadpolesp1split}
 -\tfrac12 \int G_4^{(I_1)} \wedge G_4^{(I_1)} = \int_{B_3} [\alpha]^2 \cdot \left(\bar{K} - \frac{[P]}{4}\right) = \int_{B_3} [\alpha]^2 \cdot \left(c_1(B_3) - \frac{[P]}{4}\right) \:.
\end{equation}

\

Again, we can have a flux in the Cartan of $Sp(1)$:
\begin{equation}  \label{G42sp1split}
 G_4^{(\rm Sp)} = \tfrac12 E \wedge [q] \, .
\end{equation}
This flux induces the same tadpole as before, \eqref{QG42F}.
\

In this setup we have several intersections between brane-loci: 
There is the intersection between the $Sp(1)$ locus and the remaining 7-brane, 
as well as the intersection of the remaining 7-brane with itself. This last curve produces a four-cycle as before:
\begin{equation}\label{hatC1split}
  \hat{C}_1 \,\,\, : \,\,\, \{X=0\,,\, y=0 \,,\, b_{4,1}=0 \,,\, a_{3,1}=0\,,\, v\,s=P\} \mbox{ in } X_7 \, .
\end{equation}

On the other hand, the four-cycle coming from the resolution of the singularity over the curve 
$C_2:\{P=b_{4,1}^2- b_2 a_{3,1}^2=0\}$ is the sum of two four-cycles, as explained in appendix \ref{appendixC}. 
This sum is given by
\begin{equation}
  \hat{C}_2 \,\,\, : \,\,\, \{v=0\,,\, b_{4,1}^2-b_2 a_{3,1}^2=0\} \mbox{ in } \tilde{X}_4  \, .
\end{equation}
It appears as two $\P^1$s on top of the curve $C_2$.
Thinking in terms of the weak coupling limit, we can gain some intuition about the meaning of the different
$\P^1$s of the resolution. Let us denote the two $7$-branes of the $Sp(1)$ stack by $S$ and $S'$ and the
remaining `$U(1)$-restricted' $7$-branes by $D_{I_1}$. One $\P^1$ gives M2-branes associated with strings $D_{I_1}\rightarrow S$, while 
the other $\P^1$ is the one wrapped by M2-branes associated with strings $S'\rightarrow D_{I_1}$. There exist two different four-cycles that are the fibration 
of only one of these $\P^1$s over the curve $C_2$. The four-cycle $\hat{C}_2$ is the sum of them.

Even though we can describe the sum of those two $\P^1$s as an algebraic cycle, we cannot see them independently
in an algebraic way. We can achieve this, however, by constructing an auxiliary fourfold:
We introduce a new coordinate $\xi$ living in $\bar{K}$, and a new equation $\xi^2 = b_2$ 
\footnote{From the weak coupling limit, it is clear what this means: We construct the double cover of the
base space, branched over the location of the $O7$-plane. As we have chosen $b_{6,2}$ to be a square,
this means that $\Delta_{12,2}$ describes a brane and its image (under the corresponding involution) which are 
\emph{separate} surfaces in the double cover.}. The auxiliary 
fourfold (which is no longer a CY), is a double cover of the CY fourfold given by the equations:
\begin{equation}\label{2resolvedX4Auxiliary}
\tilde{Y}_4:\,\,\left\{\begin{array}{l}
    \left(\begin{array}{cc}
	X(v\,X+\tfrac14\xi^2z^2)+ \tfrac12 b_{4,1}s\,z^4 & - Y_m \\ -Y_p & X \\
    \end{array}\right)\,
    \left(\begin{array}{c}
	\lambda_1 \\ \lambda_2 \\ 
    \end{array}\right) \,\,=\,\,
   \left(\begin{array}{c}
	0 \\ 0 \\ 
    \end{array}\right)\\    \\ v\,s = P \\ \\ \xi^2 = b_2
   \end{array}\right. \,\,\, 
\end{equation}
embedded in an ambient eightfold $X_8$. In $X_8$, the equation defining the curve $C_2$ factorizes:
\begin{equation}
 C_2\rightarrow C_{2,+} \cup C_{2,-} \qquad\mbox{ where } \qquad C_{2,\pm}: \{P=0,\,b_{4,1}=\pm \xi a_{3,1}\} \:.
\end{equation}

\begin{figure}
\begin{center}
\includegraphics[width=6cm]{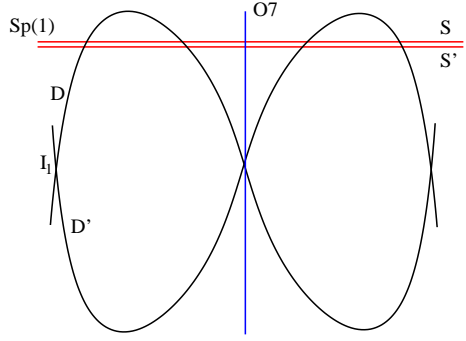}
\end{center}
\caption{Setup with an `$U(1)$-restricted' $I_1$ brane, and an $Sp(1)$-stack.}
\label{fig}
\end{figure}

In this double cover $\tilde Y_4$, the `$U(1)$-restricted' $I_1$ brane on $D_{I_1}$ lifts to two branches: 
\be
D \cup D' \mapsto D_{I_1}\,.
\ee
From the IIB perspective, these can be thought of as the brane and its orientifold image. The situation is summarized in figure \ref{fig}.

Consider the locus $\{v=0$, $b_{4,1}=\xi\,a_{3,1}\}$: 
The determinant of the $2\times 2$ matrix in \eqref{2resolvedX4Auxiliary} factorizes:
\begin{equation}
 \det\left(\begin{array}{cc}
	\tfrac14\xi^2 z^2 X+ \xi b_{4,1}s\,z^4 & - y + \tfrac12 a_{3,1}s\,z^3 \\ - y - \tfrac12 a_{3,1}s\,z^3 & X \\
    \end{array}\right) = \left( y - \Upsilon_{2,+} \right) \left(y + \Upsilon_{2,+} \right)\:
\end{equation}
where $\Upsilon_{2,+} = \tfrac12 a_{3,1}s\,z^3 + \tfrac12 \xi z\,X $. 
On each of the two branches, the two lines of the matrix become dependent and only give one more relation.
Hence we get the following two four-cycles in $X_8$ (that belong to $\tilde{Y}_4$):
\begin{eqnarray}
 \hat{C}_{2,+}^{(D\rightarrow S)} &=& \{\, v=0,\,\, b_{4,1}=\xi \,a_{3,1}, \,\, y=\Upsilon_{2,+},\,\, 
	      X\lambda_2=(\tfrac12 \xi z X+2z^3\psi s)\lambda_1 ,\,\,vs = P, \,\, \xi^2=b_2 \,\} \nonumber \\ \\
 \hat{C}_{2,+}^{(S'\rightarrow D)} &=& \{\, v=0,\,\, b_{4,1}=\xi\,a_{3,1}, \,\, y=-\Upsilon_{2,+},\,\, 
	      \lambda_2= -\tfrac12 \xi z \lambda_1 ,\,\,vs = P, \,\, \xi^2=b_2 \,\}\:. \nonumber
\end{eqnarray}
It is crucial that these two four-cycles are in different homology classes.

Let us integrate $\sigma_\alpha$ over the two four-cycles:
\begin{eqnarray}
 \int_{\hat{C}_{2,+}^{(D\rightarrow S)}} \sigma_\alpha & = &  \int_{X_8} [\alpha][y][X][\lambda_1]E[b_{4,1}-\xi a_{3,1}][vs-P][\xi^2-b_2]\nonumber \\
				     & = &   \int_{X_3} [\alpha][b_{4,1}-\xi a_{3,1}][P] =\int_{X_3} [\alpha](4\bar{K}-[P])[P] \:.
\end{eqnarray}
Where we have used that $X=0$, $Y_-=0$ and $\lambda_1=0$ in $\sigma_\alpha$ satisfy
$X\lambda_2=(\tfrac12 \xi X+2a_{3,1} s)\lambda_1$ and $y=a_{3,1} s+\tfrac12\xi X$  in $\hat{C}_1^{(D\rightarrow S)}$.

The integral on the other four-cycle vanishes:
\begin{eqnarray}
 \int_{\hat{C}_{2,+}^{(S'\rightarrow D)}} \sigma_\alpha & = &  0\:,
\end{eqnarray}
as $\lambda_1=0$ in $\sigma_\alpha$ does not intersect 
$ \lambda_2= -\tfrac12 \xi\lambda_1$ in $\hat{C}_1^{(S'\rightarrow D)}$.

\

Let us now compute the integral of the four-form $[p]\wedge E$ over the two four-cycles under study. Here $p$ is any 
polynomial in $B_3$. We claim that we get the same result from the integral of this form on the two four-cycles. 
In fact the homology classes of the two four-cycles in $X_8$ differ by the class:
\begin{equation}
 \hat{C}_{2,+}^{(D\rightarrow S)}-\hat{C}_{2,+}^{(S'\rightarrow D)} = 
      [X] [y] E [b_{4,1}-\xi a_{3,1}] [vs-P][\xi^2-b_2] \:.
\end{equation}
If we integrate $[p]\wedge E$ over it, we get 
\begin{equation}
 \int_{ \hat{C}_{2,+}^{(D\rightarrow S)}-\hat{C}_{2,+}^{(S'\rightarrow D)}}[p]\wedge E = 
\int_{X_8} [X] [y] E [P]^2 [p] [b_{4,1}-\xi a_{3,1}] [\xi^2-b_2] = 0
\end{equation}
because we have the intersection of four divisors of $B_3$\footnote{This is a bit subtle. In general the 
class $[P]$ can also contain the variables $v,s$; but when we intersect with $E[X][y]$, we are fixing 
the variables $v,s$ so that this class only describes a divisor in the base $B_3$.}.

So we are left with the integral
\begin{eqnarray}
 \int_{\hat{C}_{2,+}^{(S'\rightarrow D)}} [p]\wedge E &=& \int_{X_8} [p] [b_{4,1}-\xi a_{3,1}] [y] E [vs-P][\lambda_2] E [\xi^2-b_2]\nn \\
	    &=& \int_{X_8} [p] [b_{4,1}-\xi a_{3,1}] [y] E [vs-P][\lambda_2] (-[X]) [\xi^2-b_2]\nn \\
	    &=& - \int_{X_3} [p] (4\bar{K}-[P]) [P] 
\end{eqnarray}
where we used $[X] = -E+2F$ and the fact that $EF=E\bar{K}$.

With the help of these integrals and remembering that $\bar{K}=c_1(B_3)$, we can compute the integral of $G_4^{(I_1)}$ on the two four-cycles:
\begin{eqnarray}
 \int_{\hat{C}_{2,+}^{(D\rightarrow S)}} G_4^{(I_1)} & = & \int_{\hat{C}_{2,+}^{(D\rightarrow S)}} \sigma_\alpha - \int_{\hat{C}_{2,+}^{(D\rightarrow S)}} s \nonumber\\
	    &=& \int_{\hat{C}_{2,+}^{(D\rightarrow S)}} \sigma_\alpha + \tfrac12 \int_{\hat{C}_{2,+}^{(D\rightarrow S)}} [\alpha]E \nonumber\\
	    &=&  \tfrac12 \int_{X_3} [\alpha](4c_1(B_3)-[P])[P]\label{intG41C2+} \, ,\\
 \int_{\hat{C}_{2,+}^{(S'\rightarrow D)}} G_4^{(I_1)} & = & \int_{\hat{C}_{2,+}^{(S'\rightarrow D)}} \sigma_\alpha - \int_{\hat{C}_{2,+}^{(S'\rightarrow D)}} s \nonumber\\
	    &=& \tfrac12 \int_{\hat{C}_{2,+}^{(S'\rightarrow D)}} [\alpha]E \nonumber\\
	    &=& - \tfrac12 \int_{X_3} [\alpha](4c_1(B_3)-[P])[P]  \, .
\end{eqnarray}
In this case we see that the number of M2-branes corresponding to strings $D\rightarrow S$ is the same as those
corresponding to strings $D\rightarrow S'$. These states form a ${\bf 2}$ of $Sp(1)$. This matches with the fact 
that we have switched on a two-form flux only on the $U(1)$ 7-brane. This does not break the gauge group on
the $Sp(1)$ stack. Nonetheless, it gives rise to chiral states that are charged under the $U(1)$ and furthermore
sit in a representation of the gauge group $Sp(1)$. 

\

If we also allow for a flux $G_4^{(\rm Sp)}$ as in \eqref{G42sp1split}, the chiral indices receive the contribution
\begin{eqnarray}
 \int_{\hat{C}_{2,+}^{(D\rightarrow S)}} G_4^{(\rm Sp)} = 
 \int_{\hat{C}_{2,+}^{(S'\rightarrow D)}} G_4^{(\rm Sp)} & = & \tfrac12 \int_{\hat{C}_{2,+}^{(S'\rightarrow D)}} [q]E
	    = - \tfrac12 \int_{X_3} [q](4\bar{K}-[P])[P]\:.
\end{eqnarray}
Putting everything together, one realizes that a flux $G_4 =G_4^{(I_1)} + G_4^{(\rm Sp)} $ induces the chiralities
\begin{eqnarray}
 \int_{\hat{C}_{2,+}^{(D\rightarrow S)}} G_4 & = & \int_{X_3}\left( \tfrac{[\alpha]}{2} - \tfrac{[q]}{2}\right)(4c_1(B_3)-[P])[P] \:,\nn \\ 
 \int_{\hat{C}_{2,+}^{(S'\rightarrow D)}} G_4 & = & -\int_{X_3}\left( \tfrac{[\alpha]}{2} + \tfrac{[q]}{2}\right)(4c_1(B_3)-[P])[P]  \:.
\label{sp1splitg1g2a}
\end{eqnarray}
We will see a perfect match with the perturbative type~IIB formulae in section \ref{sectsp1splitiib}.

Finally, we note that there are two other four-cycles which come from the curve at $v=0$ and $b_4= - \xi a_3$. 
These two four-cycles are in the same homology classes as the ones studied and hence give the same result. They are 
mapped to the cycles already considered under the involution $\xi\rightarrow -\xi$.

We can also consider the singlets under $Sp(1)$ which are located at the intersection of the $U(1)$ brane with itself.
The computation of section \ref{chirsplit} carries over to the present case, so that we find that the flux $G_4^{(I_1)}$ 
gives a chiral spectrum:
\begin{equation}\label{SingletsSp1}
 \int_{\hat{C}_1}G_4^{(I_1)} = \int_{\hat{C}_1}\sigma_\alpha=\int_{B_3} (4c_1(B_3)-[P])\cdot(3c_1(B_3)-[P])\cdot [\alpha]\:.
\end{equation}
In a GUT model, we hence expect that our flux will also {\it generate chirality for GUT singlets}.

\subsubsection{$SU(N)$ GUT models and chiral matter}

In this section we want to give an example that is closer to the $SU(5)$ GUT constructions in F-theory. In those models, one has a fourfold with
an $SU(5)$ singularity along a divisor $S_{GUT}$ at $\zeta=0$. Along curves in $S_{GUT}$ the singularity can enhance to $SU(6)$ or $SO(10)$. On these curves new degrees of 
freedom arise. These loci can be seen to be the intersection between $S_{GUT}$ with the remaining 7-brane $I_1$ locus. In particular, the discriminant of the Weierstrass
equation is required to factorize 
\begin{equation}
 \Delta \sim \zeta^5 \cdot p 
\end{equation}
where $p$ is a polynomial such that $[p]+5[\zeta]=12\bar{K}$. The locus at which $\{\zeta=0\} \cap \{ p=0\}$ factorizes in two pieces, one with an $SU(6)$ 
enhancement, over which we have matter in the ${\bf 5}$ of $SU(5)$, an one with $SO(10)$ enhancement, where we have matter in the ${\bf 10}$. If we want to preserve
the GUT group, then no flux in the Cartan of the $U(5)$ stack should be switched on. On the other hand, if we want 4D chiral matter in the ${\bf \bar{5}}\oplus{\bf 10} $,
then a flux must be switched on either along the diagonal $U(1)$ of the $U(5)$ (as in \cite{Grimm:2011tb}) or along the $U(1)$ of the $U(1)$-restricted $I_1$-locus. 
We proceed with the second approach and we switch on a flux of kind $G_4^{(I_1)}$.

We now describe a simplified example: We consider a split $SU(2)$ singularity along a divisor $P=0$, instead of 
a split $SU(5)$. This case is realized when the sections $a_i$ factorize in the following way:
\begin{equation}
  a_2=P\cdot a_{2,1} \qquad a_3=P\cdot a_{3,1} \qquad a_4=P\cdot a_{4,1} \qquad a_6=P^2\cdot a_{6,2}\:. 
\end{equation}
With respect to the $Sp(1)$ case we now also have factorized $a_2$. This leads to a non-generic form of $b_2$:
\begin{equation}
 b_2 = a_1^2 + 4 a_{2,1}\cdot P \:.
\end{equation}
The discriminant of the Weierstrass equation is factorized as $\Delta= P^2(b_{4,1}^2-b_2a_{3,1}^2)$.

Let us put $a_{6,2}\equiv 0$. On the intersection with the locus $P=0$, $b_{4,1}^2-b_2a_{3,1}^2$, factorizes to
$(b_{4,1}-a_1a_{3,1})(b_{4,1}+a_1a_{3,1})$. 
This means that the matter curve $C_2$ splits into two curves in $B_3$: $C_2=C_{2,+}\cup C_{2,-}$. 
Correspondingly we expect two separate ${\bf 2}$s of $SU(2)$.

We can go further, following the $Sp(1)$ case, and resolve the singularities by one resolution along $P=0$ and one small resolution on the surviving 
singularity. We end up with the equations:
\begin{equation}\label{2resolvedX4AuxiliarySU2}
\tilde{X}_4^{s}:\,\,\left\{\begin{array}{l}
    \left(\begin{array}{cc}
	X(v\,X+\tfrac14 (a_1^2+4a_{2,1}s\,v)z^2)+ \tfrac12 b_{4,1}s\,z^4 & - Y_- \\ -Y_+ & X \\
    \end{array}\right)\,
    \left(\begin{array}{c}
	\lambda_1 \\ \lambda_2 \\ 
    \end{array}\right) \,\,=\,\,
   \left(\begin{array}{c}
	0 \\ 0 \\ 
    \end{array}\right)\\    \\ v\,s = P \\ 
   \end{array}\right. \,\,\, \subset X_7 \:.
\end{equation}
Consider the locus $\{v=0$, $b_{4,1}=a_1\,a_{3,1}\}$: 
The determinant of the $2\times 2$ matrix in \eqref{2resolvedX4AuxiliarySU2} factorizes:
\begin{equation}
 \det\left(\begin{array}{cc}
	\tfrac14 a_1^2 z^2 X+ \tfrac12 a_1 a_{3,1}s\,z^4 & - y + \tfrac12 a_{3,1}s\,z^3 \\ - y - \tfrac12 a_{3,1}s\,z^3 & X \\
    \end{array}\right) = \left( y - \Upsilon^s_{2,+} \right) \left(y + \Upsilon^s_{2,+} \right)\:
\end{equation}
where $\Upsilon^s_{2,+} = \tfrac12 a_{3,1}s\,z^3 + \tfrac12 a_1 z\,X $. 
On each of the two branches, the two lines of the matrix become dependent and then give only one more relation.
Hence we get the following two four-cycles in $X_7$ which are also inside $\tilde{X}^s_4$:
\begin{eqnarray}\label{C2pSplit}
 \hat{C}_{2,s+}^{(D\rightarrow S_1)} &=& \{\, v=0,\,\, b_{4,1}=a_1 \,a_{3,1}, \,\, y=\Upsilon^s_{2,+},\,\, 
	      X\lambda_2=(\tfrac12 a_1 z X+2z^3a_{3,1} s)\lambda_1 ,\,\,v s = P, \,\} \nonumber\:, \\ \\
 \hat{C}_{2,s+}^{(D\leftarrow S_2)} &=& \{\, v=0,\,\, b_{4,1}=a_1\,a_{3,1}, \,\, y=-\Upsilon^s_{2,+},\,\, 
	      \lambda_2= -\tfrac12 a_1 z \lambda_1 ,\,\,v\sigma = P,  \,\}\:. \nonumber
\end{eqnarray}
These two four-cycles are in different homology classes. Let us explain their origin. On top of the curve $C_{2,+}$ of the
base $B_3$ the fiber develops two $\P^1$s after the resolution. The states that make up ${\bf 2}$ of  
$SU(2)$ come from M2-branes wrapped on these $\P^1$s. If we fiber the two $\P^1$s separately over $C_{2,+}$, we get the 
two different four-cycles in \eqref{C2pSplit}. 
We have denoted the branches of the resolved $SU(2)$ singularity by $S_1$ and $S_2$ and the remaining brane by $D$.

We claim that the orientation of the M2-branes in a ${\bf 2}$ are oriented in an opposite way. To count the number of
${\bf 2}$s given by a flux on the curve $C_{2,+}$, we can either integrate $G_4^{(I_1)}$ on 
$\hat{C}_{2,s+}^{(D\rightarrow S_1)}$ or $-\hat{C}_{2,s+}^{(D\leftarrow S_2)}$, as the two integrals give the
same result.

We find
\begin{equation}
 \int_{\hat{C}_{2,s+}^{(D\rightarrow S_1)}} G_4^{(I_1)} = \tfrac12 \int_{B_3} [P]\cdot (4\bar{K}-[P])\cdot [\alpha] = \int_{C_{2,+}} \tfrac{[\alpha]}{2}
\end{equation}
i.e. we integrate the flux on the brane-imagebrane system on the intersection curve.

\subsubsection*{Orientation of the M2-branes}

Let us consider one of the curves where the $SU(2)$ singularity locus intersects the brane-imagebrane system. After the resolution,
the fiber develops an extra $\P^1$ on top of the $SU(2)$ locus. This $\P_1$ separates the two singular points of the fiber that were on top of 
each other before the resolution, creating the singularity of the full space. The M2-brane wrapped on this $\P^1$ stretches between these two 
points and gives the degrees of freedom to construct the adjoint of $SU(2)$. On top of the matter curve, this $\P^1$ splits into two $\P^1$s:
\begin{equation}
 S_1 \longrightarrow S_2   \qquad \rightsquigarrow \qquad S_1 \longrightarrow D \longrightarrow S_2    
\end{equation}
The arrows stand for oriented $\P^1$s. The M2-branes wrapping the sum of the two $\P^1$s
are still the degrees of freedom of the adjoint of $SU(2)$. The M2-branes wrapping the separate $\P^1$s make up
a doublet representation of $SU(2)$. The charges of the M2-branes with respect to the Cartan of $SU(2)$ are given
by the intersections with the sum of the two $\P^1$s (related to the gauge field coming from $C_3$ along the corresponding two-form).
Hence a doublet of $SU(2)$ is made up of two complex states with opposite charges with respect to the Cartan generator. That is why a doublet
is constructed from the states of an M2-brane wrapped on $\P^1_{S_1 \rightarrow D}$ and another M2-brane wrapped on $\P^1_{D \rightarrow S_2}$ 
with the opposite orientation.

\section{The type~IIB picture}\label{sectIIB}

In this section we want to check our F-theory results by comparing them with the ones obtained in perturbative type~IIB string theory. 

For compactifications of F-theory, there exists a region of the moduli space where the string coupling is weak everywhere over $B_3$.  
In this region, one can define a CY threefold that is a double cover of $B_3$. If this CY threefold is smooth, we can trust perturbative type~IIB 
string theory and expect the results to match those obtained from F-theory. In the case where the CY threefold is singular, this match is more 
difficult (see \cite{Donagi:2009ra}).

The weak coupling region of the CY fourfold moduli space is reached in the following way: One scales the sections
$a_i$ as (see \cite{Sen:1996vd, Donagi:2009ra})
\begin{equation}
  a_3 \rightarrow \epsilon \, a_3 \qquad   a_4 \rightarrow \epsilon \, a_4 \qquad   a_6 \rightarrow \epsilon^2 a_6\:,
\end{equation}
where $\epsilon$ is a complex constant that drives the weak coupling limit. With these redefinitions, the sections $b_i$ scale like
\begin{equation}
  b_2 \rightarrow \epsilon \, b_2 \qquad   b_4 \rightarrow \epsilon \, b_4 \qquad   b_6 \rightarrow \epsilon^2 b_6\:.
\end{equation}
At leading order in $\epsilon\rightarrow 0$, the string coupling $g_s=($Im$\tau)^{-1}$ becomes weak everywhere (except near $b_2=0$), 
while the discriminant locus of the elliptic fibration becomes:
\begin{equation}
 \Delta \approx  \epsilon^2 b_2^2(b_4^2-b_2 b_6) + {\cal O}(\epsilon^2)\:.
\end{equation}
We see that the 7-brane locus factorizes in two pieces in weak coupling limit. It turns out that the monodromies of 
the axion-dilaton $\tau$ around the two separate loci are those of an $O7$-plane and a $D7$-brane, respectively. We can identify
\begin{equation}\label{O7D7Fth}
 O7:\qquad b_2 = 0 \qquad\qquad\qquad D7: \qquad b_4^2-b_2b_6=0\:.
\end{equation}

If the polynomials $b_i$ are generic, the $D7$-brane is one connected surface. Due to the specific structure of its defining equation \eqref{O7D7Fth}, 
it is not a completely generic divisor \cite{bht08}. Furthermore, a $D7$-brane of the form \eqref{O7D7Fth} has singular points contained in its worldvolume, 
around which it has the shape of the so-called  Whitney umbrella \cite{Collinucci:2008pf}. We hence refer to a $D7$-brane described by an
equation of the type \eqref{O7D7Fth} as a \emph{Whitney type $D7$-brane}.

The weak-coupling limit of F-theory can be matched with perturbative type~IIB string theory compactified on an orientifold of
the CY threefold $X_3$:
\be
 \xi^2 = b_2 \, .
\ee
Here, we have added a new coordinate $\xi$. The orientifold involution acts on $X_3$ by sending $\xi \mapsto -\xi$, so that
$X_3$ is a double cover of $B_3$, branched over the locus $\xi=0$. The projection map
\be
\pi: X_3 \mapsto B_3
\ee
is 2-1 everywhere except at the branch loci (we assume that there are no $O3$-planes). 
We have the following relation for the Poincar\'e dual of the divisor class of the $O7$-plane:
\be [O7] = \pi^*(\bar{K}) = \pi^*(-c_1(B_3))\,.
\ee
To avoid cluttering our formulae, we will use the following notation:
\be
[O7] = -K \in H_4(X_3,\Z)\,, \quad c_1(B_3) = \bar{K}\,,  \quad \pi^*\left(K\right)=K\,, \quad \pi^*\left(c_1(B_3)\right) = -K \in H^2(X_3, \Z) \, .
\ee
We are dropping pull-back symbols and, as before, the distinction between a line bundle, its first Chern class, and its associated divisor. 

A suitable technique to describe such a background is based on Sen's tachyon condensation \cite{Collinucci:2008pf, Collinucci:2008zs, Collinucci:2009uh}.
The idea is to describe the fluxed $D7$-brane configuration by using a set of $D9$-branes and image anti-$D9$-branes with suitable vector bundles
that eventually condense to the wanted $D7$-brane configuration.

\subsection{Flux on a Whitney-type brane}\label{sectIIBwhitney}

We will now study the flux on a generic \emph{Whitney-type} $D7$-brane in perturbative type~IIB theory. 

As explained in \cite{Collinucci:2008pf}, an orientifold invariant $D7$-brane is best described as a 
tachyon condensate of two $D9$-branes and two anti-$D9$-branes with the following bundles on them:
\begin{table}[ht]\centering
\begin{tabular}{ccccccccc}
$\overline{D9_1}$&&$\overline{D9_2}$&&&&$D9_1$&&$D9_2$\\
${\mathcal{L}_a}^{-1}$&$\oplus$&${\mathcal{L}_b}^{-1}$&&$\stackrel{T}{\longrightarrow}$&&$\mathcal{L}_a$&$\oplus$&$\mathcal{L}_b$\,,
\end{tabular}\end{table}

\noindent where $\mathcal{L}_a, \mathcal{L}_b$ are holomorphic line bundles. 
The most general tachyon matrix respecting the orientifolding has the form:
\begin{eqnarray}
T(\vec{x},\xi)&=&\left(\begin{array}{cc}0&\eta(\vec{x})\\-\eta(\vec{x})&0\end{array}\right)\,+\,\xi\left(\begin{array}{cc}\rho(\vec{x})&\psi(\vec{x})\\ \psi(\vec{x})&\tau(\vec{x})\end{array}\right)\,,
\end{eqnarray}
where $\eta, \psi, \rho, \tau$ are sections of $\L_a \otimes \L_b,\, \L_a \otimes \L_b \otimes K, \,{\L_a}^2 \otimes K, \, {\L_b}^2 \otimes K$, respectively. The $D7$-brane divisor will be given by
\begin{equation} \label{whitneyfromtachyon}
{\rm det}T = \eta^2+\xi^2\,(\rho\,\tau -\psi^2) = 0\,.
\end{equation}
We see that this is the same expression as in \eqref{O7D7Fth}, once we substitute the equation of the CY threefold and identify:
\begin{equation}
  \frac{a_3}{2}\equiv \psi \qquad\qquad \frac{b_4}{2} \equiv \eta \qquad \qquad \frac{b_6}{4} \equiv  \psi^2 - \rho\,\tau \:.
\end{equation}
Note that the form of $b_6$ is not generic for non-trivial $\rho$ and $\tau$. 
As explained in \cite{Collinucci:2008pf}, this signals the presence of gauge flux on the $D7$-brane 
which fixes some of the brane moduli. In our construction of $G_4$ flux, a similar mechanism is at work.
In order to construct the flux, we need to fix some complex structure moduli of the elliptic fourfold 
such that $a_6\equiv\rho\tau$, i.e. the fourfold takes the form \eqref{X4}.

The total charge `Mukai' vector $\Gamma_{D7}$ of this system is given by the following formula:
\begin{eqnarray} \label{generalmukai}
\Gamma_{D7} &=& \Gamma_{D9_1} +\Gamma_{D9_2} - \Gamma_{\overline{D9_1}} - \Gamma_{\overline{D9_2}}\nonumber\\
&=& \Big({\rm ch}(\L_a)+{\rm ch}(\L_b)-{\rm ch}(\L_a^{-1})-{\rm ch}(\L_b^{-1}) \Big) \, \left(1+\tfrac{c_2(X_3)}{24} \right)\\
&=&2\,\big(c_1(\L_a)+c_1(\L_b)\big) + 
    \bigg[ \tfrac{1}{3}\,\left(c_1(\L_a)^3+c_1(\L_b)^3\right)+\tfrac{1}{12}\,\left( c_1(\L_a)+c_1(\L_b) \right) \cdot c_2(X_3) \bigg]\, .\nonumber
\end{eqnarray} 
The first term gives the total $D7$-charge and the second one the total $D3$-charge, which is induced both by  
curvature and gauge field-strength on the $D7$. Note that all charges in the above expression refer to
the physics in the double cover $X_3$. The projection to $B_3$ will hence half the charges.

Since we have only one $D7$-brane, its $D7$-charge must cancel the $D7$-charge of the $O7$-plane. The charge vector of the $O7$-plane is
\begin{equation}\label{GammaO7}
 \Gamma_{O7} = -8[O7] + \frac{\chi([O7])}{6}\omega \, ,
\end{equation}
where $\omega$ is the volume form on $X_3$. The cancellation of the $D7$-charge then implies 
\begin{equation}\label{c1ac1bW}
c_1(\L_a)+c_1(\L_b) = - 4 K  = 4 \bar{K} \:.
\end{equation}

For the $D3$-charge of the Whitney brane, we can easily disentangle the curvature induced one from the flux induced one. 
It was argued in \cite{Collinucci:2008pf} that the $D7$-brane resulting from the 
tachyon condensation will have zero gauge flux whenever the bundles on the $D9$ and anti-$D9$-branes are such that 
either $\rho$ or $\tau$ are constants. This is equivalent to choosing ${\L_a}^2 = K^{-1}$ or ${\L_b}^2 = K^{-1}$. 
The class of the $D7$-brane does not change, i.e. $[D7]=2(c_1(\L_a)+c_1(\L_b))$ remains the same class; what changes 
is the difference $c_1(\L_a)-c_1(\L_b)$.

We then substitute $c_1(\L_a) =- \frac12 K$ (consequently $c_1(\L_b) =\frac12 ([D7]+K)$) into the tadpole formula above\footnote{If $B_3$ is not spin, then $K$ is not divisible by two. In principle, in order to compute the curvature induced tadpole, one should resolve the Whitney brane, and compute its Euler characteristic. However, it turns out that the formal manipulation applied here still works.}:
\be
Q_{D3}^c = \tfrac{1}{24}\,\int_{X_3} [D7] \cdot \big(3\,K^2+3\,K \cdot [D7]+[D7]^2+c_2(X_3) \big)\,.
\ee
We can find the flux induced tadpole by subtracting this result from that in \eqref{generalmukai}, yielding:
\begin{eqnarray} \label{fluxtadpolektheory}
Q_{D3}^f &=& \tfrac{1}{8}\,\int_{X_3} [D7] \cdot \left(2\,c_1(\L_a)+K \right) \cdot \left(2\,c_1(\L_a)-[D7]-K \right)\\ \nonumber
&=& -\tfrac{1}{4} \int_{X_3} \left( c_1(\L_a)+c_1(\L_b) \right) \cdot \left(2\,c_1(\L_a)+K \right) \cdot \left(2\,c_1(\L_b)+K \right)\,.
\end{eqnarray}

\subsubsection*{Match with F-theory}

To make the comparison with F-theory, we need to remember that the result in \eqref{fluxtadpolektheory2}
is given in the double cover, i.e. it gives the number of $D3$-branes needed to cancel the tadpole 
before taking the orientifold projection. After the projection the number of $D3$-branes is 
halved, yielding the number of physical $D3$-branes. 

If we substitute the relation \eqref{c1ac1bW} and $[\rho]=2c_1(\L_a)+K$, so that $[\tau]=2c_1(\L_b)+K$, in \eqref{fluxtadpolektheory}
and divide by two to take into account the orientifolding, we obtain:
\begin{eqnarray} \label{fluxtadpolektheory2}
Q_{D3} &=& -\tfrac12 \int_{X_3}  \bar{K} \cdot [\rho]  \cdot [\tau]   \nonumber \\
      &=& - \int_{B_3}  c_1(B_3) \cdot [\rho]  \cdot [\tau] \, .
\end{eqnarray}
Hence we obtain precisely the same result as in the corresponding F-theory computation, \eqref{g4tadpole}.

\subsubsection*{Algebraic definition of the flux}

The flux $G_4$ in F-theory has an algebraic description. It is defined as the difference of two algebraic four-cycles of $X_4$ that are homologous 
in the ambient space $X_5$, but reside in different homology classes in $X_4$. In this section, we want to give an analogous description for the
two-form flux on a $D7$-brane.

The profile of the relic $D7$-brane after the $D9$'s and anti-$D9$'s condense is given by equation \eqref{whitneyfromtachyon}, which has 
a curve worth of double point singularities at $\eta=\xi=0$.

As described in \cite{Collinucci:2008pf}, one can resolve this singularity by blowing up $X_3$ into a new space $\hat X_3$. This entails 
introducing a new coordinate $t \in \L_a \otimes \L_b \otimes K$ and imposing the constraint $\eta = t\,\xi$. 
We will not follow this procedure here, as we do not need to make explicit computations with the flux.

The line bundle that lives on the brane corresponds to the `quotient' $E/T(F)$, where
\begin{equation}
T: F= \mathcal{L}_a^{-1} \oplus \mathcal{L}_b^{-1} \mapsto E=\mathcal{L}_a \oplus \mathcal{L}_b\,.
\end{equation}

A typical section $s = (\lambda_a, \lambda_b)$ of the quotient bundle will have its vanishing locus wherever both components are zero, 
or wherever it lies in the image of $T$. Clearly, the system of equations $\lambda_a = \lambda_b = 0$ will be satisfied along a curve that 
intersects, but does not lie on the $D7$-brane. Consequently, we will neglect this locus.

Define a matrix $\tilde T$ such that $T\, \tilde T = \det(T)$:
\begin{eqnarray}
\tilde T(\vec{x},\xi)&=&\left(\begin{array}{cc}\xi\tau(\vec{x})&-\eta-\xi\psi(\vec{x})\\ +\eta-\xi\psi(\vec{x})&\xi\rho(\vec{x})\end{array}\right)\, .
\end{eqnarray}
Hence $\tilde{T}$ is constructed such that its kernel is the image of $T$ away from the points where the rank of $T$ decreases to $0$ \footnote{
If $s$ lies in the image of $T$, i.e. $s = T \tilde s$, for some $\tilde s \in \Gamma({\L_a}^{-1} \oplus {\L_b}^{-1})$, then 
\be
\tilde T\, s|_{D7}= {\rm det}(T) \,\tilde s |_{D7}= 0\,.
\ee
Conversely, suppose we are at a locus on the $D7$, then $T$ must have rank one. The rank cannot go down to zero since that would require four 
constraints to be satisfied. Therefore, in some basis it can be rewritten as:
\begin{eqnarray}
T&=&\xi \,\left(\begin{array}{cc} P &0\\ 0 & 0\end{array}\right)\,, \quad \text{and} \quad \tilde T=\xi\,\left(\begin{array}{cc} 0 &0\\ 0 & P\end{array}\right) 
\quad \text{for some } P\,.
\end{eqnarray}
It follows that the section must lie in the image of $T$ if it is annihilated by $\tilde T$ outside $P=0$ and $\xi=0$. It is clear
that this logic only applies on the $D7$-brane, as $T$ is surjective away from there.}.
The brane wraps the divisor $S$ described by equation \eqref{whitneyfromtachyon}. Its homology class is $[S]=2(c_1(\L_a)+c_1(\L_b))$.

The locus on the brane where $s$ becomes invertible is given by the following intersection locus $\mathcal{I}$:
\begin{eqnarray}
\mathcal{I}:\qquad \tilde T \cdot \left(\begin{array}{c}
\lambda_a\\ 
\lambda_b\end{array} \right) = 0 
&\sim&\lambda_a\,\xi\tau = \lambda_b\,(\eta+\xi\psi) \quad \cap \quad \lambda_a\,(\xi\psi-\eta) = \lambda_b\,\xi\rho\, .
\end{eqnarray}
This locus contains two branches $\mathcal{I} = \mathcal{I}_\sigma \cup \mathcal{I}_{ab}$, where
\begin{eqnarray}
\mathcal{I}_\sigma\,:\,\, \tilde T \, s = 0\,, s \neq (0,0)\,, &\qquad&
\mathcal{I}_{ab}\,:\,\, s =(\lambda_a, \lambda_b)= (0,0)\,.
\end{eqnarray}

The total flux $F_2$ on the brane should correspond to the Poincar\'e dual to the branch $\mathcal{I}_\sigma$ plus the usual $+c_1(S)/2$ shift.
\be
F_2 = PD_{S}(\mathcal{I}_\sigma)+\tfrac{1}{2}\,c_1(S)\,.
\ee
This flux $F_2$ does not induce any $D5$-charge, i.e. its push-forward is trivial in the ambient threefold. In fact,
\begin{eqnarray}
PD_{X_3}\left(\mathcal{I}_\sigma \right) &=& PD_{X_3}\left(\tilde T\,s=0\right) - PD_{X_3}\left(s=(0,0)\right)\\
&=& \left(2\,c_1(\L_b)+c_1(\L_a) \right) \cdot \left(2\,c_1(\L_a)+c_1(\L_b) \right) - c_1(\L_a) \cdot c_1(\L_b)\,, \nonumber \\ \nonumber
&=& 2\left(c_1(\L_a)+c_1(\L_b) \right)^2=\tfrac12 [S] \cdot c_1(S)\:,
\end{eqnarray}
where we used $c_1(S) = -[S]$. 
Putting all the information together, we find
\begin{eqnarray}
i_*(F_2) &=& PD_{X_3}(\mathcal{I}_\sigma)+\tfrac{1}{2}\,c_1(S) \cdot [S] = 0 \, .
\end{eqnarray}
In particular, this flux respects the involution condition $\sigma^*(F_2) = -F_2$ \cite{Collinucci:2008pf}.

\subsection{Brane-Imagebrane} \label{sectIIBsplit}

Let us consider the case when the Whitney brane splits into a brane and its image. Without loss of generality we restrict to the case when
$2c_1(\L_a)- 4\bar{K}\geq 0$. We get a brane and its image in the double cover when \footnote{We want that $T$ is deformed
such that it can be brought to an off-diagonal form by a well defined gauge transformation (i.e. a gauge transformation that brings a regular 
field strength to a regular one). This forces one among $\rho$ and $\tau$ to be identically zero (depending on which one has smaller degree) 
and the other one to be proportional to $\psi$, so that it takes a factorized form. } 
$\tau=0$ and $\rho=-2\psi \beta$, where $\beta$ is a section of $\L_a\otimes \L_b^{-1}$ \cite{Collinucci:2008pf}. 
One can fix $\beta\equiv 0$ by a gauge transformation, and then also put $\rho\equiv 0$. 
In what follows, we will associate the brane/imagebrane case to the deformation that brings both 
$\rho\equiv 0$ and $\tau\equiv 0$.

The tachyon matrix becomes:
\begin{eqnarray}
T(\vec{x},\xi)&=&\left(\begin{array}{cc}0&\eta(\vec{x})+\xi\,\psi(\vec{x})\\ -\eta(\vec{x})+\xi \psi(\vec{x})&0\end{array}\right) \, .
\end{eqnarray}

The determinant of $T$ factorizes, so that the $D7$-brane splits into two branes which are identified by the orientifold involution:
\begin{equation}
 \det T = (\eta +\xi\,\psi)(\eta -\xi\,\psi) = 0 \qquad \rightarrow\qquad S_+\,:\, \eta =\xi\,\psi   \qquad S_-\, : \, \eta = -\xi\,\psi \:.
\end{equation}
Since each of these surfaces is not singular we will not need to blow up. 

In this configuration the gauge group is $U(1)$. If brane and imagebrane recombine back into a single Whitney brane, the gauge group breaks from
$U(1)$ to $O(1)$ \cite{Collinucci:2008pf,Grimm:2010ez}.

\

This system is still described by the condensate of two $D9$-branes and two anti-$D9$-branes. The action of the tachyon, however, is such that brane 
and imagebrane are described separately by the following sequences:
\begin{table}[ht]\centering
\begin{tabular}{ccccccccccccccccccccc}
&$\overline{D9_2}$&&&&$D9_1$ &\qquad\qquad\qquad&& $\overline{D9_2}$&&&&$D9_1$\\\\
$D7$:& $\L_b^{-1}$&&$\stackrel{T}{\longrightarrow}$&&$\L_{a}$ &\qquad\qquad\qquad& $D7'$:& $\L_a^{-1}$&&$\stackrel{T}{\longrightarrow}$&&$\L_{b}$
\end{tabular}
\end{table}

\noindent The `Mukai' vectors of the two branes are
\begin{eqnarray}
 \Gamma &=& \left(\ch(\L_a)-\ch(\L_b^{-1})\right)\left(1+\frac{c_2(X_3)}{24}\right) \nonumber\\
      &=& \left(c_1(\L_a)+c_1(\L_b)\right) + \left(c_1(\L_a)+c_1(\L_b)\right)\frac{c_1(\L_a)-c_1(\L_b)}{2} + 
	   Q_{D3}^{D7} \nonumber\, , \\
 \Gamma' &=& \left(\ch(\L_b)-\ch(\L_a^{-1})\right)\left(1+\frac{c_2(X_3)}{24}\right) \nonumber\\
      &=& \left(c_1(\L_a)+c_1(\L_b)\right) + \left(c_1(\L_a)+c_1(\L_b)\right)\frac{c_1(\L_b)-c_1(\L_a)}{2} + 
	   Q_{D3}^{D7'} \nonumber\, .
\end{eqnarray}
In both cases, the $D3$-charge is given by
\begin{equation}\label{D3chargeD7D7p}
 Q_{D3}^{D7} = Q_{D3}^{D7'} = \left(c_1(\L_a)+c_1(\L_b)\right)\cdot \frac{\chi(S_-)}{24} + 
			 \frac12 \left(c_1(\L_a)+c_1(\L_b)\right)\left(\frac{c_1(\L_a)-c_1(\L_b)}{2}\right)^2\: .
\end{equation}
We have used that $S_+$ and $S_-$ are homologous and that the Euler characteristic of a divisor $D$ in a CY threefold is given 
by $\chi(D) = D^3 + c_2(X_3)\cdot D$.

The first term in \eqref{D3chargeD7D7p} is the geometric contribution, while the second one is the flux $D3$-tadpole in the usual form 
$Q_{D3}^{F} = \tfrac12 \int_{D7} F_2\cdot F_2$.

\subsubsection*{Match with F-theory}
If we sum the flux contributions of the two $D7$-branes, we get
\begin{equation}\label{QD3SplW}
 Q_{D3}^{f} = \int_{X_3} \left(c_1(\L_a)+c_1(\L_b)\right)\left(\frac{c_1(\L_a)-c_1(\L_b)}{2}\right)^2  \:.
\end{equation}
The relation \eqref{c1ac1bW} still holds, so that the physical $D3$-tadpole after orientifolding is given by
\begin{equation}
 Q_{D3}^{f} = -\tfrac12 \int_{X_3} 4K \left(\frac{[\alpha]}{2}\right)^2 = \int_{B_3} c_1(B_3) \,[\alpha]^2  \: .
\end{equation}
In the above equation, $\alpha$ is a section of $\L_a\otimes \L_b^{-1}$. Hence 
\begin{equation}
[\alpha]=c_1(\L_a)-c_1(\L_b)=2c_1(\L_a)-4\bar{K}=[\rho]-3\bar{K} \, , 
\end{equation}
so that we find perfect agreement with the corresponding F-theory result, \eqref{Qalpha} and \eqref{qFsplit}.

\subsubsection*{Algebraic definition of the flux}

Let us see how the flux looks like when the Whitney brane factorizes into a brane and its image. 
Its Poincar\'e dual is still given by the locus $\mathcal{I}_\sigma=\mathcal{I}-\mathcal{I}_{ab}$.
After the transition, the locus $\mathcal{I}$ is
\begin{equation}
 \mathcal{I}\,: \tilde{T}\cdot\left(\begin{array}{c}\lambda_a\\ \lambda_b\end{array} \right)=0  \qquad\rightarrow \qquad 
    \left\{\begin{array}{l}(\eta+\xi\psi)\lambda_b=0 \\ (\eta-\xi\psi)\lambda_a =0\end{array}  \right.\:.
\end{equation}
We note that it splits into four branches:
\begin{align}
 \mathcal{I}_{ab} &:\,\, \lambda_b=0 \,\,\,\, \cap \,\,\,\, \lambda_a=0 & \qquad \mathcal{I}_{0} &:\,\, \eta = 0 \,\,\,\, \cap \,\,\,\,  \xi\psi =0 \nn\\
 \mathcal{I}_{-} &:\,\, \eta=-\xi\psi \,\,\,\, \cap \,\,\,\, \lambda_a & \qquad \mathcal{I}_{+} &:\,\, \eta=\xi\psi \,\,\,\, \cap \,\,\,\,  \lambda_b=0 \:.
\end{align}
As before, we drop the branch $\mathcal{I}_{ab}$. The branch $\mathcal{I}_{0}$ also does not contribute as the rank of $T$ goes to zero there:
On the curve $\eta=0$ inside the $D7$-brane, the quotient bundle $E/T(F)$ has rank two. Hence a generic section of this bundle does not
vanish anywhere on this curve. 

The two surviving branches define the flux on $S_-$ and the flux on $S_+$: $\mathcal{I}_{-}$ is a curve of class $c_1(\L_a)$ on $S_-$, while
$\mathcal{I}_{+}$ is a curve of class $c_1(\L_b)$ on $S_+$. To have the flux, we need to add 
half of the first Chern class of the corresponding surface: 
$F_\pm = PD_{S_\pm}(\mathcal{I}_\pm)+\tfrac12 c_1(S_\pm)$. 
We obtain
\begin{eqnarray}\label{FluxSplit}
 F_- = \tfrac12 \left(c_1(\L_a) - c_1(\L_b)\right) &\qquad&
 F_+ = \tfrac12 \left(c_1(\L_b) - c_1(\L_a)\right)
\end{eqnarray}
where we used $c_1(S_\pm) = -[S_\pm]$ and $[S_+]=[S_-] = c_1(\L_a) + c_1(\L_b)$.

\subsubsection*{Chiral states}

When the Whitney brane splits into a $D7$-brane and its image, we have a background with two $D7$-branes. At their intersection we have 6D (massless) matter. 
In presence of gauge flux on the $D7$-branes, the 4D spectrum of matter can become chiral.

Having the Mukai vector, one can compute the number of chiral states.
The DSZ intersection product between two charge vectors is given by
\begin{equation}
\langle \Gamma_1, \Gamma_2 \rangle = \int_X \Gamma_1\, \Gamma_2^*\,,
\end{equation}
where the $\Gamma^*$ is obtained in this case by flipping the two-form component of $\Gamma$. It was shown in \cite{Brunner:2003zm,Collinucci:2008pf}
that the chiral index for the spectrum between a brane and its orientifold image is given by:
\begin{equation}
I_o (\Gamma) = \tfrac{1}{2}\,\Big(\langle \Gamma', \Gamma \rangle+\tfrac{1}{4}\,\langle \Gamma_{O7}, \Gamma \rangle \Big)\,,
\end{equation}
where, in our case $\Gamma_{\rm O7} = 8\,K + \tfrac{\chi([O7])}{6}\omega$. This is an equivariant index that computes the chiral spectrum living 
at the curve where the $D7$ meets its image: $\eta = \psi=0$, \emph{away form the O7}, as opposed to the curve $\eta = \xi = 0$.
In our case, the index gives
\begin{eqnarray}
I_o(D7) &=& \int_{X_3}\left(c_1(\L_a)+c_1(\L_b)\right)\cdot\frac{c_1(\L_a)-c_1(\L_b)}{2}\cdot 
			\left(c_1(\L_a)+c_1(\L_b)+K\right) \, .\\
\end{eqnarray}
Again, we make use of the relation \eqref{c1ac1bW}, so that we can express the chiral index as
\begin{eqnarray}
 I_o(D7) &=& 12\int_{B_3} \bar{K}^2 \cdot[\alpha] = 12\int_{B_3} c_1(B_3)^2 \cdot[\alpha]\:,
\end{eqnarray}
where $\alpha$ is a section of $\L_a\otimes\L_b^{-1}$.
This is the same result we have obtained in F-theory by integrating the corresponding $G_4$
over the matter surface \eqref{ftheoryindex}.

\subsection{$Sp(1)$ stack plus a Whitney-type brane}\label{sectsp1iib}

In this section, we consider the weak coupling limit of the configuration discussed in section \ref{SecWitSp1}.
In type~IIB language, we then have an orientifold-invariant stack of two $D7$-branes, which intersect the $O7$-plane transversally. 
Each of them wraps the divisor $P=0$, where $P$ is a section of the line bundle ${\cal L}_P$. The remaining $D7$-tadpole is saturated 
by a single Whitney-type $D7$-brane. 

The corresponding system of $D9-\overline{D9}$ is given by
\begin{table}[ht]\centering
\begin{tabular}{ccccccccccccccc}
$\overline{D9_1}$&&$\overline{D9_2}$&& $\overline{D9_3}$ && $\overline{D9_4}$ \\
${\mathcal{L}_a}^{-1}$&$\oplus$&${\mathcal{L}_b}^{-1}$&$\oplus$&${\mathcal{L}_P}^{-1/2}\otimes{\mathcal{L}_q}^{-1/2}$ &$\oplus$& ${\mathcal{L}_P}^{-1/2}\otimes{\mathcal{L}_q}^{1/2}$    
\,,\\ \\
$D9_1$&&$D9_2$ &&$D9_3$ &&$D9_4$ \\
$\mathcal{L}_a$&$\oplus$&$\mathcal{L}_b$ &$\oplus$& $\mathcal{L}_P^{1/2}\otimes{\mathcal{L}_q}^{1/2}$&$\oplus$& $\mathcal{L}_P^{1/2}\otimes{\mathcal{L}_q}^{-1/2}$
\,,
\end{tabular}\end{table}

\noindent where $\mathcal{L}_a, \mathcal{L}_b,\mathcal{L}_P,\mathcal{L}_q$ are holomorphic line bundles. The divisor class of the $O7$ is equal to $\pi^*(\bar{K})$. 
The tachyon matrix of this system is given by:
\begin{equation}\label{TachSpW}
T = \left(\begin{array}{cc}
\left(\begin{array}{cc}
\xi\,\rho & \eta + \xi\,\psi \\ - \eta + \xi\,\psi & \xi\,\tau \\
\end{array}\right) & \\ 
& \left(\begin{array}{cc}
0 & P \\ - P & 0 \\
\end{array}\right)
\end{array}\right)\:,
\end{equation}
where $\eta, \psi, \rho, \tau, P$ are sections of $\L_a \otimes \L_b,\, \L_a \otimes \L_b \otimes K, \,{\L_a}^2 \otimes K, \, {\L_b}^2 \otimes K, \, \L_P$, respectively. 

The $D7$ locus has two factors:
\begin{equation}
\det T = P^2(\eta^2+\xi^2(\rho\tau-\psi^2))=0 \qquad \rightarrow \qquad S_W:\,\eta^2+\xi^2(\rho\tau-\psi^2)=0 \qquad S_S:\, P=0 \:.
\end{equation}
We see that this is the same result that we obtain by taking \eqref{O7D7Fth} and substituting the equation of the CY threefold ($b_2=\xi^2$) and
\begin{equation}
  \frac{a_3}{2}\equiv \psi \cdot P \qquad\qquad \frac{b_4}{2} \equiv \eta \cdot P \qquad \qquad \frac{b_6}{4} \equiv ( \psi^2-\rho\,\tau)\cdot P^2 \:.
\end{equation} 
If we have zero flux on the branes, the gauge group is $O(1)\times Sp(1)$, where the first factor come from the Whitney brane, and the second one from
the invariant stack of two branes\footnote{For $2n$ branes transverse to the $O7$-plane, the gauge group is $Sp(n)$; in our case we have $Sp(1)\cong SU(2)$.}.  

In the following, we denote the Whitney-type brane by $W$ and the two branes on the $Sp(1)$ stack by $S$ and $S'$.
The charge vectors are:
\begin{eqnarray}
\label{GammaNonsplW} \Gamma_{\rm W} &=& \left(\ch(\L_a)+\ch(\L_b)-\ch(\L_a^{-1})-\ch(\L_b^{-1})\right)\left(1+\tfrac{c_2(X_3)}{2}\right) \nonumber\\
      &=& 2c_1(\L_a)+2c_1(\L_b) + \left[ \tfrac13 (c_1(\L_a)^3+c_1(\L_b)^3)+\tfrac{1}{12}(c_1(\L_a)+c_1(\L_b))c_2(X_3) \right] \nonumber \\ \\
\label{GammaNonspl1} \Gamma_{S} &=& \ch(\L_P^{1/2}\otimes \L_q^{1/2})-\ch(\L_P^{-1/2}\otimes \L_q^{1/2}) \nonumber\\
      &=&  c_1(\L_P) +c_1(\L_P)\tfrac{c_1(\L_q)}{2} + 
	  \left[ \tfrac12 c_1(\L_P) \left( \tfrac{c_1(\L_q)}{2}\right)^2 +\tfrac{1}{24} \left(c_1(\L_P)^3+c_2(X_3)c_1(\L_P)\right)\right] \nonumber\\ \\
\label{GammaNonspl1p} \Gamma_{S}' &=& \ch(\L_P^{1/2}\otimes \L_q^{-1/2})-\ch(\L_P^{-1/2}\otimes \L_q^{-1/2}) \nonumber\\
      &=&  c_1(\L_P) -c_1(\L_P)\tfrac{c_1(\L_q)}{2} + 
	  \left[ \tfrac12 c_1(\L_P) \left( -\tfrac{c_1(\L_q)}{2}\right)^2 +\tfrac{1}{24} \left(c_1(\L_P)^3+c_2(X_3)c_1(\L_P)\right)\right]\:. \nonumber \\
\end{eqnarray}
We see that the flux on the $Sp(1)$ stack is given by $F_S=\frac{c_1(\L_q)}{2}$. 
The cancellation of Freed-Witten anomalies says that we must choose $\L_q$ such that the difference $c_1(\L_P)-c_1(\L_q)$ is
even.

Depending on the choice of $\L_a,\L_b,\L_q$, we can have a non-zero flux on all the branes. 
We will consider the case in which we have a flux on the Sp(1) stack, as this is the only flux that can generate chiral states 
on the intersections of the $D7$-branes. As one can see from the charge vector, the flux is proportional to the 
zero locus of a section of $\L_q$ in this case. Such a flux breaks the gauge group $Sp(1)$ to the $U(1)$ generated by the Cartan element.

From the charge vectors we can read off the $D3$-tadpole of the flux in the same way as in the previous sections. For the Whitney brane, we have 
exactly the same result as before:
\begin{eqnarray}
Q_{D3}^{\rm W} &=& -\tfrac{1}{4} \int_{X_3} \left( c_1(\L_a)+c_1(\L_b) \right) \cdot \left(2\,c_1(\L_a)+K \right) \cdot \left(2\,c_1(\L_b)+K \right)\:,
\end{eqnarray}
where now $c_1(\L_a)+c_1(\L_b)=4\bar{K}-c_1(\L_P)$. As before, we pull the integral down to the quotient $B_3$ and divide the charges we have
computed before orientifolding by $2$. Our result is
\begin{eqnarray} \label{fluxtadpolektheoryWitLn}
Q_{D3}^{\rm W} &=& -  \int_{B_3} \left( \bar{K} -\tfrac{1}{4}[P] \right) \cdot [\rho] \cdot \left(6\bar{K}-[\rho]-2[P] \right)\:.
\end{eqnarray}
which matches with the $D3$-charge of the flux $G_4^{(I_1)}$ obtained from F-theory \eqref{TadWSpG41bis} .

Both branes of the Sp(1) stack contribute the same $D3$-tadpole. After orientifolding it is given by
\begin{equation} \label{fluxtadpolektheorysp1W}
Q_{D3} \,=\, Q_{D3}^{S} + Q_{D3}^{S'} \,=\, \tfrac12\int_{X_3}  c_1(\L_P)\cdot  \left( \frac{c_1(\L_q)}{2}\right)^2 
      \,=\, \int_{B_3}  [P]\cdot  \left( \frac{[q]}{2}\right)^2  \:.
\end{equation}
This matches with the charge of $G_4^{(\rm Sp)}$ in \eqref{QG42F}.

\subsubsection*{Chiral states}

After switching on the flux on the $Sp(1)$ stack, the unbroken gauge group is $U(1)$ \footnote{We do not write the irrelevant $O(1)$ factor.}. 
The flux along this $U(1)$ allows for a net number of chiral states living at the intersection with the Whitney brane.

The number of chiral states stretching between one brane on $P=0$ with flux $F_S=\frac{c_1(\L_q)}{2}$ and the Whitney brane is given by
\begin{eqnarray}\label{aWitChir}
 \langle \Gamma_S,\Gamma_{\rm W}\rangle &=&  \int_{X_3} \Gamma_S\Gamma_{\rm W}^* = -\int_{[{\rm W}]\cap [P]}F_S \nonumber  \\
	  &=& -\int_{X_3} \left( 2c_1(\L_a)+2c_1(\L_b)\right)\cdot c_1(\L_P)\cdot \frac{c_1(\L_q)}{2} \:.
\end{eqnarray}
The intersection of the second brane in the stack with the Whitney brane is the orientifold image of the one considered.
As the Whitney brane is an invariant brane and the orientifold transformation changes the orientation of the strings, the 
strings $S\rightarrow W$ are mapped to the strings $S'\leftarrow W$. In fact, as expected, the number of chiral states 
is the same in the two intersections: $\langle \Gamma_W,\Gamma_{S}'\rangle = \langle \Gamma_S,\Gamma_{\rm W}\rangle$.

Using the fact that $c_1(\L_a)+c_1(\L_b)=4\bar{K}-[P]$ and that $\bar{K}=c_1(B_3)$, we obtain
\begin{eqnarray}
\langle \Gamma_S,\Gamma_{\rm W}\rangle &=& -  \int_{B_3} (8c_1(B_3)-2[P])\cdot [P]\cdot [q] \:. 
\end{eqnarray}
This is the same chiral index we have computed in F-theory by integrating the flux $G_4^{(\rm Sp)}$ along the curve $\hat{C}$,
\eqref{chirfsp(1)}.

\

Other chiral states come from the bulk of the $Sp(1)$ stack. In general, the number of such states is given by \cite{Collinucci:2008pf} 
\begin{eqnarray}\label{ChirD7D7}
 \tfrac12\left(\langle \Gamma_{D7}',\Gamma_{D7}\rangle +\tfrac14 \langle \Gamma_{\rm O7},\Gamma_{D7}\rangle\right)
	      &=& - \int_{X_3} [D7]\cdot F\cdot 
			\left(c_1([D7])-K\right) \nonumber   \\ 
	      &=& - \int_{[D7]} F\cdot 
			\left(c_1([D7])-K\right)\:, \nonumber
\end{eqnarray}
where $\Gamma_{\rm O7}=8K+\tfrac{\chi([O7])}{6}\omega$.

In our case we obtain
\begin{eqnarray}\label{ChirSpSp}
 \tfrac12\left(\langle \Gamma_S',\Gamma_S\rangle +\tfrac14 \langle \Gamma_{\rm O7},\Gamma_S\rangle\right)
	      &=&\int_{X_3}c_1(\L_P)\cdot\frac{c_1(\L_q)}{2}\cdot 
			\left(c_1(\L_P)+K\right) \nonumber   \\ 
	      &=& \int_{B_3}[P]\cdot [q]\cdot 
			\left([P]-\bar{K}\right) \:. \nonumber
\end{eqnarray}
i.e. the result we obtained in F-theory, using the conjectured formula \eqref{bulkchi}.

\subsection{$Sp(1)$ singularity plus brane/imagebrane}\label{sectsp1splitiib}

In this section we split the Whitney brane of the last section into separate brane and imagebrane while keeping the $Sp(1)$ stack.
We can obtain this configuration by considering a tachyon matrix \eqref{TachSpW} with $\tau\equiv 0 \equiv \rho$. 
In this case, the determinant of $T$ factorizes into three branches:
\begin{equation}
\det T = P^2(\eta^2-\xi^2\psi^2)=0 \quad \rightarrow \quad S_{D}:\,\eta=\xi\psi \qquad S_{D'}:\,\eta=-\xi\psi  \qquad S_S:\, P=0 \:.
\end{equation}

Again, we denote the two branes on the $Sp(1)$ stack by $S$ and $S'$. The two $D7$-branes coming from the Whitney one are now called $D$ and $D'$.
The charge vectors of the $D7$-brane and its image are
\begin{eqnarray}\label{ChVect22pSp}
 \Gamma_{D} &=& \ch(\L_a)-\ch(\L_b^{-1}) \nonumber\\
      &=& \left(c_1(\L_a)+c_1(\L_b)\right) + \left(c_1(\L_a)+c_1(\L_b)\right)\frac{c_1(\L_a)-c_1(\L_b)}{2} + 
	   Q_{D3}^{(D)} \nonumber \\ \\
 \Gamma_{D}' &=& \ch(\L_b)-\ch(\L_a^{-1}) \nonumber\\
      &=& \left(c_1(\L_a)+c_1(\L_b)\right) + \left(c_1(\L_a)+c_1(\L_b)\right)\frac{c_1(\L_b)-c_1(\L_a)}{2} + 
	   Q_{D3}^{(D')} \:.\nonumber
\end{eqnarray}
$\Gamma_S$ and $\Gamma_S'$ remain the same as before (see eq. \eqref{GammaNonspl1} and \eqref{GammaNonspl1p}).

The $D3$-charge of the flux on the $Sp(1)$ stack is the same as computed in \eqref{fluxtadpolektheorysp1W}. The $D3$-tadpole of the remaining branes
can be read from the charge vectors \eqref{ChVect22pSp}. It is equal to \eqref{QD3SplW}, where now $c_1(\L_a)+c_1(\L_b)=4\bar{K}-c_1(\L_P)$.
After orientifolding we find 
\begin{eqnarray}\label{QD3SplWSp}
 Q_{D3}^{D}& =& \tfrac12 \int_{X_3} \left(4\bar{K}-c_1(\L_P)\right)\left(\frac{c_1(\L_a)-c_1(\L_b)}{2}\right)^2  \nonumber\\
	  &=&\int_{B_3} \left(4c_1(B_3)-[P]\right)\left(\frac{[\alpha]}{2}\right)^2  \:.
\end{eqnarray}
This matches with the corresponding F-theory result, \eqref{G1tadpolesp1split}.

\subsubsection*{Chiral states - Flux preserving the non-abelian gauge group}

First, we want to consider the case of a flux which allows for chiral states, but does \emph{not} break the non-abelian part of the 
gauge group. Hence we should only switch on a flux on the $D7$ and its image, but not on the $Sp(1)$ stack. 
In the case when there is zero flux on the $Sp(1)$ stack, the number of chiral states coming from strings $D\rightarrow S$ is:
\begin{eqnarray}\label{chirSp1U1}
 \int_{X_3} \Gamma_S \Gamma_D^* &=&\int_{X_3} \left(4\bar{K}-[P]\right)\cdot [P] \cdot \frac{c_1(\L_a)-c_1(\L_b)}{2} \:.
\end{eqnarray}
The same result is obtained for $\int_{X_3} \Gamma_{S}' \Gamma_D^*$, which counts the chirality of strings stretching $D\rightarrow S'$. 
In fact, the chiral states coming from the two different intersections 
form a ${\bf 2}$ of $Sp(1)$. The other intersections ($\Gamma_S\Gamma_{D}^{'*}$ and $\Gamma_{S}'\Gamma_{D}^{'*}$) 
are the images of the studied ones under the orientifold involution.

Taking into account that $[\alpha]=c_1(\L_a)-c_1(\L_b)$ and that $\bar{K}=c_1(B_3)$, the expression \eqref{chirSp1U1} is exactly the same result obtained in 
F-theory, \eqref{intG41C2+}.

\subsubsection*{Chiral states - Flux along a Cartan element of the non-abelian group}

We now allow for a non-zero flux $F_S=\frac{c_1(\L_q)}{2}$ on the $Sp(1)$ stack. The overall gauge group is then broken to $U(1)\times U(1)$. 
We have a different number of chiral states on the two different intersections $\Gamma_S\Gamma_D^*$ 
(strings $D\rightarrow S$) and $\Gamma_D\Gamma_{S'}^*$ (strings $S'\rightarrow D$):
\begin{eqnarray}\label{chirSp1U12fl}
 \int_{X_3} \Gamma_S \Gamma_D^* &=&\int_{X_3} \left(4c_1(B_3)-[P] \right)\cdot [P] \cdot \left( \frac{[\alpha]}{2} - \frac{[q]}{2}\right)\:,\nn \\
 \label{abpInters}\int_{X_3} \Gamma_D \Gamma_{S'}^* &=&- \int_{X_3}  \left(4c_1(B_3)-[P] \right)\cdot [P] \cdot \left( \frac{[\alpha]}{2} + \frac{[q]}{2}\right) \:.
\end{eqnarray}
We get the same results for the orientifold images of the chiral intersections, i.e. for the strings $S'\rightarrow D'$ and $D'\rightarrow S$.
It is again easy to see that the expressions \eqref{abpInters} match with the F-theory ones, \eqref{sp1splitg1g2a}.

\

The flux $F_S$ also induces chiral bulk states on the $Sp(1)$ stack, as in the previous case. The number of chiral states in the symmetric 
representation of $U(1)$ is still be given by \eqref{ChirSpSp}.

\subsubsection*{Chiral states from the intersections of brane and imagebrane}

Finally, we compute the chirality induced on the intersection between the $D7$-brane and its image. 
Using $\Gamma_{\rm O7}=8 K+\tfrac{\chi([O7])}{6}\omega$ we compute
\begin{eqnarray}\label{chirBrImbr}
 \tfrac12\left(\langle \Gamma_{D}',\Gamma_D\rangle +\tfrac14 \langle \Gamma_{\rm O7},\Gamma_D\rangle \right)
	      &=&\int_{X_3}\left(c_1(\L_a)+c_1(\L_b)\right)\cdot\frac{c_1(\L_a)-c_1(\L_b)}{2}\cdot 
			\left(c_1(\L_a)+c_1(\L_b)-\bar{K}\right) \nonumber\\   \\ 
	      &=& \int_{B_3}\left(c_1(\L_a)+c_1(\L_b)\right)\cdot\left(c_1(\L_a)-c_1(\L_b)\right)\cdot 
			\left(c_1(\L_a)+c_1(\L_b)-\bar{K}\right) \nonumber \\
	      &=& \int_{B_3}\left(4\bar{K}-[P]\right)[\alpha]\left(3\bar{K}-[P]\right) \:.
\end{eqnarray}
This index is reproduced on the F-theory side by integrating the flux $G_4^{(I_1)}$, \eqref{g41split}, over the four-cycle $\hat{C}_1$, \eqref{SingletsSp1}.

\section{The Weierstrass equation as a Pfaffian}\label{sectpfaff}
\subsection{$G_4$-flux via vector bundles}

The results of this paper rely on the existence of additional holomorphic four-cycles in the CY fourfold that cannot be written as complete intersections of two divisors with the Weierstrass equation, but as three equations in an ambient fivefold that imply the Weierstrass equation. This is only possible if the complex structure moduli of the fourfold are appropriately tuned. We repeat the main structure here for convenience. 

If we impose that the Weierstrass model have a factorizable $a_6 = \rho\,\tau$, then we can write it as:
\be \label{weierstrassasquadric}
W \equiv Y_-\,Y_+-z^6\,\rho\,\tau-X\,Q = 0\,,
\ee 
where
\be
Y_\pm = y \pm \tfrac{1}{2}\,a_3\,z^3\,,
\ee
and the other variables are polynomials defined in appendix \ref{appweier}. Then, the crucial new four-cycles are:
\begin{eqnarray} \label{surfacesrhotau}
\sigma_\rho: \quad \{Y_- &=& 0\} \quad \cap \quad \{X=0\} \quad \cap \quad \{\rho=0\}\,,\\
\sigma_\tau: \quad \{Y_- &=& 0\} \quad \cap \quad \{X=0\} \quad \cap \quad \{\tau=0\}\,. \nonumber
\end{eqnarray}
As we have seen in section \ref{sectIIBwhitney}, this factorizability of $a_6$ also has physical significance when we take Sen's weak coupling limit. It creates new holomorphic two-cycles on the divisor where the $D7$ is wrapped. In terms of the `Tate model' variables, the divisor is now:
\be \label{divisorequation}
b_4^2+\xi^2\,(4\,\rho\,\tau -a_3^2) = 0\,,
\ee
and the new two-cycles are
\begin{eqnarray} \label{curvesrhotau}
C_\rho: \quad \{b_4 &=& \pm \xi\,a_3 \}\quad  \cap \quad \{\rho=0\}\\
C_\tau: \quad \{b_4 &=& \pm \xi\,a_3 \} \quad   \cap \quad \{\tau=0\}\,.
\end{eqnarray}
These two-cycles imply the existence of a flux that restricts the divisor moduli via the superpotential in \cite{Martucci:2006ij, Martucci:2005ht}.

In the IIB setting, we can understand this phenomenon of the enhancement of the Picard lattice of the $D7$ divisor in a more systematic way: The divisor equation \eqref{divisorequation} is the determinant of a two by two matrix that we physically interpret as encoding the tachyon modes between $D9$/anti-$D9$ stacks:
\begin{eqnarray} \label{tatetachyonmatrix}
T&=&\tfrac12 \left(\begin{array}{cc}2\,\xi\,\rho&b_4+\xi\,a_3\\-b_4+\xi\,a_3&2\,\xi\,\tau \end{array}\right)\,.
\end{eqnarray}
This matrix can be understood as a map between two rank-two holomorphic vector bundles $T: E_2 \rightarrow F_2 $, such that the following short exact sequence of sheaves:
\be \label{shortexactiib}
0 \rightarrow E_2 \stackrel{T}\longrightarrow F_2 \rightarrow \cale_1 \rightarrow 0
\ee
defines a sheaf $\cale_1$ that corresponds to a line bundle $L_1$ with support on the $D7$ divisor given by:
\be
{\rm det}(T) = b_4^2+\xi^2\,(4\,\rho\,\tau -a_3^2) = 0\,,
\ee
The holomorphic curves \eqref{curvesrhotau} are nothing but the vanishing loci of some typical sections of this line bundle. As explained in \cite{Collinucci:2008pf} and section \ref{sectIIBwhitney}, a section $s$ of the sheaf $\cale$ can be written as a section $s_E$ of $F_2$ modulo the image $T(E_2)$, hence the vanishing locus of $s$ is simply the locus where 
\be
s_E  = T(s_F)
\ee
for some $s_F \in \Gamma(F_2)$. Searching for such loci corresponds to solving the equation:
\begin{eqnarray}
\tilde T\cdot \left(\begin{array}{c} \lambda_a\\ \lambda_b \end{array}\right) &=&\left(\begin{array}{cc}2\,\xi\,\tau&-b_4-\xi\,a_3\\b_4-\xi\,a_3&2\,\xi\,\rho \end{array}\right) \cdot \left(\begin{array}{c} \lambda_a\\ \lambda_b \end{array}\right) = 0\,.
\end{eqnarray}
in the CY threefold. The solutions will be holomorphic curves in the threefold that are automatically contained in the divisor.  The curves in \eqref{curvesrhotau} are a special case of this, and can be obtained roughly by choosing one of the $\lambda$'s to be a constant. 

This $D7$ divisor is what is known in mathematics as a \emph{determinantal variety}. Such varieties have been studied for a long time (see \cite{Harris, Beauville}). The point is that whenever a divisor can be written as the determinant of a map, then the divisor will admit a holomorphic line bundle on it defined via the exact sequence \eqref{shortexactiib}. 

The fundamental reason for the presence of such curves is the following: The locus of the $D7$ is by definition located wherever the determinant of the two by two matrix is less than two. However, unlike a polynomial, a two by two matrix of polynomials can have loci where it has rank that is not maximal, but also not zero. This is where the special curves are located.

\vskip 3mm
Given the similarity of the equations defining the curves on the $D7$-brane in \eqref{curvesrhotau} and the four-cycles \eqref{surfacesrhotau} in our restricted Weierstrass equation, one cannot help but wonder whether the Weierstrass equation might not also admit some structure like that of a determinantal variety. This turns out \emph{not} to be the case.

However, this is good news. If the fourfold \emph{were} determinantal, then it would admit a new class of line bundles, and their corresponding six-cycles. However, we are interested in new holomorphic \emph{four-cycles}. More precisely, we would like the fourfold to admit a holomorphic \emph{rank-two} vector bundle $V_2$, such that its second Chern class is related to the flux
\be
G_4 = c_2(V_2) - \delta  \,,
\ee
where $\delta$ is some subtraction term that ensures the Poincar\'e invariance of $G_4$. It will be of the form $\delta  = \omega_1 \wedge \omega_2$, where the $\omega$ are two-forms in $X_4$.
Since the second Chern class must be a quantized $(2,2)$-form, a rank two holomorphic vector bundle would have the right properties to describe a $G_4$-flux, modulo the issue of semi-integral quantization.

It turns out that there is a suitable structure available. If the Weierstrass equation can be written as the \emph{Pfaffian of an anti-symmetric matrix} $M$, then, in the ambient fivefold $X_5$, we can write an exact sequence:
\be \label{ftheoryshortexactseq}
0 \rightarrow E_4 \stackrel{M}\longrightarrow F_4 \rightarrow \cale_2 \rightarrow 0\,,
\ee
where $E_4, F_4$ are holomorphic vector bundles of rank four on $X_5$, and $\cale_2$ is a coherent sheaf corresponding to a rank two holomorphic vector bundle $V_2$ with support over the CY fourfold. Indeed, our Weierstrass equation can be written as the Pfaffian of the following matrix:
\begin{equation} \label{weierstrasstachyon}
 M = \left( 
\begin{array}{cccc}
 0 & X & \rho\,z^3 & Y_+ \\ -X & 0 & -Y_- & \tau\,z^3 \\ -\rho\,z^3 & Y_- &  0 & Q \\ -Y_+ & -\tau\,z^3 & -Q & 0
\end{array}
 \right)\,.
\end{equation}
Notice the striking similarities with the IIB picture. If one makes the substitution $y \mapsto \tfrac{1}{2}\,b_4$, then one can recognize the tachyon matrix \eqref{tatetachyonmatrix} as being imbedded into $M$ in the off-diagonal blocks.

The fact that the CY fourfold is a \emph{Pfaffian} variety gives similar properties to those of determinantal varieties. However, a four by four anti-symmetric matrix has two pairs of equal and opposite eigenvalues. Therefore, depending on the locus, it can only have ranks four, two and zero, but not three or one. The CY fourfold is located wherever the rank is less than four. However, special four-cycles will appear wherever its rank is two. They can be found (analogously to the case of the IIB curves) by solving equations of the form:

\begin{equation}
 \Theta: \,\, \widetilde{M}\cdot \left(\begin{array}{c} v_1 \\ v_2 \\ v_3 \\ v_4 \end{array}
 \right)=0  \:,
\qquad \mbox{where} \qquad \vec{v} \equiv \left(\begin{array}{c} v_1 \\ v_2 \\ v_3 \\ v_4 \end{array}
 \right) \in \Gamma(F_4)
\end{equation}
and

\begin{equation} \label{tildeweierstrasspfaffian}
 \widetilde{M} \equiv {\rm det}(M) \cdot M^{-1}= \left( 
\begin{array}{cccc}
 0 & -\tau\,z^3 & Q & -Y_- \\ \tau\,z^3 & 0 & -Y_+ & X \\ -Q & Y_+ &  0 & -\rho\,z^3 \\ Y_- & -X &\rho\,z^3 & 0\\
\end{array}
 \right)\:.
\end{equation}

The special four-cycles \eqref{surfacesrhotau} can now be rediscovered through an appropriate choice of $F_4$.

\subsection{D3-tadpole computation}

To show the power of this re-formulation of the Weierstrass model that incorporates a rank two vector bundle, let us compute the $D3$-charge induced by the flux and compare it to our calculations in \eqref{fluxtadpolektheory} and \eqref{g4tadpole}.

In the short exact sequence \eqref{ftheoryshortexactseq}, let us choose the following rank four vector bundles:
\begin{eqnarray} \label{rankfourbundles}
E_4 &=& \left(\call_a^{-1} \otimes K^{-2} \otimes F^{-4}\right) \oplus \left(\call_a \otimes K^{2} \otimes F^{-4}\right) \oplus \left(\call_a^{-1} \otimes K^{-2}\otimes F^{-5}\right) \oplus \left(\call_a \otimes K^{2} \otimes F^{-5}\right) \nonumber \\
F_4 &=& \left(\call_a \otimes K^{2} \otimes F^{-2}\right) \oplus \left(\call_a^{-1} \otimes K^{-2} \otimes F^{-2} \right) \oplus \left(\call_a \otimes K^2 \otimes F^{-1}\right) \oplus \left(\call_a^{-1} \otimes K^{-2} \otimes F^{-1} \right)\:. \nonumber \\
\end{eqnarray}
One can easily check that this choice is consistent with the entries of the matrix $M$ in \eqref{weierstrasstachyon}. 

The short exact sequence of these bundles defines a rank two vector bundle $V_2$ localized on the CY$_4$ hypersurface $X_4$ defined by Pfaff$(M)=0$, such that $ G_4 = c_2(V_2) -\delta$. Hence, the flux-induced $D3$-tadpole should be equal to
\be
Q_{D3}^f = -\tfrac{1}{2}\,\int_{X_4} \left( c_2(V_2)-\delta \right)^2\,.
\ee
In order to compute this, we will make use of the \emph{Grothendieck-Riemann-Roch} applied to the push-forward of the inclusion map $\imath$ of the fourfold into the ambient fivefold $\imath: X_4 \hookrightarrow X_5$. The theorem relates the Chern character of $V_2$ to the Chern character of its push-forward $\imath_*(V_2) = \cale_2$
\be
\imath_*\big({\rm ch}(V_2) \cdot {\rm Td}(X_4) \big) = {\rm ch}(\cale_2) \cdot {\rm Td}(X_5)\,,
\ee
where the push-forward on the lhs acts on forms by Poincar\'e dualizing them to cycles, then pushing them forward, and then re-dualizing into forms. The Chern character of the sheaf $\cale_2$ is given by
\be
{\rm ch}(\cale_2) = {\rm ch}(F_4)-{\rm ch}(E_4)\:.
\ee
Since the Chern classes of $X_4$ are made of forms pulled back from $X_5$, we can rewrite this more conveniently as follows:
\be
\imath_*\big({\rm ch}(V_2)\big) = {\rm ch}(\cale_2) \cdot {\rm Td}(F^6)\,,
\ee
whereby the Weierstrass equation is a section of the line-bundle $F^6$.

Substituting our choices in \eqref{rankfourbundles}, we find the following for the push-forward of the Chern character of the bundle $V_2$
\be
\imath_*\big({\rm ch}(V_2) \big)  = 6\,F \cdot \mathcal{I} \:,
\ee
where 
\begin{eqnarray}
\mathcal{I} &=& 2+ \big(-2\,F^2+(c_1(\call_a)-2\,\bar K)^2 \big)\\&&+\big(\tfrac{11}{6}\,F^4+(c_1(\call_a)-2\,\bar K)^2 \cdot \left(\tfrac{1}{12}\,(c_1(\call_a)-2\,\bar K)^2-F^2 \right) \big)\,, \nonumber
\end{eqnarray}
and $6\,F$ is Poincar\'e dual to the $X_4$ hypersurface in $X_5$.
Since the one-form component of $\mathcal{I}$ is zero, we deduce that the first Chern character of its preimage is zero, hence
\begin{eqnarray}
\imath_*\big({\rm ch}(V_2) \big)  &=& 2+ \imath_*({\rm ch}_2(V_2))+ \imath_*({\rm ch}_4(V_2)) \nonumber\\
&=&  2 - \imath_*(c_2(V_2))+ \tfrac{1}{12}\,\imath_*(c_2(V_2)^2)\,.
\end{eqnarray}
Let us decompose $c_2(V_2)$ into a part that lies in the kernel of the pushforward map, and a part that is orthogonal to it:
\be
c_2(V_2) = \left( c_2(V_2)-\delta \right) + \delta
\ee
where $\delta  = \imath^* \left(\tilde \delta \right)$, is the pullback of some four-form in $X_5$, such that
\be\label{c2orthogr}
\imath_*\left( c_2(V_2)-\delta \right)=0 \quad \text{and} \quad \left( c_2(V_2)-\delta \right) \cdot \delta=0\,.
\ee
Note that with $\delta$ defined in this way, the flux $G_4$ is Poincar\'e invariant.

Using relations \eqref{c2orthogr}, we deduce the following:
\be
\imath_* \big(c_2(V_2)^2 \big) = \imath_* \big( \left( c_2(V_2)-\delta \right)^2-\delta^2 \big) = \imath_* \big( \left( c_2(V_2)-\delta \right)^2\big)-\imath_*\left(\delta^2 \right)\,.
\ee
Therefore, we can compute the flux-induced $D3$-tadpole as follows:
\be
Q_{D3}^f = -\tfrac{1}{2}\,\int_{X_4} G_4^2 = -\tfrac{1}{2}\,\int_{X_5} \imath_* \big( \left( c_2(V_2)-\delta \right)^2\big)\,.
\ee
In terms of the $i$-form components $\mathcal{I}_i$ of $\mathcal{I}$, this is given by

\begin{eqnarray}
Q_{D3}^f  &=& -\tfrac{1}{2}\,\int_{X_5} 6\,F \cdot \left(12\,\mathcal{I}_4 -\mathcal{I}_2^2 \right) \nonumber \\
&=& 6\,\int_{X_4} F^2 \cdot \bar K \cdot \left(2\,c_1(\call_a)-\bar K \right) \cdot \left(2\,c_1(\call_a)-7\,\bar K \right) \\ \nonumber
&=& \int_{B_3} c_1(B_3) \cdot \left(2\,c_1(\call_a)-c_1(B_3) \right) \cdot \left(2\,c_1(\call_a)-7\,c_1(B_3) \right)\,. \nonumber
\end{eqnarray}
as predicted in\footnote{Remember that $\rho$ is a section of $\L_a^2\otimes K$ and $\tau$ is a section of $\L_b^2\otimes K$. Moreover we have the relation $\L_a\otimes\L_b=K^{-4}$.} \eqref{g4tadpole}.

Hence, as we see, this description of our $G_4$-flux in terms of coherent sheaves gives a prescription for computing the induced tadpole that is much simpler than a direct calculation of the self-intersection of the four-form.

\subsection{Matrix Factorizations}

In this section, we point out some interesting links between our construction and those that appear in the context of matrix factorization both in algebraic geometry and string theory.

The phenomenon we have observed in this section is known in mathematics. Whenever a hypersurface can be written as the determinant or the Pfaffian of a matrix, then new sheaves appear on it. More generally, given a hypersurface equation $W=0$, if one can find two $n \times n$ matrices $M_1$, and $M_2$ such that
\be
M_1 \cdot M_2 = W \cdot {\rm id}_n\,,
\ee
then there exist special sheaves $S^{(1)}$ and $S^{(2)}$ with support on the hypersurface, defined by the sequences on the ambient space:
\begin{eqnarray}
&0& \rightarrow E_n^{(1)} \stackrel{M_1}{\longrightarrow} F_n^{(1)} \rightarrow S^{(1)} \rightarrow 0 \\
&0& \rightarrow E_n^{(2)} \stackrel{M_2}{\longrightarrow} F_n^{(2)} \rightarrow S^{(2)} \rightarrow 0
\end{eqnarray}
whereby $E_{1,2}$ and $F_{1,2}$ are vector bundles of rank $n$. In our case, $M_1$ is the matrix defined in \eqref{weierstrasstachyon}, and $M_2$ is the matrix $\tilde M$ defined in \eqref{tildeweierstrasspfaffian}.

In \cite{Achinger} and \cite{Addington}, matrix factorizations are exploited in detail for the class of hypersurfaces given by quadrics in $\mathbb{P}^N$. Our CY hypersurface is not in general a quadric per se, however the form \eqref{weierstrassasquadric} is a quadric with respect to the polynomials $Y_\pm, \rho, \tau, X$ and $Q$, and hence shares similar structures to those described in these articles.

The technology of matrix factorizations has found its incarnation in string theory, in the context of D-branes, through the works \cite{Kapustin:2002bi} and \cite{Brunner:2003dc}. More specific applications were found, where codimension two objects in CY threefolds, i.e. D2-branes on holomorphic curves, could be described via matrix factorizations, in \cite{Baumgartl:2007an, Herbst:2008jq}. Those situations are analogous to ours, since they treat the construction of holomorphic codimension two objects that cannot be written as the intersection of two divisors with the CY hypersurface, but must be written as the intersection of three specially chosen divisors in the ambient space.

Finally, in \cite{Brunner:2006tc}, the matrix factorization techniques are applied to study holomorphic curves in K3 surfaces. In particular, ways for enhancing the Picard group are investigated. In our case, we are interested in enhancing the analogous group $H^{2,2}_H(X_4) \cap H^4(X_4, \mathbb{Z})$.

It is curious and suggestive that such techniques, which have proven useful in perturbative string theory, should make an appearance in our F-theory setups. One wonders, whether the K-theoretic treatment of M-theory of \cite{Diaconescu:2000wy} is related to our construction, despite the fact that ours lives in a thirteen-dimensional spacetime ambient to the M-theory spacetime, as opposed to their twelve-dimensional one. It would be interesting to pursue this connection further.

\section{Conclusions and Outlook}

In this paper, we introduced a new way of looking at the problem of $G_4$ fluxes in F-theory. We provide a construction technique that is very direct, 
as it describes them in terms of Poincar\'e duals of holomorphic four-cycles. The resulting fluxes are automatically quantized, of $(2,2)$-type, and 
have one leg on the fiber, as required for 4D Poincar\'e invariance.

Because of their algebraic description, we can directly compute the induced D3-tadpole as the self-intersection of a four-cycle. In simple setups with 
intersecting branes, we computed chirality indices by integrating our fluxes on four-cycles obtained from resolutions of the singularities along matter curves.

Whenever a weak coupling limit was available, we were able to map our fluxes to worldvolume fluxes of $D7$-branes, and succesfully match all of our results 
on $D3$-tadpoles and chirality.

Our construction also shows how our fluxes lift some of the complex structure moduli of the fourfold, much like worldvolume fluxes lift D7 divisor moduli.
Recently, there has been a lot of progress in computing flux-induced potentials for compactifications of F-theory on Calabi-Yau fourfolds 
\cite{Alim:2009rf, Alim:2009bx, Grimm:2009ef, Aganagic:2009jq, Grimm:2009sy, Jockers:2009ti, Alim:2010za}. 
It would be very interesting to use these methods to explicitly see how our fluxes fix some of the complex structure moduli such that the section $a_6$ of 
the Weierstrass model factorizes.

Finally, we found a curious incarnation of the treatment of $D7$-branes in terms of coherent sheaves (i.e. tachyon condensation of $D9$ and anti-$D9$-branes), 
in our treatment of CY fourfolds with their fluxes. In this picture, our $G_4$ fluxes appear as rank two vector bundles on the fourfolds.

Our results are succinctly summarized in section \ref{sec:mainideaandsummary}.

We believe that the constructions presented in this paper will open up new and interesting ways of studying F-theory models. They will also provide a check 
for phenomenologically oriented models that rely on the spectral cover construction.

\subsection*{Acknowledgements}
We have benefited from discussions with Ilka Brunner, Michael Kay, Luca Martucci, Daniel Plencner, Raffaele Savelli and Nils-Ole Walliser.

The work of A. P. Braun was supported by the FWF under grant I192. 
The work of R. Valandro was supported by the German Science Foundation (DFG) under the Collaborative Research Center (SFB) 676. 
The work of A. Collinucci is supported in part by the Cluster of Excellence Origin and Structure of the Universe in Munich, Germany, and by a EURYI award of the European Science Foundation.

\appendix

\section{Rewriting the Weierstrass model}\label{appweier}

It is paramount to our construction to cast the familiar Weierstrass model in 
a specific form. The ordinary Weierstrass form,
\begin{equation}\label{weierapp}
 y^2=x^3+xz^4 f+z^6 g \, ,
\end{equation}
is related to the Tate form
\begin{equation}\label{weiertateapp}
Y^2+a_1XYZ+a_3YZ^3=X^3+a_2X^2Z^2+a_4XZ^4+a_6Z^6 \, 
\end{equation}
by shifting the coordinates $x,y,z$. The $a_n$ are sections of the $n$th power
of the anticanonical bundle, so their divisor class is $n\bar{K}$.

Introducing the quantities
\begin{align}\label{tate1app}
b_2&=a_1^2+4a_2  \nn\\
b_4&=a_1a_3+2a_4 \nn\\
b_6&=a_3^2+4a_6 
\end{align}
the relations between the sections $f_4$, $g_6$ and the discriminant $\Delta$ can be written as
\begin{align}\label{fgdeltaapp}
f&=-\frac{1}{48}\left(b_2^2-24b_4\right) \nn\\
g&=\frac{1}{864}\left(b_2^3-36b_2b_4+216b_6\right) \nn\\
\Delta&=4f^3+27g^2=\frac{1}{64}b_2^2\left(b_2 b_6-b_4^2\right)-\frac{9}{16}b_2b_4b_6+\frac{1}{2}b_4^3+\frac{27}{16}b_6^2 \, .
\end{align}
If we use the parametrization \eqref{tate1app} in the standard Weierstrass model \eqref{weierapp}, we can rewrite it in the
following way:
\begin{align}
 y^2=&x^3-z^4x\frac{1}{48}(b_2^2-24b_4)+z^6\frac{1}{864}\left(b_2^3-36b_4 b_2+216b_6\right)  \nn \\
=& (x-\frac{1}{12}z^2b_2)\left((x-\frac{1}{12}z^2b_2)(x+\frac{1}{6}z^2b_2)+\frac{1}{2}z^4b_4\right) + \frac{1}{4}z^6b_6 \, .
\end{align}
If we also use that $b_6=a_3^2+4a_6$ we obtain 
\begin{align}
(y-\frac{1}{2}z^3a_3)(y+\frac{1}{2}z^3a_3)-z^6a_6=(x-\frac{1}{12}z^2b_2)\left((x-\frac{1}{12}z^2b_2)(x+\frac{1}{6}z^2b_2)+\frac{1}{2}z^4b_4\right) \, . 
\end{align}
To flesh out the structure of this expression, we introduce the quantities 
\begin{align}\label{defYXQapp}
Y_\pm&=y\pm\frac{1}{2}z^3a_3 \\
X&=x-\frac{1}{12}z^2b_2 \\
Q&=(x-\frac{1}{12}z^2b_2)(x+\frac{1}{6}z^2b_2)+ \frac{1}{2}z^4b_4 \nn\\
&=X(X+\frac{1}{4}z^2b_2)+\frac{1}{2}z^4b_4 \, ,
\end{align}
so that we can write the Weierstrass model in the simple form
\begin{equation}\label{weierformapp}
Y_-Y_+-z^6a_6= X Q \, .
\end{equation}
This form of the Weierstrass equation is the starting point of our investigations of $G_4$ fluxes and algebraic cycles.

\section{The intersection form of an embedded submanifold}\label{appendixB}

Given a smooth map $f$ between two spaces, $f:X\mapsto Y$, there are induced maps on cohomology and homology:
\begin{equation}
f^*:H^\bullet(Y)\mapsto H^\bullet(X) \, ,\qquad f_*:H_\bullet(X)\mapsto H_\bullet(Y) \, .
\end{equation}

In the present case, we have the embedding $i:X_4 \hookrightarrow X_A$ of our elliptic Calabi-Yau fourfold $X_4$
into some ambient space $X_A$. We would like to show that the intersection product between two-forms on $X_4$
vanishes if the dual cycle of one is in the kernel of the pushforward $i_*\gamma=0$ (`trivial in the ambient space') while 
the other is in the image of the pullback $\alpha=i^*\beta$ (`descends from the ambient space').

We compute
\begin{eqnarray}
 \int_{X_4}\gamma\wedge\alpha&=&  \int_{\gamma}\alpha
= \int_{\gamma}i^* \beta 
= \int_{i_*\gamma}\beta  = 0 \, ,
\end{eqnarray}
where we have used naturalness of the cap product \cite{DoPa1997}.
As $i_*\gamma=0$ holds by assumption, the integral vanishes.

\section{The exceptional curves of an $Sp(1)$ singularity}\label{appendixC}

We are interested in the exceptional curve that is present when resolve a fourfold that describes an $Sp(1)$
brane situated at $P=0$, as well as the remaining 7-brane. The resolution of the singularity over the $Sp(1)$
brane is described by
\begin{align}\label{sp1protrans}
 (y+ \tfrac 12 s a_{3,1}z^3)(y-\tfrac12 s a_{3,1}z^3)-s^2a_{6,2}z^6-X\left(X(Xv+\tfrac14 b_2z^2)+\tfrac12 sb_{4,1}z^4\right)&=0 \nn\\
vs&=P \, .
\end{align}
The weights of the coordinates are \\
\begin{center}
 \begin{tabular}[c]{ccccc}
 $X$ & $y$ & $z$ & $v$ & $s$ \\
 0 & 0 & $2K$ & 0 & $[P]$ \\
2 & 3 & 1 & 0 & 0 \\
1 & 1 & 0 & -1 & 1
\end{tabular}
\end{center}
To make the following expressions simple, we omit the coordinate $z$ in the following. It can be reinstated
at any point. 

To find the exceptional curve after the resolution, we consider the proper transform
over the locus $P=0$. This means we have two branches, $v=0$ and $s=0$. The $s=0$
branch is given by
\begin{equation}
 y^2-X^2(Xv+\tfrac14 b_2)=0 \, .
\end{equation}
As explained in \cite{Collinucci:2010gz}, this is a $\P^1$ for any point on $P=0$, which
correspond to the exceptional root of the Dynkin diagram of $Sp(1)$.
\par
The $v=0$ branch is more interesting, \eqref{sp1protrans} becomes
\begin{equation}\label{v0branch}
 y^2-(\tfrac14 a_{3,1}^2 +a_{6,2})s^2 -\tfrac14 b_2 X^2-X\tfrac12 sb_{4,1}=0 \, .
\end{equation}
For any point in the base, this is a $\P^1$ embedded into the $\P^2$ with homogeneous
coordinates $y,x,s$. It represent the Cartan node of the $Sp(1)$ stack.
As it is embedded by a quartic equation, it wraps the $\P^1$ in $\P^2$ twice. The two bits 
are connected by some ``throat'', so they have the topology of a sphere. This throat can, 
however, shrink to produce two separate $\P^1$s. This is exactly what happens over special 
points in the base: \eqref{v0branch} factorizes if it is the sum of two squares. 

The simplest way to determine over which points in the base this happens is the
following: Over the locus where the Cartan $\P^1$ of $Sp(1)$ degenerates into two 
$\P^1$s, \eqref{v0branch}, considered as a hypersurface in the $\P^2$ with homogeneous 
coordinates $X,y,\sigma$, becomes singular. The gradient of \eqref{v0branch} is
\begin{align}\label{v0branchgrad}
 2ydy=&0\nn\\
\left(\tfrac12 X b_{4,1} +2s (\tfrac14 a_{3,1}^2 +a_{6,2})\right)d s= & 0\nn \\
\left(\tfrac12 s b_{4,1}+\tfrac12 b_2 X\right)dX = & 0 \, . 
\end{align}
Note that solving those three equations implies \eqref{v0branch}. The last two equations 
can be rewritten as
\begin{equation}\label{v0branchgradmatrix}
 \left(\begin{array}{cc}
                  b_{4,1} &  a_{3,1}^2 +4a_{6,2} \\ 
	  b_2  &   b_{4,1}   \end{array} \right) \left(\begin{array}{c}X\\ s \end{array} 
 \right)=0\, .
\end{equation}
Hence all three equations in \eqref{v0branchgrad} can only have a simultaneous 
solution if 
\begin{equation}
 \det  \left(\begin{array}{cc}
                  b_{4,1} &  a_{3,1}^2 +4a_{6,2} \\ 
	  b_2  &   b_{4,1}   \end{array} \right) = b_{4,1}^2-b_2(a_{3,1}^2+4a_{6,2})= 0 \, .
\end{equation}
The singularity occurs at $y=0$ and the $X,\sigma$ that solve \eqref{v0branchgradmatrix}.
In summary, we have a $\P^1$, which we call $S$, over generic points of $P$. 
This $\P^1$ is split into two $\P^1$s over $b_{4,1}^2-b_2(a_{3,1}^2+4a_{6,2})= 0$. 
Denoting these by $S_1^a$ and $S_1^b$, it must be in homology that $S=S_1^a+S_1^b$.

The discriminant locus of the configuration we have considered is
\begin{equation}
 \Delta=P^2\left(\tfrac{1}{64}b_2^2(b_2 b_{6,2}-b_{4,1})+P(-\tfrac{1}{16}b_2b_{4,1}b_{6,2}+\tfrac12 b_{4,1}^3+
P\tfrac{27}{16}b_{6,2}^2) \right) \, .
\end{equation}
The intersection between the curve $P=0$ and the remaining part hence has two branches:
\begin{equation}
b_2^2(b_2 b_{6,2}-b_{4,1})=0 \, .
\end{equation}
As we have seen by resolving the $Sp(1)$ singularity, only the second branch leads to an enhancement of
the singularity. Consequently, there is only charged matter at this branch.


\begin{thebibliography}{10%
0}

\bibitem{Beasley:2008dc}
C.~Beasley, J.~J. Heckman, and C.~Vafa, ``{GUTs and Exceptional Branes in
  F-theory - I},'' \href{http://dx.doi.org/10.1088/1126-6708/2009/01/058}{{\em
  JHEP} {\bf 01} (2009)  058},
\href{http://arxiv.org/abs/0802.3391}{{\tt arXiv:0802.3391 [hep-th]}}.

\bibitem{Beasley:2008kw}
C.~Beasley, J.~J. Heckman, and C.~Vafa, ``{GUTs and Exceptional Branes in
  F-theory - II: Experimental Predictions},''
  \href{http://dx.doi.org/10.1088/1126-6708/2009/01/059}{{\em JHEP} {\bf 01}
  (2009)  059},
\href{http://arxiv.org/abs/0806.0102}{{\tt arXiv:0806.0102 [hep-th]}}.

\bibitem{Donagi:2008ca}
R.~Donagi and M.~Wijnholt, ``{Model Building with F-Theory},''
\href{http://arxiv.org/abs/0802.2969}{{\tt arXiv:0802.2969 [hep-th]}}.

\bibitem{Donagi:2008kj}
R.~Donagi and M.~Wijnholt, ``{Breaking GUT Groups in F-Theory},''
\href{http://arxiv.org/abs/0808.2223}{{\tt arXiv:0808.2223 [hep-th]}}.

\bibitem{Donagi:2009ra}
R.~Donagi and M.~Wijnholt, ``{Higgs Bundles and UV Completion in F-Theory},''
\href{http://arxiv.org/abs/0904.1218}{{\tt arXiv:0904.1218 [hep-th]}}.

\bibitem{Hayashi:2008ba}
H.~Hayashi, R.~Tatar, Y.~Toda, T.~Watari, and M.~Yamazaki, ``{New Aspects of
  Heterotic--F Theory Duality},''
  \href{http://dx.doi.org/10.1016/j.nuclphysb.2008.07.031}{{\em Nucl. Phys.}
  {\bf B806} (2009)  224--299},
\href{http://arxiv.org/abs/0805.1057}{{\tt arXiv:0805.1057 [hep-th]}}.

\bibitem{Hayashi:2009ge}
H.~Hayashi, T.~Kawano, R.~Tatar, and T.~Watari, ``{Codimension-3 Singularities
  and Yukawa Couplings in F- theory},''
  \href{http://dx.doi.org/10.1016/j.nuclphysb.2009.07.021}{{\em Nucl. Phys.}
  {\bf B823} (2009)  47--115},
\href{http://arxiv.org/abs/0901.4941}{{\tt arXiv:0901.4941 [hep-th]}}.

\bibitem{Blumenhagen:2008zz}
R.~Blumenhagen, V.~Braun, T.~W. Grimm, and T.~Weigand, ``{GUTs in Type IIB
  Orientifold Compactifications},''
  \href{http://dx.doi.org/10.1016/j.nuclphysb.2009.02.011}{{\em Nucl. Phys.}
  {\bf B815} (2009)  1--94},
\href{http://arxiv.org/abs/0811.2936}{{\tt arXiv:0811.2936 [hep-th]}}.

\bibitem{Blumenhagen:2009yv}
R.~Blumenhagen, T.~W. Grimm, B.~Jurke, and T.~Weigand, ``{Global F-theory
  GUTs},'' \href{http://dx.doi.org/10.1016/j.nuclphysb.2009.12.013}{{\em Nucl.
  Phys.} {\bf B829} (2010)  325--369},
\href{http://arxiv.org/abs/0908.1784}{{\tt arXiv:0908.1784 [hep-th]}}.

\bibitem{Giddings:2001yu}
S.~B. Giddings, S.~Kachru, and J.~Polchinski, ``{Hierarchies from fluxes in
  string compactifications},''
  \href{http://dx.doi.org/10.1103/PhysRevD.66.106006}{{\em Phys. Rev.} {\bf
  D66} (2002)  106006},
\href{http://arxiv.org/abs/hep-th/0105097}{{\tt arXiv:hep-th/0105097}}.

\bibitem{Dasgupta:1999ss}
K.~Dasgupta, G.~Rajesh, and S.~Sethi, ``{M theory, orientifolds and G-flux},''
  {\em JHEP} {\bf 08} (1999)  023,
\href{http://arxiv.org/abs/hep-th/9908088}{{\tt arXiv:hep-th/9908088}}.

\bibitem{Vafa:1996xn}
C.~Vafa, ``{Evidence for F-Theory},''
  \href{http://dx.doi.org/10.1016/0550-3213(96)00172-1}{{\em Nucl. Phys.} {\bf
  B469} (1996)  403--418},
\href{http://arxiv.org/abs/hep-th/9602022}{{\tt arXiv:hep-th/9602022}}.

\bibitem{Sen:1996vd}
A.~Sen, ``{F theory and orientifolds},''
  \href{http://dx.doi.org/10.1016/0550-3213(96)00347-1}{{\em Nucl.Phys.} {\bf
  B475} (1996)  562--578},
\href{http://arxiv.org/abs/hep-th/9605150}{{\tt arXiv:hep-th/9605150
  [hep-th]}}.

\bibitem{Denef:2008wq}
F.~Denef, ``{Les Houches Lectures on Constructing String Vacua},''
\href{http://arxiv.org/abs/0803.1194}{{\tt arXiv:0803.1194 [hep-th]}}.

\bibitem{Weigand:2010wm}
T.~Weigand, ``{Lectures on F-theory compactifications and model building},''
  \href{http://dx.doi.org/10.1088/0264-9381/27/21/214004}{{\em Class. Quant.
  Grav.} {\bf 27} (2010)  214004},
\href{http://arxiv.org/abs/1009.3497}{{\tt arXiv:1009.3497 [hep-th]}}.

\bibitem{Blumenhagen:2010at}
R.~Blumenhagen, ``{Basics of F-theory from the Type IIB Perspective},''
  \href{http://dx.doi.org/10.1002/prop.201000030}{{\em Fortsch. Phys.} {\bf 58}
  (2010)  820--826},
\href{http://arxiv.org/abs/1002.2836}{{\tt arXiv:1002.2836 [hep-th]}}.

\bibitem{Braun:2010ff}
A.~P. Braun, ``{F-Theory and the Landscape of Intersecting D7-Branes},''
\href{http://arxiv.org/abs/1003.4867}{{\tt arXiv:1003.4867 [hep-th]}}.

\bibitem{Ganor:1996pe}
O.~J. Ganor, ``{A note on zeroes of superpotentials in F-theory},''
  \href{http://dx.doi.org/10.1016/S0550-3213(97)00311-8}{{\em Nucl. Phys.} {\bf
  B499} (1997)  55--66},
\href{http://arxiv.org/abs/hep-th/9612077}{{\tt arXiv:hep-th/9612077}}.

\bibitem{Blumenhagen:2006xt}
R.~Blumenhagen, M.~Cvetic, and T.~Weigand, ``{Spacetime instanton corrections
  in 4D string vacua - the seesaw mechanism for D-brane models},''
  \href{http://dx.doi.org/10.1016/j.nuclphysb.2007.02.016}{{\em Nucl. Phys.}
  {\bf B771} (2007)  113--142},
\href{http://arxiv.org/abs/hep-th/0609191}{{\tt arXiv:hep-th/0609191}}.

\bibitem{Ibanez:2006da}
L.~E. Ibanez and A.~M. Uranga, ``{Neutrino Majorana masses from string theory
  instanton effects},''
  \href{http://dx.doi.org/10.1088/1126-6708/2007/03/052}{{\em JHEP} {\bf 03}
  (2007)  052},
\href{http://arxiv.org/abs/hep-th/0609213}{{\tt arXiv:hep-th/0609213}}.

\bibitem{Florea:2006si}
B.~Florea, S.~Kachru, J.~McGreevy, and N.~Saulina, ``{Stringy Instantons and
  Quiver Gauge Theories},''
  \href{http://dx.doi.org/10.1088/1126-6708/2007/05/024}{{\em JHEP} {\bf 05}
  (2007)  024},
\href{http://arxiv.org/abs/hep-th/0610003}{{\tt arXiv:hep-th/0610003}}.

\bibitem{Argurio:2007qk}
R.~Argurio, M.~Bertolini, S.~Franco, and S.~Kachru, ``{Metastable vacua and
  D-branes at the conifold},''
  \href{http://dx.doi.org/10.1088/1126-6708/2007/06/017}{{\em JHEP} {\bf 06}
  (2007)  017},
\href{http://arxiv.org/abs/hep-th/0703236}{{\tt arXiv:hep-th/0703236}}.

\bibitem{Argurio:2007vqa}
R.~Argurio, M.~Bertolini, G.~Ferretti, A.~Lerda, and C.~Petersson, ``{Stringy
  Instantons at Orbifold Singularities},''
  \href{http://dx.doi.org/10.1088/1126-6708/2007/06/067}{{\em JHEP} {\bf 06}
  (2007)  067},
\href{http://arxiv.org/abs/0704.0262}{{\tt arXiv:0704.0262 [hep-th]}}.

\bibitem{Blumenhagen:2007sm}
R.~Blumenhagen, S.~Moster, and E.~Plauschinn, ``{Moduli Stabilisation versus
  Chirality for MSSM like Type IIB Orientifolds},''
  \href{http://dx.doi.org/10.1088/1126-6708/2008/01/058}{{\em JHEP} {\bf 01}
  (2008)  058},
\href{http://arxiv.org/abs/0711.3389}{{\tt arXiv:0711.3389 [hep-th]}}.

\bibitem{Blumenhagen:2010ja}
R.~Blumenhagen, A.~Collinucci, and B.~Jurke, ``{On Instanton Effects in
  F-theory},'' \href{http://dx.doi.org/10.1007/JHEP08(2010)079}{{\em JHEP} {\bf
  08} (2010)  079},
\href{http://arxiv.org/abs/1002.1894}{{\tt arXiv:1002.1894 [hep-th]}}.

\bibitem{Cvetic:2011gp}
M.~Cvetic, J.~Halverson, and I.~Garcia-Etxebarria, ``{Three Looks at Instantons
  in F-theory -- New Insights from Anomaly Inflow, String Junctions and
  Heterotic Duality},''
\href{http://arxiv.org/abs/1107.2388}{{\tt arXiv:1107.2388 [hep-th]}}.

\bibitem{Bianchi:2011qh}
M.~Bianchi, A.~Collinucci, and L.~Martucci, ``{Magnetized E3-brane instantons
  in F-theory},''
\href{http://arxiv.org/abs/1107.3732}{{\tt arXiv:1107.3732 [hep-th]}}.

\bibitem{Becker:1996gj}
K.~Becker and M.~Becker, ``{M-Theory on Eight-Manifolds},''
  \href{http://dx.doi.org/10.1016/0550-3213(96)00367-7}{{\em Nucl. Phys.} {\bf
  B477} (1996)  155--167},
\href{http://arxiv.org/abs/hep-th/9605053}{{\tt arXiv:hep-th/9605053}}.

\bibitem{Haack:2001jz}
M.~Haack and J.~Louis, ``{M-theory compactified on Calabi-Yau fourfolds with
  background flux},''
  \href{http://dx.doi.org/10.1016/S0370-2693(01)00464-6}{{\em Phys. Lett.} {\bf
  B507} (2001)  296--304},
\href{http://arxiv.org/abs/hep-th/0103068}{{\tt arXiv:hep-th/0103068}}.

\bibitem{Grimm:2010ks}
T.~W. Grimm, ``{The N=1 effective action of F-theory compactifications},''
  \href{http://dx.doi.org/10.1016/j.nuclphysb.2010.11.018}{{\em Nucl. Phys.}
  {\bf B845} (2011)  48--92},
\href{http://arxiv.org/abs/1008.4133}{{\tt arXiv:1008.4133 [hep-th]}}.

\bibitem{Valandro:2008zg}
R.~Valandro, ``{Type IIB Flux Vacua from M-theory via F-theory},''
  \href{http://dx.doi.org/10.1088/1126-6708/2009/03/122}{{\em JHEP} {\bf 03}
  (2009)  122},
\href{http://arxiv.org/abs/0811.2873}{{\tt arXiv:0811.2873 [hep-th]}}.

\bibitem{Braun:2008pz}
A.~P. Braun, A.~Hebecker, C.~Ludeling, and R.~Valandro, ``{Fixing D7 Brane
  Positions by F-Theory Fluxes},''
  \href{http://dx.doi.org/10.1016/j.nuclphysb.2009.02.025}{{\em Nucl. Phys.}
  {\bf B815} (2009)  256--287},
\href{http://arxiv.org/abs/0811.2416}{{\tt arXiv:0811.2416 [hep-th]}}.

\bibitem{Katz:1996xe}
S.~H. Katz and C.~Vafa, ``{Matter from geometry},''
  \href{http://dx.doi.org/10.1016/S0550-3213(97)00280-0}{{\em Nucl. Phys.} {\bf
  B497} (1997)  146--154},
\href{http://arxiv.org/abs/hep-th/9606086}{{\tt arXiv:hep-th/9606086}}.

\bibitem{Heckman:2008es}
J.~J. Heckman, J.~Marsano, N.~Saulina, S.~Schafer-Nameki, and C.~Vafa,
  ``{Instantons and SUSY breaking in F-theory},''
\href{http://arxiv.org/abs/0808.1286}{{\tt arXiv:0808.1286 [hep-th]}}.

\bibitem{Heckman:2008qa}
J.~J. Heckman and C.~Vafa, ``{Flavor Hierarchy From F-theory},''
  \href{http://dx.doi.org/10.1016/j.nuclphysb.2010.05.009}{{\em Nucl. Phys.}
  {\bf B837} (2010)  137--151},
\href{http://arxiv.org/abs/0811.2417}{{\tt arXiv:0811.2417 [hep-th]}}.

\bibitem{Heckman:2009mn}
J.~J. Heckman, A.~Tavanfar, and C.~Vafa, ``{The Point of $E_8$ in F-theory
  GUTs},'' \href{http://dx.doi.org/10.1007/JHEP08(2010)040}{{\em JHEP} {\bf 08}
  (2010)  040},
\href{http://arxiv.org/abs/0906.0581}{{\tt arXiv:0906.0581 [hep-th]}}.

\bibitem{Marsano:2009gv}
J.~Marsano, N.~Saulina, and S.~Schafer-Nameki, ``{Monodromies, Fluxes, and
  Compact Three-Generation F-theory GUTs},''
  \href{http://dx.doi.org/10.1088/1126-6708/2009/08/046}{{\em JHEP} {\bf 08}
  (2009)  046},
\href{http://arxiv.org/abs/0906.4672}{{\tt arXiv:0906.4672 [hep-th]}}.

\bibitem{Cecotti:2009zf}
S.~Cecotti, M.~C.~N. Cheng, J.~J. Heckman, and C.~Vafa, ``{Yukawa Couplings in
  F-theory and Non-Commutative Geometry},''
\href{http://arxiv.org/abs/0910.0477}{{\tt arXiv:0910.0477 [hep-th]}}.

\bibitem{Hayashi:2009bt}
H.~Hayashi, T.~Kawano, Y.~Tsuchiya, and T.~Watari, ``{Flavor Structure in
  F-theory Compactifications},''
  \href{http://dx.doi.org/10.1007/JHEP08(2010)036}{{\em JHEP} {\bf 08} (2010)
  036},
\href{http://arxiv.org/abs/0910.2762}{{\tt arXiv:0910.2762 [hep-th]}}.

\bibitem{Tatar:2009jk}
R.~Tatar, Y.~Tsuchiya, and T.~Watari, ``{Right-handed Neutrinos in F-theory
  Compactifications},''
  \href{http://dx.doi.org/10.1016/j.nuclphysb.2009.07.020}{{\em Nucl. Phys.}
  {\bf B823} (2009)  1--46},
\href{http://arxiv.org/abs/0905.2289}{{\tt arXiv:0905.2289 [hep-th]}}.

\bibitem{Hayashi:2010zp}
H.~Hayashi, T.~Kawano, Y.~Tsuchiya, and T.~Watari, ``{More on Dimension-4
  Proton Decay Problem in F-theory -- Spectral Surface, Discriminant Locus and
  Monodromy},'' \href{http://dx.doi.org/10.1016/j.nuclphysb.2010.07.011}{{\em
  Nucl. Phys.} {\bf B840} (2010)  304--348},
\href{http://arxiv.org/abs/1004.3870}{{\tt arXiv:1004.3870 [hep-th]}}.

\bibitem{Dudas:2009hu}
E.~Dudas and E.~Palti, ``{Froggatt-Nielsen models from E8 in F-theory GUTs},''
  \href{http://dx.doi.org/10.1007/JHEP01(2010)127}{{\em JHEP} {\bf 01} (2010)
  127},
\href{http://arxiv.org/abs/0912.0853}{{\tt arXiv:0912.0853 [hep-th]}}.

\bibitem{Dudas:2010zb}
E.~Dudas and E.~Palti, ``{On hypercharge flux and exotics in F-theory GUTs},''
  \href{http://dx.doi.org/10.1007/JHEP09(2010)013}{{\em JHEP} {\bf 09} (2010)
  013},
\href{http://arxiv.org/abs/1007.1297}{{\tt arXiv:1007.1297 [hep-ph]}}.

\bibitem{Conlon:2009qq}
J.~P. Conlon and E.~Palti, ``{Aspects of Flavour and Supersymmetry in F-theory
  GUTs},'' \href{http://dx.doi.org/10.1007/JHEP01(2010)029}{{\em JHEP} {\bf 01}
  (2010)  029},
\href{http://arxiv.org/abs/0910.2413}{{\tt arXiv:0910.2413 [hep-th]}}.

\bibitem{Font:2008id}
A.~Font and L.~E. Ibanez, ``{Yukawa Structure from U(1) Fluxes in F-theory
  Grand Unification},''
  \href{http://dx.doi.org/10.1088/1126-6708/2009/02/016}{{\em JHEP} {\bf 02}
  (2009)  016},
\href{http://arxiv.org/abs/0811.2157}{{\tt arXiv:0811.2157 [hep-th]}}.

\bibitem{Font:2009gq}
A.~Font and L.~E. Ibanez, ``{Matter wave functions and Yukawa couplings in
  F-theory Grand Unification},''
  \href{http://dx.doi.org/10.1088/1126-6708/2009/09/036}{{\em JHEP} {\bf 09}
  (2009)  036},
\href{http://arxiv.org/abs/0907.4895}{{\tt arXiv:0907.4895 [hep-th]}}.

\bibitem{Aparicio:2011jx}
L.~Aparicio, A.~Font, L.~E. Ibanez, and F.~Marchesano, ``{Flux and Instanton
  Effects in Local F-theory Models and Hierarchical Fermion Masses},''
\href{http://arxiv.org/abs/1104.2609}{{\tt arXiv:1104.2609 [hep-th]}}.

\bibitem{Chen:2010tp}
C.-M. Chen and Y.-C. Chung, ``{Flipped SU(5) GUTs from $E_8$ Singularities in
  F-theory},'' \href{http://dx.doi.org/10.1007/JHEP03(2011)049}{{\em JHEP} {\bf
  03} (2011)  049},
\href{http://arxiv.org/abs/1005.5728}{{\tt arXiv:1005.5728 [hep-th]}}.

\bibitem{Dolan:2011iu}
M.~J. Dolan, J.~Marsano, N.~Saulina, and S.~Schafer-Nameki, ``{F-theory GUTs
  with U(1) Symmetries: Generalities and Survey},''
\href{http://arxiv.org/abs/1102.0290}{{\tt arXiv:1102.0290 [hep-th]}}.

\bibitem{Oikonomou:2011ba}
V.~K. Oikonomou, ``{F-theory and the Witten Index},''
  \href{http://dx.doi.org/10.1016/j.nuclphysb.2011.04.021}{{\em Nucl. Phys.}
  {\bf B850} (2011)  273--286},
\href{http://arxiv.org/abs/1103.1289}{{\tt arXiv:1103.1289 [hep-th]}}.

\bibitem{Cordova:2009fg}
C.~Cordova, ``{Decoupling Gravity in F-Theory},''
\href{http://arxiv.org/abs/0910.2955}{{\tt arXiv:0910.2955 [hep-th]}}.

\bibitem{Ludeling:2011en}
C.~Ludeling, H.~P. Nilles, and C.~C. Stephan, ``{The Potential Fate of Local
  Model Building},'' \href{http://dx.doi.org/10.1103/PhysRevD.83.086008}{{\em
  Phys. Rev.} {\bf D83} (2011)  086008},
\href{http://arxiv.org/abs/1101.3346}{{\tt arXiv:1101.3346 [hep-th]}}.

\bibitem{Marsano:2009ym}
J.~Marsano, N.~Saulina, and S.~Schafer-Nameki, ``{F-theory Compactifications
  for Supersymmetric GUTs},''
  \href{http://dx.doi.org/10.1088/1126-6708/2009/08/030}{{\em JHEP} {\bf 08}
  (2009)  030},
\href{http://arxiv.org/abs/0904.3932}{{\tt arXiv:0904.3932 [hep-th]}}.

\bibitem{Marsano:2009wr}
J.~Marsano, N.~Saulina, and S.~Schafer-Nameki, ``{Compact F-theory GUTs with
  $U(1)_PQ$},'' \href{http://dx.doi.org/10.1007/JHEP04(2010)095}{{\em JHEP}
  {\bf 04} (2010)  095},
\href{http://arxiv.org/abs/0912.0272}{{\tt arXiv:0912.0272 [hep-th]}}.

\bibitem{Grimm:2009yu}
T.~W. Grimm, S.~Krause, and T.~Weigand, ``{F-Theory GUT Vacua on Compact
  Calabi-Yau Fourfolds},''
  \href{http://dx.doi.org/10.1007/JHEP07(2010)037}{{\em JHEP} {\bf 07} (2010)
  037},
\href{http://arxiv.org/abs/0912.3524}{{\tt arXiv:0912.3524 [hep-th]}}.

\bibitem{Chen:2010ts}
C.-M. Chen, J.~Knapp, M.~Kreuzer, and C.~Mayrhofer, ``{Global SO(10) F-theory
  GUTs},'' \href{http://dx.doi.org/10.1007/JHEP10(2010)057}{{\em JHEP} {\bf 10}
  (2010)  057},
\href{http://arxiv.org/abs/1005.5735}{{\tt arXiv:1005.5735 [hep-th]}}.

\bibitem{Chung:2010bn}
Y.-C. Chung, ``{On Global Flipped SU(5) GUTs in F-theory},''
  \href{http://dx.doi.org/10.1007/JHEP03(2011)126}{{\em JHEP} {\bf 03} (2011)
  126},
\href{http://arxiv.org/abs/1008.2506}{{\tt arXiv:1008.2506 [hep-th]}}.

\bibitem{Chen:2010tg}
C.-M. Chen and Y.-C. Chung, ``{On F-theory $E_6$ GUTs},''
  \href{http://dx.doi.org/10.1007/JHEP03(2011)129}{{\em JHEP} {\bf 03} (2011)
  129},
\href{http://arxiv.org/abs/1010.5536}{{\tt arXiv:1010.5536 [hep-th]}}.

\bibitem{Cvetic:2010rq}
M.~Cvetic, I.~Garcia-Etxebarria, and J.~Halverson, ``{Global F-theory Models:
  Instantons and Gauge Dynamics},''
  \href{http://dx.doi.org/10.1007/JHEP01(2011)073}{{\em JHEP} {\bf 01} (2011)
  073},
\href{http://arxiv.org/abs/1003.5337}{{\tt arXiv:1003.5337 [hep-th]}}.

\bibitem{Knapp:2011wk}
J.~Knapp, M.~Kreuzer, C.~Mayrhofer, and N.-O. Walliser, ``{Toric Construction
  of Global F-Theory GUTs},''
  \href{http://dx.doi.org/10.1007/JHEP03(2011)138}{{\em JHEP} {\bf 03} (2011)
  138},
\href{http://arxiv.org/abs/1101.4908}{{\tt arXiv:1101.4908 [hep-th]}}.

\bibitem{Knapp:2011ip}
J.~Knapp and M.~Kreuzer, ``{Toric Methods in F-theory Model Building},''
\href{http://arxiv.org/abs/1103.3358}{{\tt arXiv:1103.3358 [hep-th]}}.

\bibitem{Friedman:1997ih}
R.~Friedman, J.~W. Morgan, and E.~Witten, ``{Vector bundles over elliptic
  fibrations},''
\href{http://arxiv.org/abs/alg-geom/9709029}{{\tt arXiv:alg-geom/9709029}}.

\bibitem{Curio:1998bva}
G.~Curio and R.~Y. Donagi, ``{Moduli in N = 1 heterotic/F-theory duality},''
  \href{http://dx.doi.org/10.1016/S0550-3213(98)00185-0}{{\em Nucl. Phys.} {\bf
  B518} (1998)  603--631},
\href{http://arxiv.org/abs/hep-th/9801057}{{\tt arXiv:hep-th/9801057}}.

\bibitem{Marsano:2010ix}
J.~Marsano, N.~Saulina, and S.~Schafer-Nameki, ``{A Note on G-Fluxes for
  F-theory Model Building},''
  \href{http://dx.doi.org/10.1007/JHEP11(2010)088}{{\em JHEP} {\bf 11} (2010)
  088},
\href{http://arxiv.org/abs/1006.0483}{{\tt arXiv:1006.0483 [hep-th]}}.

\bibitem{Marsano:2011nn}
J.~Marsano, N.~Saulina, and S.~Schafer-Nameki, ``{On G-flux, M5 instantons, and
  U(1)s in F-theory},''
\href{http://arxiv.org/abs/1107.1718}{{\tt arXiv:1107.1718 [hep-th]}}.

\bibitem{Grimm:2011tb}
T.~W. Grimm, M.~Kerstan, E.~Palti, and T.~Weigand, ``{Massive Abelian Gauge
  Symmetries and Fluxes in F-theory},''
\href{http://arxiv.org/abs/1107.3842}{{\tt arXiv:1107.3842 [hep-th]}}.

\bibitem{Grimm:2010ez}
T.~W. Grimm and T.~Weigand, ``{On Abelian Gauge Symmetries and Proton Decay in
  Global F- theory GUTs},''
  \href{http://dx.doi.org/10.1103/PhysRevD.82.086009}{{\em Phys. Rev.} {\bf
  D82} (2010)  086009},
\href{http://arxiv.org/abs/1006.0226}{{\tt arXiv:1006.0226 [hep-th]}}.

\bibitem{Sen:1997kw}
A.~Sen, ``{F theory and the Gimon-Polchinski orientifold},''
  \href{http://dx.doi.org/10.1016/S0550-3213(97)00262-9}{{\em Nucl.Phys.} {\bf
  B498} (1997)  135--155},
\href{http://arxiv.org/abs/hep-th/9702061}{{\tt arXiv:hep-th/9702061
  [hep-th]}}.

\bibitem{Sen:1997gv}
A.~Sen, ``{Orientifold limit of F-theory vacua},''
  \href{http://dx.doi.org/10.1103/PhysRevD.55.R7345}{{\em Phys. Rev.} {\bf D55}
  (1997)  7345--7349},
\href{http://arxiv.org/abs/hep-th/9702165}{{\tt arXiv:hep-th/9702165}}.

\bibitem{Collinucci:2008pf}
A.~Collinucci, F.~Denef, and M.~Esole, ``{D-brane Deconstructions in IIB
  Orientifolds},'' \href{http://dx.doi.org/10.1088/1126-6708/2009/02/005}{{\em
  JHEP} {\bf 02} (2009)  005},
\href{http://arxiv.org/abs/0805.1573}{{\tt arXiv:0805.1573 [hep-th]}}.

\bibitem{Braun:2009wh}
A.~P. Braun, R.~Ebert, A.~Hebecker, and R.~Valandro, ``{Weierstrass meets
  Enriques},'' \href{http://dx.doi.org/10.1007/JHEP02(2010)077}{{\em JHEP} {\bf
  1002} (2010)  077},
\href{http://arxiv.org/abs/0907.2691}{{\tt arXiv:0907.2691 [hep-th]}}.

\bibitem{Aldazabal:1996du}
G.~Aldazabal, A.~Font, L.~E. Ibanez, and A.~M. Uranga, ``{New branches of
  string compactifications and their F- theory duals},''
  \href{http://dx.doi.org/10.1016/S0550-3213(96)00699-2}{{\em Nucl. Phys.} {\bf
  B492} (1997)  119--151},
\href{http://arxiv.org/abs/hep-th/9607121}{{\tt arXiv:hep-th/9607121}}.

\bibitem{Klemm:1996hh}
A.~Klemm, P.~Mayr, and C.~Vafa, ``{BPS states of exceptional non-critical
  strings},''
\href{http://arxiv.org/abs/hep-th/9607139}{{\tt arXiv:hep-th/9607139}}.

\bibitem{Klemm:1996ts}
A.~Klemm, B.~Lian, S.~S. Roan, and S.-T. Yau, ``{Calabi-Yau fourfolds for M-
  and F-theory compactifications},''
  \href{http://dx.doi.org/10.1016/S0550-3213(97)00798-0}{{\em Nucl. Phys.} {\bf
  B518} (1998)  515--574},
\href{http://arxiv.org/abs/hep-th/9701023}{{\tt arXiv:hep-th/9701023}}.

\bibitem{Candelas:1997pv}
P.~Candelas, E.~Perevalov, and G.~Rajesh, ``{Comments on A, B, C chains of
  heterotic and type II vacua},''
  \href{http://dx.doi.org/10.1016/S0550-3213(97)00374-X}{{\em Nucl. Phys.} {\bf
  B502} (1997)  594--612},
\href{http://arxiv.org/abs/hep-th/9703148}{{\tt arXiv:hep-th/9703148}}.

\bibitem{Bershadsky:1998vn}
M.~Bershadsky, T.~Pantev, and V.~Sadov, ``{F-theory with quantized fluxes},''
  {\em Adv. Theor. Math. Phys.} {\bf 3} (1999)  727--773,
\href{http://arxiv.org/abs/hep-th/9805056}{{\tt arXiv:hep-th/9805056}}.

\bibitem{Berglund:1998va}
P.~Berglund, A.~Klemm, P.~Mayr, and S.~Theisen, ``{On type IIB vacua with
  varying coupling constant},''
  \href{http://dx.doi.org/10.1016/S0550-3213(99)00420-4}{{\em Nucl. Phys.} {\bf
  B558} (1999)  178--204},
\href{http://arxiv.org/abs/hep-th/9805189}{{\tt arXiv:hep-th/9805189}}.

\bibitem{Lust:2005bd}
D.~Lust, P.~Mayr, S.~Reffert, and S.~Stieberger, ``{F-theory flux,
  destabilization of orientifolds and soft terms on D7-branes},''
  \href{http://dx.doi.org/10.1016/j.nuclphysb.2005.09.011}{{\em Nucl. Phys.}
  {\bf B732} (2006)  243--290},
\href{http://arxiv.org/abs/hep-th/0501139}{{\tt arXiv:hep-th/0501139}}.

\bibitem{bht08}
A.~Braun, A.~Hebecker, and H.~Triendl, ``{D7-Brane Motion from M-Theory Cycles
  and Obstructions in the Weak Coupling Limit},''
  \href{http://dx.doi.org/10.1016/j.nuclphysb.2008.03.021}{{\em Nucl.Phys.}
  {\bf B800} (2008)  298--329},
\href{http://arxiv.org/abs/0801.2163}{{\tt arXiv:0801.2163 [hep-th]}}.

\bibitem{Greene:1993vm}
B.~R. Greene, D.~R. Morrison, and M.~R. Plesser, ``{Mirror manifolds in higher
  dimension},'' \href{http://dx.doi.org/10.1007/BF02101657}{{\em Commun. Math.
  Phys.} {\bf 173} (1995)  559--598},
\href{http://arxiv.org/abs/hep-th/9402119}{{\tt arXiv:hep-th/9402119}}.

\bibitem{Strominger:1990pd}
A.~Strominger, ``{SPECIAL GEOMETRY},''
\href{http://dx.doi.org/10.1007/BF02096559}{{\em Commun. Math. Phys.} {\bf 133}
  (1990)  163--180}.

\bibitem{Esole:2011sm}
M.~Esole and S.-T. Yau, ``{Small resolutions of SU(5)-models in F-theory},''
\href{http://arxiv.org/abs/1107.0733}{{\tt arXiv:1107.0733 [hep-th]}}.

\bibitem{Collinucci:2010gz}
A.~Collinucci and R.~Savelli, ``{On Flux Quantization in F-Theory},''
\href{http://arxiv.org/abs/1011.6388}{{\tt arXiv:1011.6388 [hep-th]}}.

\bibitem{Collinucci:2008zs}
A.~Collinucci, ``{New F-theory lifts},''
  \href{http://dx.doi.org/10.1088/1126-6708/2009/08/076}{{\em JHEP} {\bf 08}
  (2009)  076},
\href{http://arxiv.org/abs/0812.0175}{{\tt arXiv:0812.0175 [hep-th]}}.

\bibitem{Collinucci:2009uh}
A.~Collinucci, ``{New F-theory lifts II: Permutation orientifolds and enhanced
  singularities},'' \href{http://dx.doi.org/10.1007/JHEP04(2010)076}{{\em JHEP}
  {\bf 04} (2010)  076},
\href{http://arxiv.org/abs/0906.0003}{{\tt arXiv:0906.0003 [hep-th]}}.

\bibitem{Brunner:2003zm}
I.~Brunner and K.~Hori, ``{Orientifolds and mirror symmetry},''
  \href{http://dx.doi.org/10.1088/1126-6708/2004/11/005}{{\em JHEP} {\bf 11}
  (2004)  005},
\href{http://arxiv.org/abs/hep-th/0303135}{{\tt arXiv:hep-th/0303135}}.

\bibitem{Martucci:2006ij}
L.~Martucci, ``{D-branes on general N = 1 backgrounds: Superpotentials and
  D-terms},'' \href{http://dx.doi.org/10.1088/1126-6708/2006/06/033}{{\em JHEP}
  {\bf 06} (2006)  033},
\href{http://arxiv.org/abs/hep-th/0602129}{{\tt arXiv:hep-th/0602129}}.

\bibitem{Martucci:2005ht}
L.~Martucci and P.~Smyth, ``{Supersymmetric D-branes and calibrations on
  general N = 1 backgrounds},''
  \href{http://dx.doi.org/10.1088/1126-6708/2005/11/048}{{\em JHEP} {\bf 11}
  (2005)  048},
\href{http://arxiv.org/abs/hep-th/0507099}{{\tt arXiv:hep-th/0507099}}.

\bibitem{Harris}
J.~Harris, ``{Algebraic Geometry: A first course},'' {\em Springer Verlag}
  (1992)  .

\bibitem{Beauville}
A.~Beauville, ``{Determinantal hypersurfaces},'' {\em Michigan Math J. Volume 48, Issue 1 (2000), 39-64}.

\bibitem{Achinger}
P.~Achinger, ``{Frobenius Push-Forwards on Quadrics},'' arXiv: 1005.0594v1 [math].


\bibitem{Addington}
N.~Addington, ``{Spinor sheaves on singular quadrics},''
  \href{http://arxiv.org/abs/0904.1766}{{\tt arXiv:0904.1766 [math]}}.

\bibitem{Kapustin:2002bi}
A.~Kapustin and Y.~Li, ``{D-Branes in Landau-Ginzburg Models and Algebraic
  Geometry},'' {\em JHEP} {\bf 12} (2003)  005,
\href{http://arxiv.org/abs/hep-th/0210296}{{\tt arXiv:hep-th/0210296}}.

\bibitem{Brunner:2003dc}
I.~Brunner, M.~Herbst, W.~Lerche, and B.~Scheuner, ``{Landau-Ginzburg
  realization of open string TFT},''
  \href{http://dx.doi.org/10.1088/1126-6708/2006/11/043}{{\em JHEP} {\bf 11}
  (2006)  043},
\href{http://arxiv.org/abs/hep-th/0305133}{{\tt arXiv:hep-th/0305133}}.

\bibitem{Baumgartl:2007an}
M.~Baumgartl, I.~Brunner, and M.~R. Gaberdiel, ``{D-brane superpotentials and
  RG flows on the quintic},''
  \href{http://dx.doi.org/10.1088/1126-6708/2007/07/061}{{\em JHEP} {\bf 07}
  (2007)  061},
\href{http://arxiv.org/abs/0704.2666}{{\tt arXiv:0704.2666 [hep-th]}}.

\bibitem{Herbst:2008jq}
M.~Herbst, K.~Hori, and D.~Page, ``{Phases Of N=2 Theories In 1+1 Dimensions
  With Boundary},''
\href{http://arxiv.org/abs/0803.2045}{{\tt arXiv:0803.2045 [hep-th]}}.

\bibitem{Brunner:2006tc}
I.~Brunner, M.~R. Gaberdiel, and C.~A. Keller, ``{Matrix factorisations and
  D-branes on K3},''
  \href{http://dx.doi.org/10.1088/1126-6708/2006/06/015}{{\em JHEP} {\bf 06}
  (2006)  015},
\href{http://arxiv.org/abs/hep-th/0603196}{{\tt arXiv:hep-th/0603196}}.

\bibitem{Diaconescu:2000wy}
D.-E. Diaconescu, G.~W. Moore, and E.~Witten, ``{E(8) gauge theory, and a
  derivation of K-theory from M- theory},'' {\em Adv. Theor. Math. Phys.} {\bf
  6} (2003)  1031--1134,
\href{http://arxiv.org/abs/hep-th/0005090}{{\tt arXiv:hep-th/0005090}}.

\bibitem{Alim:2009rf}
M.~Alim, M.~Hecht, P.~Mayr, and A.~Mertens, ``{Mirror Symmetry for Toric Branes
  on Compact Hypersurfaces},''
  \href{http://dx.doi.org/10.1088/1126-6708/2009/09/126}{{\em JHEP} {\bf 09}
  (2009)  126},
\href{http://arxiv.org/abs/0901.2937}{{\tt arXiv:0901.2937 [hep-th]}}.

\bibitem{Alim:2009bx}
M.~Alim {\em et al.}, ``{Hints for Off-Shell Mirror Symmetry in type
  II/F-theory Compactifications},''
  \href{http://dx.doi.org/10.1016/j.nuclphysb.2010.06.017}{{\em Nucl. Phys.}
  {\bf B841} (2010)  303--338},
\href{http://arxiv.org/abs/0909.1842}{{\tt arXiv:0909.1842 [hep-th]}}.

\bibitem{Grimm:2009ef}
T.~W. Grimm, T.-W. Ha, A.~Klemm, and D.~Klevers, ``{Computing Brane and Flux
  Superpotentials in F-theory Compactifications},''
  \href{http://dx.doi.org/10.1007/JHEP04(2010)015}{{\em JHEP} {\bf 04} (2010)
  015},
\href{http://arxiv.org/abs/0909.2025}{{\tt arXiv:0909.2025 [hep-th]}}.

\bibitem{Aganagic:2009jq}
M.~Aganagic and C.~Beem, ``{The Geometry of D-Brane Superpotentials},''
\href{http://arxiv.org/abs/0909.2245}{{\tt arXiv:0909.2245 [hep-th]}}.

\bibitem{Grimm:2009sy}
T.~W. Grimm, T.-W. Ha, A.~Klemm, and D.~Klevers, ``{Five-Brane Superpotentials
  and Heterotic/F-theory Duality},''
  \href{http://dx.doi.org/10.1016/j.nuclphysb.2010.06.011}{{\em Nucl. Phys.}
  {\bf B838} (2010)  458--491},
\href{http://arxiv.org/abs/0912.3250}{{\tt arXiv:0912.3250 [hep-th]}}.

\bibitem{Jockers:2009ti}
H.~Jockers, P.~Mayr, and J.~Walcher, ``{On N=1 4d Effective Couplings for
  F-theory and Heterotic Vacua},''
\href{http://arxiv.org/abs/0912.3265}{{\tt arXiv:0912.3265 [hep-th]}}.

\bibitem{Alim:2010za}
M.~Alim {\em et al.}, ``{Type II/F-theory Superpotentials with Several
  Deformations and N=1 Mirror Symmetry},''
  \href{http://dx.doi.org/10.1007/JHEP06(2011)103}{{\em JHEP} {\bf 06} (2011)
  103},
\href{http://arxiv.org/abs/1010.0977}{{\tt arXiv:1010.0977 [hep-th]}}.

\bibitem{DoPa1997}
{Dodson, C. T. J. and Parker, Phillip E.}, {\em A user's guide to algebraic
  topology}, vol.~387 of {\em Mathematics and its Applications}.
\newblock Kluwer Academic Publishers Group, Dordrecht, 1997.

\end{thebibliography}
\begingroup\raggedright\endgroup
\end{document}